\documentclass[10pt,aps,floats,floatfix,amssymb,amsmath,prb,nofootinbib,twocolumn,longbibliography]{revtex4-2}

\usepackage[caption=false]{subfig}
\usepackage{physics}
\usepackage{psfrag}
\usepackage{graphicx}
\usepackage{color}
\usepackage{esint} 
\usepackage[dvipsnames]{xcolor}
\usepackage[unicode=true,pdfusetitle,
 bookmarks=true,bookmarksnumbered=false,bookmarksopen=false,
 breaklinks=false,pdfborder={0 0 0},backref=false,colorlinks=true,
 allcolors=blue]
 {hyperref}
\usepackage{geometry}
\usepackage{comment}
\usepackage{subfig}
\newcommand{\poc}{\hspace*{8pt}}

\newcommand{\nph}{n_\mathrm{ph}}

\begin{document}
\title{Applicability of the cumulant expansion method for the calculation of transport properties in electron-phonon systems}

\author{Petar Mitri\'c}
\affiliation{Institute of Physics Belgrade, University of Belgrade, 
Pregrevica 118, 11080 Belgrade, Serbia}
\author{Veljko Jankovi\'c}
\affiliation{Institute of Physics Belgrade, University of Belgrade, 
Pregrevica 118, 11080 Belgrade, Serbia}
\author{Darko Tanaskovi\'c}
\affiliation{Institute of Physics Belgrade, University of Belgrade, 
Pregrevica 118, 11080 Belgrade, Serbia}
\author{Nenad Vukmirovi\'c}
\affiliation{Institute of Physics Belgrade, University of Belgrade, 
Pregrevica 118, 11080 Belgrade, Serbia}

\begin{abstract}
We assess the accuracy of the cumulant expansion (CE) method, combined with the independent-particle approximation (IPA), for calculating charge mobility in electron–phonon systems. As representative testbeds, we consider the Peierls and Fr\"ohlich models, which serve as simplified frameworks where accurate or numerically exact benchmarks are available.  These are used to compare the CE results with those obtained using the Boltzmann formalism, the one-shot Migdal approximation, and its self consistent extension—approaches that are presently the most commonly employed alternatives for transport calculations. Supported by analytical arguments based on spectral sum rules and by our previous results for the Holstein model, we argue that, for weak to moderate coupling strengths and not-too-low temperatures, the CE within the IPA framework yields accurate results. In the case of the Peierls model, the role of vertex corrections is also discussed.
\end{abstract}

\maketitle

\section{\label{sec:intro} Introduction}
%
First-principles calculations of charge carrier mobility in semiconductors with weak electron-phonon interaction have reached predictive power, mostly due to the advances in calculation of momentum-dependent electron-phonon coupling constants \cite{2017_Giustino, 2020_Ponce}. In this regime the mobility calculation is typically performed within the Boltzmann semiclassical approach \cite{2016_Zhou, 2018_Ponce, 2021_Vukmirovic}.
However, its underlying assumptions tend to break down at stronger couplings or higher temperatures, and the calculated mobility can thus strongly deviate from the experimental values \cite{2018_Zhou_PRL}.
The cases in which the Boltzmann approach is believed to break down include various materials where charge transport is governed by small polaron hopping \cite{Banday_2017,Snyder_2000,Adelstein_2014,Yildiz_2009}, perovskite oxides where charges must be treated beyond the quasiparticle picture \cite{2018_Zhou_PRL,Quan_2024}, and crystalline small-molecule organic semiconductors \cite{2006_Troisi}.  The limitations of the Boltzmann approach have been specifically demonstrated by recent calculations of spectral functions in real materials and the subsequent comparison of calculated mobilities to Boltzmann-derived results \cite{2025_Lihm_nonperturbative,2025_Lihm}.
In these cases, a significant improvement can be achieved by using the independent-particle approximation (IPA) within the Kubo formalism \cite{1957_Kubo}, combined with an appropriate method for calculating the single-particle Green’s function. One such choice is the one-shot Migdal approximation (MA), which retains only the lowest-order electron–phonon diagram, while 
higher-order effects can also be captured in the self-consistent version of this method (SCMA) \cite{2025_Lihm_nonperturbative}. 

A promising alternative, 
which has recently become more popular in the literature, is the so-called cumulant expansion (CE) method \cite{2019_Zhu, 2022_Chang, 2014_Kas, 2022_Robinson_1, 2022_Robinson_2, 2014_Lischner, 2016_Gumhalter}: it has been reported that this method can include effects beyond the lowest order perturbation theory (i.e., MA) and simple quasiparticle picture, while--unlike SCMA--it avoids the computational cost of self-consistent iterations \cite{2018_Nery, 2020_Antonius, 2014_Story, 2022_Kandolf}. However, there have also been reports of unphysical features appearing in CE predictions \cite{2025_Lihm_nonperturbative}. In addition, the IPA entirely neglects vertex corrections to mobility, whose contributions are difficult to estimate a priori and even harder to compute exactly. It is therefore essential to carefully assess both the advantages and limitations of CE, as well as its range of applicability.

An effective strategy for such an assessment is to apply CE to well-controlled, simplified model systems \cite{1959_Holstein, 1979_SSH, 1954_Frolih}, where accurate solutions are available and can serve as benchmarks \cite{2006_Weiss, 2005_Schubert, 2020_Li, 2022_Ge, 2022_Jansen, 1994_Jaklic, 2000_Jaklic, 2013_Prelovsek, 2023_Miladic, 2015_Mishchenko, 2024_QC_Mitric, 2025_Mitric, 2025_Miladic, 2022_Wang, 2021_Li_nature}. This is precisely the approach we adopted in our earlier studies, where we evaluated the accuracy of the CE method \cite{2023_Mitric} and analyzed the role of vertex corrections \cite{2024_Jankovic} within the Holstein model \cite{1959_Holstein}.
 This was made possible by the availability of several highly accurate approximate and entirely real-axis numerically exact methods for this model \cite{2022_Mitric, 2022_Jankovic, 2023_Jankovic}. We found that, despite its simplicity, the CE method provided surprisingly accurate results for quasiparticle properties, spectral functions, and mobilities—not only at weak coupling, but also at intermediate strengths. Furthermore, in these regimes, vertex corrections were found to be relatively small \cite{2024_Jankovic}. However, it should be emphasized that the Holstein model assumes local (i.e., momentum-independent) electron-phonon coupling—a simplification that, while useful for gaining insight, limits the model’s physical realism. This highlights the importance of testing CE in more realistic models with momentum-dependent interactions.

In this work, we apply the CE to systems with nonlocal electron–phonon coupling, focusing on the Peierls \cite{1979_SSH} and Fröhlich models \cite{1954_Frolih}. For the Peierls model, we compare the CE mobility predictions with MA, SCMA, Boltzmann, as well as the numerically exact  hierarchical equations of motion (HEOM) results that have become available only recently \cite{2025_Jankovic_I, 2025_Jankovic_II}.
Moreover, the HEOM method for spectral functions developed in Ref.~\cite{2022_Jankovic} enables a systematic comparison between different methods also at the level of single-particle excitations.
These can also be used for the calculation of HEOM IPA mobility results, which, by comparing them with full HEOM results reveal how large 
the contributions of vertex corrections are.
Unfortunately, for the Fröhlich model, numerically exact results following from an entirely real-axis method are currently not available, so we present only the CE, SCMA, and Boltzmann results.
However, in the regimes relevant for this paper we argue that SCMA should provide quite accurate results.
Finally, by combining our findings for the Peierls and Fröhlich models with general analytical considerations based on spectral sum rules, as well as our previous results on the Holstein model \cite{2023_Mitric, 2024_Jankovic}, we draw a model-independent conclusion that CE can in fact be used for accurate mobility predictions, within IPA framework, if the interaction is not too strong, and temperature is not too low. 

The remainder of this paper is organized as follows. Section~\ref{sec:model_method} summarizes the independent-particle approximation and the CE, MA, and SCMA approaches within a general single-band electron–phonon Hamiltonian with one optical phonon branch. Numerical details, regarding method implementation, specific to the Peierls and Fröhlich models are given in Sec.~\ref{sec:num_implementation}.
Our main results are presented in Sec.~\ref{sec:results}: Secs.~\ref{sec:peierls_model-53232} and~\ref{sec:frohlich_model-53232} discuss the Peierls and Fröhlich cases, respectively, while Sec.~\ref{sec:spectral_sum_rules} provides additional analytical insights based on spectral sum rules. A discussion and concluding remarks are given in Sec.~\ref{Sec:discussion}. Additional mobility data and spectral functions are included in the Supplementary Material (SM) \cite{SuppMat}.

\section{\label{sec:model_method} {Theoretical framework behind the} CE in Electron–Phonon Systems}
\subsection{Hamiltonian} \label{sec:hamiltonian}
\vspace*{-0.3cm}
\poc
The Hamiltonian describing an electron in a single band linearly coupled to a dispersionless optical phonon mode of frequency $\omega_0$, defined on a lattice with $N$ sites (or, equivalently, in the continuum with system volume $V$, with the correspondence $N \leftrightarrow V$), can be written in the general form
\begin{align} \label{eq:1}
    H &= \sum_{\bf k} \varepsilon_{\bf k} c_{\bf k}^\dagger c_{\bf k}
    + \frac{1}{\sqrt{N}} \sum_{{\bf k}, {\bf q}} g_{{\bf k}, {\bf q}} \, c_{{\bf k+q}}^\dagger c_{\bf k}
    X_{\bf q} + \omega_0 \sum_{\bf q}   a_{\bf q}^\dagger a_{\bf q}.
\end{align}
%
Here, $c_{\bf k}$ and $a_{\bf k}$ are electron and phonon annihilation operators, $\varepsilon_{\bf k}$ is the noninteracting electron dispersion, $g_{\bf k,q}$ is the electron-phonon coupling satisfying $g_{\bf{k},\bf{q}}^*=g_{\bf{k}+\bf{q},-\bf{q}}$, and $X_{\bf q} = a_{\bf q} + a^{\dagger}_{\bf -q}$.  We set $\hbar$, $k_B$, and elementary charge to unity. To model a weakly doped semiconductor, we focus on the thermodynamic limit $N \to \infty$ together with $\mu_F \to -\infty$, where $\mu_F$ is the chemical potential.

\subsection{Single particle properties in the limit $\mu_F \to -\infty$}

Following Refs.~\cite{2024_Jankovic, 2022_Mitric}, to ensure that the spectral weight is nonzero at finite frequencies, the electron single-particle quantities—such as the spectral function $\tilde{A}_{\bf k}(\omega)$, the Green’s function $\tilde{G}_{\bf k}(\omega)$, and the self-energy $\tilde{\Sigma}_{\bf k}(\omega)$—must be redefined by shifting them by $\mu_F$: $A_{\bf k}(\omega) = \tilde{A}_{\bf k}(\omega- \mu_F)$, $G_{\bf k}(\omega) = \tilde{G}_{\bf k}(\omega- \mu_F)$, $\Sigma_{\bf k}(\omega) = \tilde{\Sigma}_{\bf k}(\omega- \mu_F)$. Then, elementary single-particle correlation functions can be written as \cite{2024_Jankovic}
\begin{align}
    \langle c_{\bf k}(t) c_{\bf q}^\dagger \rangle &= \delta_{\bf k,q} \, e^{i \mu_F t} 
    \int_{-\infty}^\infty d\omega A_{\bf k}(\omega) e^{-i \omega t}, \label{eq:corrF_1} \\
    \langle c_{\bf k}^\dagger c_{\bf q}(t) \rangle &= \delta_{\bf k,q} \, e^{i \mu_F t} e^{\beta \mu_F}
    \int_{-\infty}^\infty d\omega A_{\bf k}(\omega) e^{-\beta \omega} e^{-i \omega t}, \label{eq:corrF_2}
\end{align}
where $\beta = 1/T$ is the inverse temperature. For instance, the number of electrons can be expressed as
\begin{equation}
    \tilde{n}_e = \sum_{\bf k} \langle c_{\bf k}^\dagger c_{\bf k} \rangle = 
    e^{\beta \mu_F} \sum_{\bf k} \int_{-\infty}^\infty d\omega A_{\bf k}(\omega) e^{-\beta\omega}
    \equiv e^{\beta \mu_F} n_e,
    \label{eq:def_ne_4342}
\end{equation}
where we introduced the convenient quantity $n_e$, which will be useful later.

A significant simplification, occurs also for phononic variables, as it is known that the phonon propagator
\begin{equation}
    iD_{\bf q_2, q_1}(t) = \langle X_{\bf q_2}(t) X_{\bf q_1} \rangle = \delta_{\bf q_1, -q_2} iD(t)
\end{equation}
remains unrenormalized in the limit $\mu_F \to -\infty$. Hence \cite{2000_Mahan}
\begin{equation}
    iD(t) = \nph e^{i \omega_0 t} + (\nph + 1) e^{-i\omega_0 t}, \label{eq:phon_prop_D}
\end{equation}
where $\nph = 1/(e^{\beta \omega_0} - 1)$.

\subsection{Kubo formula and independent particle approximation} 

\begin{figure}[t!]
  \centering
  \includegraphics[width=\linewidth]{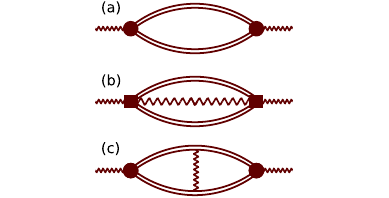}
  \caption{Feynman diagrams for (a) $\mu_e$, (b) $\mu_{ph}$ within the independent-particle approximation, and (c) corrections beyond the independent-particle approximation.}
  \label{fig:dijag}
\end{figure}

According to the Kubo formula, the electron mobility (along $x$ direction) can be calculated as 
\begin{equation} \label{eq:kubo_form}
    \mu = \frac{\beta}{2\tilde{n}_e}
    \int_{-\infty}^\infty \mathrm{Re}\;\langle j_x(t) j_x \rangle \, \mathrm{d}t.
\end{equation}
Here, $j_x$ is the $x$ component of the current operator $\bf j$, which for the Hamiltonian~\eqref{eq:1} reads as
\begin{equation} \label{eq:current}
    {\bf j} = \sum_{\bf k} (\nabla_{\bf k} \varepsilon_{\bf k} ) c_{\bf k}^\dagger c_{\bf k} +
    \frac{1}{\sqrt{N}}\sum_{\bf q} {X}_{\bf q} 
     \sum_{\bf k} (\nabla_{\bf k} \, g_{{\bf k}, {\bf q}}) c_{{\bf k+q}}^\dagger c_{\bf k}.
\end{equation}
The first and second terms correspond to the electronic and phonon-assisted contributions to the current, respectively. Substituting Eq.~\eqref{eq:current} into Eq.~\eqref{eq:kubo_form} yields three naturally occurring contributions to the total mobility \cite{2025_Jankovic_I}
\begin{equation} \label{eq:total_mobility_4342}
    \mu = \mu_e +  \mu_x + \mu_{ph},
\end{equation}
where $\mu_e$ arises purely from electronic terms, $\mu_{ph}$ purely from  phonon-assisted terms, and $\mu_x$ denotes the cross term. Each of these contributions is notoriously difficult to calculate, particularly in realistic systems, which is why approximate methods are commonly employed. In this work, we focus on the independent-particle approximation. This approach consists of applying Wick’s decoupling to the current–current correlation function, treating the operators as if they were in the interaction picture, and then expressing the single-particle correlation functions using Eqs.~\eqref{eq:corrF_1}~and~\eqref{eq:corrF_2}. In the literature, it is usually only the electronic part that is taken into account \cite{2019_Zhu}, leading to a well-known formula \cite{2024_Jankovic}
%
\begin{equation} \label{muedc}
    \mu_e = 
    \frac{\beta\pi}{n_e} \sum_{ \bf k} 
    |(\nabla_{\bf k} \varepsilon_{\bf k} )_x|^2
    \int_{-\infty}^\infty d\nu \, 
    e^{-\beta \nu} {A}_{\bf k}(\nu)^2,
\end{equation}
where $(\nabla_{\bf k} \varepsilon_{\bf k} )_x$ denotes the $x$ component of vector $(\nabla_{\bf k} \varepsilon_{\bf k} )$. The same logic can in fact be also used for the calculation of $\mu_{ph}$ and $\mu_x$. The cross term $\mu_x$, within this approximation, is actually zero as there are no operators with which we could contract $X_{\bf q}$
\begin{equation} \label{eq:mux}
    \mu_x = 0.
\end{equation}
On the other hand, the expression for $\mu_{ph}$ can be straightforwardly written as
\begin{multline}
    \mu_{ph} = \frac{\beta}{2N \tilde{n}_e} \mathrm{Re}
      \sum_{\bf k_1, q_1, k_2, q_2}  
    (\nabla_{\bf k_1} g_{\bf k_1, q_1})_x    (\nabla_{\bf k_2} g_{\bf k_2, q_2})_x \\ \times \int_{-\infty}^\infty 
    \langle c^\dagger_{\bf k_2 + q_2}(t) c_{\bf k_1} \rangle
    \langle c_{\bf k_2 }(t) c_{\bf k_1 + q_1}^\dagger \rangle
    \langle X_{\bf q_2}(t) X_{\bf q_1} \rangle dt.
    \label{eq:muph_aux}
\end{multline}
One additional Wick contraction could, in principle, be written. This term, however, does not contribute: it vanishes in the limit $\mu_F \to -\infty$, and even at finite $\mu_F$ corresponds to an unconnected Feynman diagram, which does not contribute according to the linked-cluster theorem \cite{2000_Mahan}. Using Eqs.~\eqref{eq:corrF_1}--\eqref{eq:phon_prop_D}, it is easy to show that Eq.~\eqref{eq:muph_aux} simplifies to
%
\begin{multline}
    \mu_{ph} = \frac{\beta \pi}{N n_e} 
     \sum_{\bf k, q}
    |(\nabla_{\bf k} g_{\bf k, q})_x|^2
    \int_{-\infty}^\infty d\omega \,    e^{-\beta\omega}   A_{\bf k}(\omega) \\ 
    \left[ 
    \nph A_{\bf k + q}(\omega + \omega_0)
    +
    (\nph+1) A_{\bf k + q}(\omega - \omega_0)
    \right]  \label{eq:muph}.
\end{multline}
As we see, Eq.~\eqref{eq:muph} contains a prefactor that scales quadratically with the electron-phonon interaction strength, in contrast to Eq.~\eqref{muedc}. Thus, at least in the weak-coupling limit, Eq.~\eqref{eq:muph} can be interpreted as a higher-order correction to Eq.~\eqref{muedc}. Nevertheless, we calculate both contributions in this work, as they arise naturally from the application of Wick’s recipe and together constitute what is referred to as the independent-particle approximation. These two terms are diagrammatically shown in Fig.~\ref{fig:dijag}(a)~and~Fig.\ref{fig:dijag}(b).

It should be noted that Eq.~\eqref{eq:muph} is not the only higher-order contribution that scales quadratically with the electron-phonon coupling; another such contribution is illustrated in Fig.~\ref{fig:dijag}(c). In fact, it is to be expected that even the higher-order ladder diagrams should also be of the same order in the weak-coupling limit \cite{1966_Mahan_mobility}. Nevertheless, within CE, we focus only on the two terms that constitute the IPA, since our goal is to assess whether this approach—which is not extremely difficult to apply in realistic materials—yields accurate results. The question of whether all higher-order terms give significant corrections will also be addressed in this work, for the case of the Peierls model. This will be done using the hierarchical equations of motion (HEOM) method, which is briefly reviewed in Sec.~\ref{sec:HEOM}. The same questions were already previously answered in the case of the Holstein model in Refs. \cite{2023_Mitric, 2024_Jankovic}.

\subsection{One-shot Migdal and self-consistent Migdal approximation} 
\label{subsec_general_migldal_scma}

For the calculation of the spectral functions $A_{\bf k}(\omega)$, present in Eqs.~\eqref{muedc}~and~\eqref{eq:muph}, one could proceed diagrammatically to calculate the self-energy $\Sigma_{\bf k}(\omega)$, which is related to the Green's function $G_{\bf k}(\omega)$ and $A_{\bf k}(\omega)$ as follows
\begin{align}
    G_{\bf k}(\omega) &= \frac{1}{\omega - \varepsilon_{\bf k} - \Sigma_{\bf k}(\omega)}, \label{eq:dyson_eq} \\
    A_{\bf k}(\omega) &= -\frac{1}{\pi} \mathrm{Im} G_{\bf k}(\omega).
    \label{eq:spectral_def}
\end{align}
\begin{figure}[t!]
  \centering
  \includegraphics[width=0.9\linewidth]{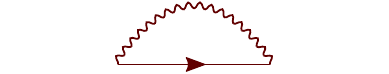}
  \caption{Self-energy in the one-shot Migdal approximation.}
  \label{fig:dijag_MA}
\end{figure}
Various approximations are possible, depending on the number of terms taken into account in $\Sigma_{\bf k}(\omega)$. The simplest approach is to retain only the lowest-order term (see Fig.~\ref{fig:dijag_MA}), which, after summing over the Matsubara frequencies and analytically continuing to the real axis, can be written as \cite{2017_Giustino}
\begin{multline} \label{eq:MAselfen}
    \Sigma^\mathrm{MA}_{{\bf k}}(\omega) = \frac{1}{N} \sum_{\bf q}
    |g_{{\bf k}, {\bf q}}|^2
     \Big[
     \\
     \frac{\nph + 1}{\omega - \omega_0 - \varepsilon_{\bf k+q} + i0^+}
    +  \frac{\nph}{\omega + \omega_0 - \varepsilon_{\bf k+q} + i0^+}
    \Big].
\end{multline}
This is known as the one-shot Migdal approximation. Alternatively, a more accurate result for $\Sigma_{\bf k}(\omega)$ can be obtained by replacing the bare Green’s function $G_{\bf k}^0 (\omega) = 1/(\omega - \varepsilon_{\bf k} + i0^+)$ in Fig.~\ref{fig:dijag_MA} with the full interacting Green’s function $G_{\bf k}(\omega)$. The corresponding expression for the self-energy reads as
\begin{multline} \label{eq:selfenSCMA}
    \Sigma^\mathrm{SCMA}_{{\bf k}}(\omega) = \frac{1}{N} \sum_{\bf q}
    |g_{{\bf k}, {\bf q}}|^2
     \Big[ \\ (\nph + 1)G_{\bf k+q}(\omega-\omega_0)
    + \nph G_{\bf k+q}(\omega+\omega_0)
    \Big].
\end{multline}
Equation~\eqref{eq:selfenSCMA} must be solved self-consistently together with the Dyson’s equation~\eqref{eq:dyson_eq}, which is why this approach is referred to as the self-consistent Migdal approximation (SCMA).

\subsection{Second order cumulant expansion method: general theory} 
One increasingly popular alternative to the approach described in the previous subsection is the cumulant expansion (CE) method. In this framework, the retarded Green’s function $G_{\bf k}(t)$ is written as
\begin{equation} \label{eq:green_def_cumulant}
    G_{\bf k}(t) = G_{\bf k}^0(t) e^{C_{\bf k}(t)} = 
    -i \theta(t) e^{-i\varepsilon_{\bf k}t} e^{C_{\bf k}(t)},
\end{equation}
so that the corresponding spectral function, using Eq.~\eqref{eq:spectral_def}, is given by
\begin{equation}
    A_{\bf k}(\omega + \varepsilon_{\bf k}) = 
    \frac{1}{\pi} \mathrm{Re} \int_0^\infty dt e^{i\omega t} 
    e^{C_{\bf k}(t)}.
\end{equation}
Here, $G_{\bf k}^0(t)$ is the noninteracting Green’s function, $\theta(t)$ the Heaviside step function, and $C_{\bf k}(t)$ is the cumulant function that needs to be determined.

Within the so-called second order cumulant expansion method, which is the central topic of this work, $C_{\bf k} (t)$ is uniquelly determined as the lowest (second) order expression with respect to the electron-phonon coupling, such that obtained $C_{\bf k} (t)$ gives the exact Green's function (see Eq.~\eqref{eq:green_def_cumulant}) in the weak-coupling limit $g_{\bf k,q}\to 0$. Such $C_{\bf k}(t)$ can actually be quite generally related to $\Sigma_{\bf k}^\mathrm{MA}(\omega)$ \cite{2023_Mitric, 2014_Kas}
\begin{equation} \label{Eq:Cumulant_expression}
C_{\bf k}(t) = \frac{1}{\pi} \int_{-\infty}^\infty
d\omega \frac{|\mathrm{Im}\Sigma^\mathrm{MA}_{\bf k}(\omega + \varepsilon_{\bf k})|}
{\omega^2} (e^{-i\omega t} + i\omega t -1).
\end{equation}
Let us note that although CE and MA are, by construction, equivalent when $g_{\bf k,q} \to 0$, they should produce different results away from this strict limit.

As shown in Ref.~\cite{2023_Mitric}, the numerical instability in Eq.~\eqref{Eq:Cumulant_expression}, around $\omega =0$, can be circumvented by rewriting the expression as
\begin{equation}
    C_{\bf k}(t) = \int_0^t dx (t-x) \int_{-\infty}^\infty \frac{d\omega}{\pi}
    \mathrm{Im} \Sigma_{\bf k}^\mathrm{MA}(\omega+\varepsilon_{\bf k}) e^{-i\omega x}.
\end{equation}
Inserting the expression for $\Sigma_{\bf k}^\mathrm{MA}(\omega)$ from Eq.~\eqref{eq:MAselfen}, and using the Plemelj–Sokhotski relation $(x+i0^+)^{-1} = \mathcal{P}(1/x) - i\pi \delta(x)$ (where $\mathcal{P}$ denotes the Cauchy principal value, while $\delta$ is the Dirac delta function), we obtain
%
\begin{align} \label{eq:general_cumulant_expression}
    C_{\bf k}(t) &=  -\int_0^t dx (t-x) iD(x) \sum_{\bf q} 
     \frac{|g_{{\bf k}, {\bf q}}|^2}{N}
      e^{i(\varepsilon_{\bf k} - \varepsilon_{\bf k+q}) x} \nonumber \\
      &\equiv -\int_0^t dx (t-x) iD(x) \; e^{i \varepsilon_{\bf k}x} B_{\bf k}(x).
\end{align}
Here, we introduced a convenient shorthand notation $B_{\bf k}(x)$, which in the case of $d$-dimensional system can also be written as
\begin{equation} \label{eq:definition_Bkx}
     B_{\bf k}(x) 
     = \int \frac{d{\bf q}}{(2\pi)^d} 
     {|g_{{\bf k}, {\bf q}}|^2} e^{-i \varepsilon_{\bf k+q} x}.
\end{equation}

The formulation in Eq.~\eqref{eq:general_cumulant_expression} has the key advantage that it does not require artificial broadening, unlike Eq.\eqref{Eq:Cumulant_expression}, which is crucial for transport calculations.


\section{Numerical implementation of CE, MA, SCMA and other methods for specific model hamiltonians}
\label{sec:num_implementation}

In this section, we present numerical details concerning the implementation of the different methods for specific model Hamiltonians. The most prominent electron–phonon models of the form~\eqref{eq:1} are: 
\newline 
i) $d$-dimensional Holstein (lattice) model
\begin{equation}
  \varepsilon_k = -2J \sum_{j=1}^d\cos k_j, \; g_{k,q} = g = \text{const.}, 
\end{equation}
where $J$ is the hopping parameter; 
\newline ii) 1D Peierls (lattice) model
\begin{equation} \label{eq:def_peierls}
  \varepsilon_k = -2J\cos k, \; g_{k,q} = 2ig[\sin (k+q) - \sin k],
\end{equation}
where, as in the Holstein model, $J$ is the hopping parameter, while $g$ is the electron-phonon coupling strength. In practical numerical implementations, in addition to the convention of Sec.~\ref{sec:hamiltonian}, one typically sets $J$ and the lattice constant equal to unity, and introduces a dimensionless coupling strength $\lambda = 2g^2/(J\omega_0)$;
\newline iii) 3D Fr\"ohlich (continuum) model 
\begin{equation} \label{eq:def_frohlich}
    \varepsilon_{\bf k} = \frac{{\bf k}^2}{2m_0}, \;
    g_{\bf k,q} = \frac{\mathcal{M}_0}{|{\bf q}|}, \; 
    \mathcal{M}_0 = \sqrt{\frac{\omega_0}{2\varepsilon_0} \left( 
    \frac{1}{\varepsilon_r^\infty} - \frac{1}{\varepsilon_r^{st}}
    \right)},
\end{equation}
where $m_0$ is the band mass, $\varepsilon_0$ is the vacuum permittivity, and $\varepsilon_r^\infty$ and $\varepsilon_r^{st}$ are the high-frequency and static relative dielectric constants, respectively. For numerical applications, in addition to the convention introduced in  Sec.~\ref{sec:hamiltonian}, it is convenient to set $m_0=\omega_0=1$. In this system of units, the Fr\"ohlich Hamiltonian depends on a single free parameter, commonly defined as
\begin{equation}
\alpha = \frac{1}{\sqrt{2}}\frac{1}{4 \pi \varepsilon_0}
\left( \frac{1}{\varepsilon_r^\infty} - \frac{1}{\varepsilon_r^{\mathrm{st}}} \right).
\end{equation}
In these units, the kinetic term reduces to $\varepsilon_{\bf k} = {\bf k}^2/2$, while $\mathcal{M}_0 = \sqrt{2\sqrt{2} \, \alpha\pi}$.

\subsection{CE method}

In Ref.~\cite{2023_Mitric}, we thoroughly investigated CE within the Holstein model. Here, we therefore turn to the Peierls and Fr\"ohlich models, where we demonstrate that $B_{\bf k}(x)$ from Eq.~\eqref{eq:general_cumulant_expression} can be evaluated analytically. This will allow us to derive a numerically stable and efficient expression, well suited for practical computations.

\subsubsection{The case of the (1D) Peierls model}

In Eq.~\eqref{eq:definition_Bkx} let us shift the momentum $q\to q-k$ 
\begin{equation}
B_{ k}(x) 
     =
     \int \frac{d{ q}}{2\pi} 
     {|g_{k,  q-k}|^2} e^{-i\varepsilon_{q} x},
\end{equation}
and (considering Eq.~\eqref{eq:def_peierls}) notice that only an even part of $|g_{k,q-k}|$, with respect to $q$, contributes
\begin{equation} \label{eq:symmetric_gkq}
    \frac{1}{2} \left( |g_{k,q-k}|^2 + |g_{k,-q-k}|^2  \right)= g^2 \left[4\sin^2k + 4 - \frac{\varepsilon_q^2}{J^2} \right].
\end{equation}
Since the whole $q$ dependence only appears through $\varepsilon_q$, we can switch  from momentum $q$ to energy $\varepsilon$ integration
\begin{multline} \label{eq:reghi45353}
B_k(x) = g^2 \int_{-\infty}^\infty d\varepsilon \rho(\varepsilon)
 \left[
    4\sin^2 k + 4 - \frac{\varepsilon^2}{J^2}
    \right] e^{-i\varepsilon x},
\end{multline}
where we introduced the noninteracting density of states
\begin{equation} \label{eq:DOS_nonint}
    \rho(\varepsilon) = \frac{1}{2\pi} \int_0^{2\pi} \dd k \: \delta(\varepsilon - \varepsilon_k) = \frac{\theta(4J^2 - \varepsilon^2)}{\pi \sqrt{4J^2 - \varepsilon^2}}.
\end{equation}
The integral over $\varepsilon$, in Eq.~\eqref{eq:reghi45353}, can actually be calculated analytically. To see that, let us note that 
\begin{align}
\int_{-\infty}^\infty d\varepsilon \; \rho(\varepsilon) e^{-i\varepsilon x}
 &= J_0(2Jx), \\
    \int_{-\infty}^\infty d\varepsilon \; \rho(\varepsilon) \varepsilon^2 e^{-i\varepsilon x}
    &= - \frac{d^2}{dx^2} \int_{-\infty}^\infty d\varepsilon \rho(\varepsilon) e^{-i\varepsilon x} \\
    &= 4 J^2 J_0(2 J x)-\frac{2 J J_1(2 J x)}{x},
\end{align}
where $J_0$ and $J_1$ are the Bessel function of the first kind of order zero and one, respectively. Hence, we get
\begin{equation}
 B_k(x) = 
 2 g^2 \left[2 \sin ^2k \cdot J_0(2 Jx) +\frac{J_1(2 J x)}{J x}\right],
\end{equation}
implying
\begin{multline}
    C_k(t) = -2g^2 \int_0^t dx (t-x)  iD(x) e^{i\varepsilon_k x}
    \Big[ \\
    2\sin^2 k \cdot J_0(2J x) + \frac{J_1(2J x)}{J x}
    \Big].
\end{multline}
As the subintegral expression is a combination of linear, trigonometric and Bessel terms, it can be calculated in a stable and efficient way using the Levin's collocation scheme; see Refs. \cite{1996_Levin, 2023_Mitric}.

\subsubsection{The case of the Frohlich model}
\label{subsec:frohlich_ce}

In Eq.~\eqref{eq:definition_Bkx}, using the explicit functional forms from Eq.~\eqref{eq:def_frohlich}, it is easy to perform an integral over the angular variables of ${\bf q}$
\begin{equation}
      B_{\bf k}(x) = 
     \frac{e^{-i \varepsilon_{\bf k}x} \mathcal{M}_0^2}{2\pi^2} 
    \int_0^\infty dq e^{-i\frac{q^2 x}{2m}} \frac{\sin \left( \frac{|{\bf k}|qx}{m} \right)}{\frac{|{\bf k}|qx}{m} }.
\end{equation}
Since the integrand is even with respect to $q$, the range of integration can be extended to $(-\infty, \infty)$ with an additional prefactor of $1/2$. Furthermore, the denominator can be eliminated using Feynman’s trick
\begin{equation}
     B_{\bf k}(x) = 
     \frac{e^{-i \varepsilon_{\bf k}x} \mathcal{M}_0^2}{4\pi^2 |{\bf k}|} 
     \int_0^{|{\bf k}|} dk'
    \int_{-\infty}^\infty dq e^{-i\frac{q^2 x}{2m}}  \cos \left( 
    \frac{k' q x}{m}
    \right).
\end{equation}
This can also be written as 
\begin{equation}
     B_{\bf k}(x) = 
     \frac{e^{-i \varepsilon_{\bf k}x} \mathcal{M}_0^2}{4\pi^2 |{\bf k}|} 
     \int_0^{|{\bf k}|} dk'
    \int_{-\infty}^\infty dq e^{-i\frac{q^2 x}{2m}+i\frac{k' q x}{m}},
\end{equation}
as the odd part with respect to $q$ does not contribute. By completing the square in the exponent, the last integral takes the form of the Fresnel integral that has an analytic solution
\begin{equation}
     B_{\bf k}(x) = 
     \frac{e^{-i \varepsilon_{\bf k}x} \mathcal{M}_0^2}{4\pi^2 |{\bf k}|} 
     \sqrt{\frac{2\pi m}{ix}}
     \int_0^{|{\bf k}|} dk' e^{i \frac{k'^2x}{2m}}.
\end{equation}
The remaining integral can be written in terms of the Fresnel function $E(x) = \int_0^x e^{i u^2} du$. Hence, we finally obtain
\begin{multline}
    C_{\bf k}(t) = -\frac{m \mathcal{M}_0^2}{2\pi^2 |{\bf k}|}
    \sqrt{\frac{\pi}{i}} 
    \int_0^t dx (t-x) iD(x)\frac{E\left( |{\bf k}|\sqrt{\frac{x}{2m}} \right)}{x}.
\end{multline}

\subsection{Migdal approximation in the Peierls model} \label{sec:mig_app_in_peierls}
\subsubsection{One shot Migdal approximation}

As we now show, the self-energy within the one-shot Migdal approximation (MA), given by Eq.~\eqref{eq:MAselfen}, admits an analytic solution in the Peierls model. We begin with Eq.~\eqref{eq:MAselfen}, perform the momentum shift $q \to q-k$, and note that only the even part of $|g_{k,q-k}|^2$ with respect to $q$ (see Eq.~\eqref{eq:symmetric_gkq}) contributes. Consequently, the entire $q$-dependence in Eq.~\eqref{eq:MAselfen} comes from $\varepsilon_q$, so we can switch from momentum summation to energy integration
\begin{multline} \label{eq:ma_selfen_analytic_form_123}
    \Sigma^{\mathrm{MA}}_k (\omega) = 4g^2 \int_{-\infty}^\infty d\varepsilon\,
    \rho(\varepsilon) 
    \left[
    1+\sin^2 k - \frac{\varepsilon^2}{4J^2}
    \right] \Big[ \\
    \frac{\nph + 1}{\omega - \omega_0 - \varepsilon + i0^+}
    + \frac{\nph}{\omega + \omega_0 - \varepsilon + i0^+} \Big],
\end{multline}
where $\rho(\varepsilon)$ is the noninteracting density of states, given by Eq.~\eqref{eq:DOS_nonint}. To proceed, we use the substitution $\varepsilon =2J\sin x$, and define a convenient shorthand notation
\begin{align} \label{eq:definition_of_B}
    B_j^\pm (\omega) &= \frac{1}{2J}(\omega \pm\omega_0+i0^+), \\
    \mathcal{P}(g,J,B) &= \frac{g^2}{\pi J} \int_{-\pi}^\pi dx
    \frac{1+\sin^2 k - \sin^2 x}{B-\sin x},
\end{align}
which enables us to write
\begin{equation} \label{eq:dsada1239454}
    \Sigma^{\mathrm{MA}}_k (\omega) =  
    (\nph+1) \mathcal{P}(g,J,B_J^- (\omega)) + 
    \nph \mathcal{P}(g,J,B_J^+ (\omega)).
\end{equation}
Thus, everything comes down to the calculation of $\mathcal{P}(g,J,B)$. To accomplish this, we use another substitution $z=e^{ix}$, giving
\begin{equation}
    \mathcal{P}(g,J,B) = -\frac{g^2}{2\pi J} \oint_{|z|=1} dz
    \frac{(2+4\sin^2k + z^2)z^2+1}{z^2(z-z_+)(z-z_-)},
\end{equation}
where $z_{\pm} = iB\pm\sqrt{1-B^2}$. Using the residue method, and the fact that $|z_+|<1$, while $|z_-|>1$, we obtain
\begin{equation} \label{eq:solution_PgjB}
     \mathcal{P}(g,J,B) =\frac{2g^2}{J} \left[ 
     \frac{1+\sin^2k-B^2}{i\sqrt{1-B^2}} + B
     \right].
\end{equation}
This completes the exact analytic solution for $ \Sigma^{\mathrm{MA}}_k (\omega)$. It is important to note, however, that one necessarily needs to keep the $i0^+$ term in Eq.~\eqref{eq:definition_of_B} to obtain the correct result when using the  solution given by Eqs.~\eqref{eq:dsada1239454}~and~\eqref{eq:solution_PgjB}. A more numerically stable expression, where $i0^+$ term is not needed, is obtained by substituting the following expression in Eq.~\eqref{eq:solution_PgjB}
\begin{multline}
    \frac{1}{i\sqrt{1-B^2}} = \mathrm{Re} \frac{1}{B\sqrt{1-\frac{1}{B^2}}}
    + i \mathrm{Im} \frac{1}{i\sqrt{1-B^2}}.
\end{multline}

\subsubsection{Self-consistent Migdal approximation}

In the case of the self-consistent Migdal approximation (SCMA), formulated through Eqs.~\eqref{eq:selfenSCMA} and \eqref{eq:dyson_eq}, an analytic solution is generally not available. However, for the Peierls model, we will show that instead of using Eq.~\eqref{eq:selfenSCMA}, there is a more convenient approach, which not only improves numerical, and memorical efficiency, as well as stability, but also yields a solution directly in the thermodynamic limit $N \to \infty$. Much of the derivation largely parallels the steps in Sec.~\ref{sec:mig_app_in_peierls}: first, we start from Eq.~\eqref{eq:selfenSCMA}, shift the momentum $q \to q-k$ and note that only the even part of $|g_{k,k-q}|^2$ with respect to $q$ (see Eq.~\eqref{eq:symmetric_gkq}) contributes. This implies that the momentum part of the self-energy is analytically solved
\begin{equation} \label{eq:functional_form_selfen_scma}
    \Sigma_k^{\mathrm{SCMA}}(\omega) = \Sigma^{(1)}(\omega) + 
    \sin^2 k \cdot\Sigma^{(2)}(\omega),
\end{equation}
so we need to store only two functions of frequency
\begin{align}
     \Sigma^{(1)}(\omega) &= 4g^2 \left[ 
     (\nph +1) F(\omega-\omega_0) + \nph F(\omega + \omega_0)
     \right], \nonumber \\ 
     \Sigma^{(2)}(\omega) &= 4g^2 \left[ 
     (\nph +1) G(\omega-\omega_0) + \nph G(\omega + \omega_0)
     \right].
\end{align}
Here, $F(\omega) = \frac{1}{N} \sum_q G_q(\omega) \sin^2 q$, and $ G(\omega) = \frac{1}{N} \sum_q G_q(\omega)$. They can also be written as
\begin{align} \label{eq:def_F_function}
    F(\omega)
    &= \frac{1}{N}\sum_q \frac{1-\frac{\varepsilon_q^2}{4J^2}}{\omega - \varepsilon_q - \Sigma^{(1)}(\omega) - \sin^2q \cdot \Sigma^{(2)}(\omega)},
    \\
    G(\omega) 
    &= \frac{1}{N}
    \sum_q \frac{1}{\omega - \varepsilon_q - \Sigma^{(1)}(\omega) - \sin^2q \cdot \Sigma^{(2)}(\omega)},
    \label{eq:def_G_function}
\end{align}
as a consequence of the Dyson equation~\eqref{eq:dyson_eq} and Eq.~\eqref{eq:functional_form_selfen_scma}. Therefore, implementation of SCMA rests upon efficient calculation of Eqs.~\eqref{eq:def_F_function}~and~\eqref{eq:def_G_function}. Again, we exploit the same ideas as in Sec.~\ref{sec:mig_app_in_peierls}: we switch from the summation over $q$ to integral over energies (by introducing the density of states $\rho(\varepsilon)$), and use the substitutions $\varepsilon = 2J \sin x$ and $z=e^{ix}$. We get
\begin{multline} \label{eq:G_final_expression}
    G(\omega) = \frac{1}{2\pi i} \oint_{|z|=1} dz\; 4z \Big[ 
    -\Sigma^{(2)}(\omega) z^4 + 4iJz^3 + \\ \left( 
    4\omega - 4\Sigma^{(1)}(\omega) - 2\Sigma^{(2)}(\omega)
    \right)z^2 - 4iJz - \Sigma^{(2)}(\omega) \Big]^{-1},
\end{multline}
and 
\begin{multline} \label{eq:F_final_expression}
    F(\omega) = -\frac{1}{\Sigma^{(2)}(\omega)} - 
    \frac{\Sigma^{(1)}(\omega)-\omega}{\Sigma^{(2)}(\omega)} G(\omega) + 
    \frac{4iJ}{\Sigma^{(2)}(\omega)} \\ \times \frac{1}{2\pi i} \oint dz 
    (z^2 - 1)
    \Big[ 
    -\Sigma^{(2)}(\omega) z^4 + 4iJz^3 + \\  \left( 
    4\omega - 4\Sigma^{(1)}(\omega) - 2\Sigma^{(2)}(\omega)
    \right)z^2 - 4iJz - \Sigma^{(2)}(\omega) \Big]^{-1}.
\end{multline}
Let us note that although these expressions look cumbersome, they are quite easy to evaluate using the residue theorem. In fact, the polynomials in the denominators, in both Eqs.~\eqref{eq:G_final_expression}~and~\eqref{eq:F_final_expression}, are of fourth order. Hence, the zeros of such polynomials (and thus the residues as well) can be straightforwardly obtained using the Ferrari-Cardano's method.

\subsection{Boltzmann transport equation}

The approach based on the solution of the Boltzmann equation is presently the dominant approach for calculations of electron and hole mobility in realistic semiconducting materials. For this reason, we evaluate the mobility based on this approach in this work as well. This enables us to perform checks for the limit when the Boltzmann approach is expected to be exact and to gain insight into the necessity of using more elaborate approaches beyond this limit.

Within the Boltzmann approach, the populations of electronic states $n_{\vb{k}}$ for a homogeneous system in steady state satisfy
\begin{equation}\label{eq:at824}
\begin{split}
-\vb{E}\cdot \frac{\partial n_{\vb k}}{\partial {\vb k}}=&
-\sum_{\vb{k'}}W_{\vb{k}\to\vb{k'}}n_{\vb{k}} \qty(1-n_{\vb{k'}}) \\
&+\sum_{\vb{k'}}W_{\vb{k'}\to\vb{k}}n_{\vb{k'}} \qty(1-n_{\vb{k}}),
\end{split}
\end{equation}
where $\vb{E}$ is the external electric field and $W_{\vb{k}\to\vb{k'}}$ is the transition probability between the electronic state of momentum $\vb{k}$ and the state of momentum $\vb{k'}$ given as
\begin{equation}
\begin{split}
W_{\vb{k}\to\vb{k'}}=
2\pi \qty|g_{\vb{k},\vb{k'}-\vb{k}}|^2 &
\left[
\qty(n_{\mathrm{ph}}+1)
\delta\qty(\varepsilon_{\vb{k}}-\omega_0-\varepsilon_{\vb{k'}})
+\right. \\ 
& \left. n_{\mathrm{ph}}
\delta\qty(\varepsilon_{\vb{k}}+\omega_0-\varepsilon_{\vb{k'}})
\right ].
\end{split}
\end{equation}
In the case of weak electric field that we are interested in, the solution is sought in the form
\begin{equation}
n_{\vb{k}}=n_{\vb{k}}^{(0)}+\vb{E}\cdot\frac{\partial n_{\vb{k}}}{\partial \vb{E}},
\end{equation}
where $n_{\vb{k}}^{(0)}$ is the equilibrium population of state $\vb{k}$ given as $n_{\vb{k}}^{(0)}\propto e^{-\beta\varepsilon_{\vb{k}}}$ in the case of low carrier concentration that we are considering. By replacing this solution in Eq.~\eqref{eq:at824} one obtains the linearized (with respect to electric field) Boltzmann equation. The ansatz
\begin{equation}
\frac{\partial n_{\vb{k}}}{\partial \vb{E}}=
\vb{v}_{\vb{k}}
\frac{\partial n_{\vb{k}}^{(0)}}{\partial \varepsilon_{\vb{k}}}
\tilde{\tau}_{\vb {k}}
\end{equation}
where $\vb{v}_{\vb{k}}=\nabla_{\vb{k}}{\varepsilon}_{\vb{k}}$
yields the exact solution of linearized Boltzmann equation
when $\tilde{\tau}_{\vb{k}}$ satisfy the self-consistent set of equations
\begin{equation}\label{eq:bo12}
 \sum_{\vb{k}'} W_{\vb{k}\to\vb{k}'}
 \left( 1 -
 \cos\theta_{\vb{k},\vb{k}'}
 \frac{ \left|\vb{v}_{\vb{k}'}\right| \tilde{\tau}_{\vb{k}'}}
      { \left|\vb{v}_{\vb{k}}\right|  \tilde{\tau}_{\vb{k}}}
  \right)
 = \frac{1}{\tilde{\tau}_{\vb{k}}}
\end{equation}
with $\cos\theta_{\vb{k},\vb{k}'}=\frac{\vb{v}_{\vb{k}}\cdot \vb{v}_{\vb{k}'}}{\left|\vb{v}_{\vb{k}}\right|\cdot \left|\vb{v}_{\vb{k}'}\right|}$. The mobility in the $x-$ direction is then given as
\begin{equation}\label{eq:bo18}
 \mu_{xx}=\beta\frac{\sum_{\vb{k}}n_{\vb{k}}^{(0)}\left(\vb{v}_{\vb{k}}\right)_x^2
 \tilde{\tau}_{\vb{k}}}{\sum_{\vb{k}}n_{\vb{k}}^{(0)}}.
\end{equation}
We refer to the mobility obtained from the self-consistent solution of Eq.~\eqref{eq:bo12} and using Eq.~\eqref{eq:bo18} as the full Boltzmann mobility. One can also neglect the second term in the bracket in Eq. ~\eqref{eq:bo12}. The time $\tilde{\tau}_{\vb{k}}$ then reduces to the relaxation time stemming from Migdal self-energy given in Eq.~\eqref{eq:serta416}. This result will therefore be referred to as the Boltzmann SERTA (self-energy relaxation time approximation) mobility.

Explicit expression for $\tilde{\tau}_{{k}}$ for the Peierls model reads
\begin{equation}
\begin{split}
\frac{1}{\tilde{\tau}_{{k}}}=&
\sum_{{q}={q}_1^\pm}
\frac{H_{{k},{q}} n_{\mathrm{ph}} \theta\qty(1+\cos k-\frac{\omega_0}{2J})}{2J\sqrt{1-\qty(\cos k-\frac{\omega_0}{2J})^2}}  +\\
&\sum_{{q}={q}_2^\pm}
\frac{H_{{k},{q}} \qty(n_{\mathrm{ph}}+1)
\theta\qty(1-\cos k-\frac{\omega_0}{2J})}{2J\sqrt{1-\qty(\cos k+\frac{\omega_0}{2J})^2}},
\end{split}
\end{equation}
where
$q_1^\pm=-k\pm\arccos\qty(\cos k-\frac{\omega_0}{2J})$,
$q_2^\pm=-k\pm\arccos\qty(\cos k+\frac{\omega_0}{2J})$,
$H_{{k},{q}}=\qty|g_{kq}|^2\qty(1-F_{kq})$ and
 $F_{kq}=0$ for Boltzmann SERTA mobility, while $F_{kq}= \cos\theta_{k,k+q}
 \frac{ \left|\vb{v}_{k+q}\right| \tilde{\tau}_{k+q}}
      { \left|\vb{v}_{k}\right|  \tilde{\tau}_{k}}$
for full Boltzmann mobility.
Explicit expression for $\tilde{\tau}_{\vb{k}}$ for the Fr{\"o}hlich model is given in Sec.~XVII of Supplemental Material in Ref.~\cite{2021_Vukmirovic} (referred to as the scattering time therein), while the expression for mobility is given by Eq. (18) in Ref. \cite{2025_Vukmirovic}.

The Boltzmann mobility gives the exact result for the mobility in the limit of weak electron-phonon interaction. This statement comes from the fact that Boltzmann equation can be derived from fully quantum kinetic equations and the approximation that electronic spectral function is a delta-like function (see, for example, Ref.~\cite{2020_Ponce}), which is valid in the limit of weak electron-phonon interaction. On the other hand, the Boltzmann SERTA mobility corresponds to the limit of weak electron-phonon interaction within the IPA. This can be derived starting from the Kubo formula and employing the IPA and the approximation that electronic spectral function is a delta-like function (see also, for example, Ref.~\cite{2020_Ponce}).

\subsection{HEOM method} \label{sec:HEOM}
The HEOM method provides a numerically exact density-matrix framework to study the dynamics of a system of interest [electrons in Eq.~\eqref{eq:1}] interacting with a collection of harmonic oscillators [phonons in Eq.~\eqref{eq:1}]~\cite{2007_Xu, 2020_Tanimura}.
The method has enabled numerically exact computations of both single-electron~\cite{2022_Jankovic} and transport~\cite{2023_Jankovic,2024_Jankovic,2025_Jankovic_I,2025_Jankovic_II} properties in interacting electron--phonon models on a lattice in the limit of vanishing electron concentration.
For completeness, here we give the actual equations that we have numerically implemented to obtain HEOM results for the Peierls model (Sec.~\ref{sec:results}).
We refer the reader to the original papers for details on more formal aspects of the method, such as its formulation in the single-electron subspace~\cite{2024_Jankovic,2022_Mitric} or the generalized Wick's theorem~\cite{2025_Jankovic_I}.

\subsubsection{Single-particle properties}
Although we originally formulated the HEOM method for single-electron properties considering the one-dimensional Holstein model~\cite{2022_Jankovic}, there is no formal obstacle to deriving similar equations for the more general model in Eq.~\eqref{eq:1}.
The HEOM for both the greater $G^>_k(t)=-ie^{-i\mu_Ft}\langle c_k(t)c_k^\dagger(0)\rangle$ and lesser $G^<_k(t)=ie^{-i\mu_Ft}\langle c_k^\dagger(0)c_k(t)\rangle/\widetilde{n}_e$ Green's functions [see Eqs.~\eqref{eq:corrF_1}--\eqref{eq:def_ne_4342} and Ref.~\cite{2024_Jankovic}] assumes the following form~\cite{2022_Jankovic}:
\begin{equation}
\label{Eq:HEOM-GF}
\begin{split}
 &\partial_t G_{\mathbf{n},k-k_\mathbf{n}}^{(\gtrless,n)}(t)=-i(\varepsilon_{k-k_\mathbf{n}}+\mu_\mathbf{n})G_{\mathbf{n},k-k_\mathbf{n}}^{(\gtrless,n)}(t)\\
 &-\frac{i}{\sqrt{N}}\sum_{qm}\sqrt{1+n_{qm}}\sqrt{c_m}\:g_{k-k_\mathbf{n},-q}^*G_{\mathbf{n}_{qm}^+,k-k_\mathbf{n}-q}^{(\gtrless,n+1)}(t)\\
 &-\frac{i}{\sqrt{N}}\sum_{qm}\sqrt{n_{qm}}\sqrt{c_m}\:g_{k-k_\mathbf{n},q}^*G_{\mathbf{n}_{qm}^-,k-k_\mathbf{n}+q}^{(\gtrless,n-1)}(t)\\
 &-\delta_{n,D}\frac{1}{2}\tau_{k-k_\mathbf{n}}^{-1}G_{\mathbf{n},k-k_\mathbf{n}}^{(\gtrless,n)}(t).
\end{split}
\end{equation}
In Eq.~\eqref{Eq:HEOM-GF}, the physical Green's functions $G^{(\gtrless,0)}_{\mathbf{0},k}(t)\equiv G^\gtrless_k(t)$ are at the root of the hierarchy formed by auxiliary Green's functions $G_{\mathbf{n},k-k_\mathbf{n}}^{(\gtrless,n)}$, which are enumerated by the vector $\mathbf{n}=[n_{qm}]$ of nonnegative integers $n_{qm}$.
These count the number of single phonon-assisted processes involving emission ($m=0,c_0=1+n_\mathrm{ph}$) or absorption ($m=1,c_1=n_\mathrm{ph}$) of a phonon with wave vector $q$ [more compactly, $c_m=n_\mathrm{ph}+\frac{1+(-1)^m}{2}$].
The total number, energy, and momentum of exchanged phonons are $n=\sum_{qm}n_{qm}$, $\mu_\mathbf{n}=\omega_0\sum_{qm}(n_{q0}-n_{q1})$, and $k_\mathbf{n}=\sum_{qm} qn_{qm}$, respectively.

The hierarchy in Eq.~\eqref{Eq:HEOM-GF} is, in principle, infinite, and we truncate it at a sufficiently large maximum depth $D$.
Although not strictly necessary in the context of single-particle properties~\cite{2022_Jankovic}, the closing term [the last term on the right-hand side of Eq.~\eqref{Eq:HEOM-GF}, with $\tau_{k-k_\mathbf{n}}^{-1}$ defined in Eq.~\eqref{Eq:tau_k}] at depth $D$ removes long-time small-amplitude oscillations of $G^\gtrless$ around zero~\cite{2022_Jankovic} without affecting its dynamics on short-to-intermediate timescales.
This ensures that the spectral function $A_k(\omega)=-\frac{1}{2\pi}\mathrm{Im}\:G^>_k(\omega)$ does not exhibit pronounced high-frequency tails (see Fig.~\ref{fig:specf}).
We have checked that the presence or absence of the closing term, as well as its particular form, do not affect the IPA mobility computed using Eqs.~\eqref{Eq:mu_e_IPA} and~\eqref{Eq:mu_ph_IPA}.

The hierarchy for $G^>$ is propagated with the initial condition $G^{(>,n)}_{\mathbf{n},k-k_\mathbf{n}}(0)=-i\delta_{n,0}\delta_{\mathbf{n},\mathbf{0}}$.
Meanwhile, the initial condition for the hierarchy for $G^<$ follows from the imaginary-time HEOM~\cite{2022_Jankovic}
\begin{equation}
\label{Eq:i-HEOM}
\begin{split}
 &\partial_\tau\sigma^{(n)}_{\mathbf{n},k-k_\mathbf{n}}(\tau)=-(\varepsilon_{k-k_\mathbf{n}}+\mu_\mathbf{n})\sigma^{(n)}_{\mathbf{n},k-k_\mathbf{n}}(\tau)\\
 &-\frac{1}{\sqrt{N}}\sum_{qm}\sqrt{1+n_{qm}}\sqrt{c_m}\:g_{k-k_\mathbf{n},-q}^*\sigma_{\mathbf{n}_{qm}^+,k-k_\mathbf{n}-q}^{(n+1)}(\tau)\\
 &-\frac{1}{\sqrt{N}}\sum_{qm}\sqrt{n_{qm}}\sqrt{c_m}\:g_{k-k_\mathbf{n},q}^*\sigma_{\mathbf{n}_{qm}^-,k-k_\mathbf{n}+q}^{(n-1)}(\tau).
\end{split}
\end{equation}
The HEOM in Eq.~\eqref{Eq:i-HEOM} is also truncated at depth $D$ and propagated from $\tau=0$ to $\tau=\beta$ starting from the infinite-temperature initial condition $\sigma_{\mathbf{n},k-k_\mathbf{n}}^{(n)}(0)=\delta_{n,0}\delta_{\mathbf{n},\mathbf{0}}$.
The initial condition for the hierarchy for $G^<$ then reads
\begin{equation}
 G^{(<,n)}_{\mathbf{n},k-k_\mathbf{n}}(0)=i\frac{\sigma^{(n)}_{\mathbf{n},k-k_\mathbf{n}}(\beta)}{\sum_k\sigma^{(0)}_{\mathbf{0},k}(\beta)}.
\end{equation}

\subsubsection{Transport properties}
Within the HEOM method, the electron mobility (either fully numerically exact or in the IPA) is most conveniently computed in the time domain [see Eq.~\eqref{eq:kubo_form}].

The evaluation of Eq.~\eqref{eq:kubo_form} in the IPA using the numerically exact Green's functions $G_k^\gtrless(t)$ yields the following purely electronic contribution to mobility~\cite{2024_Jankovic}:
\begin{equation}
\label{Eq:mu_e_IPA}
\begin{split}
 &\mu_{e,\mathrm{HEOM}}^\mathrm{IPA}=\\&-\frac{1}{T}\int_0^{+\infty}dt\sum_k(\partial_k\varepsilon_k)^2G_k^>(t)[G_k^<(t)]^*.
\end{split}
\end{equation}
The corresponding phonon-assisted contribution reads [$D(t)$ is the phonon propagator]
\begin{equation}
\label{Eq:mu_ph_IPA}
\begin{split}
 &\mu_{ph,\mathrm{HEOM}}^\mathrm{IPA}=\\&-\frac{i}{NT}\int_0^{+\infty}dt\sum_{kq}|\partial_k g_{k,q}|^2G_{k+q}^>(t)[G^<_k(t)]^*D(t).
\end{split}
\end{equation}

To obtain the numerically exact mobility, it is convenient to recast the current autocorrelation function in Eq.~\eqref{eq:kubo_form} as $C_{jj}(t)=\langle j_x(t)j_x(0)\rangle/\widetilde{n}_e=\mathrm{Tr}_\mathrm{1e}\left\{j_x\iota(t)\right\}$, where $\mathrm{Tr}_\mathrm{1e}$ denotes the trace over the subspace containing a single electron (and an arbitrary number of phonons)~\cite{2024_Jankovic}, while
\begin{equation}
\label{Eq:def_iota_t}
\begin{split}
    \iota(t)&=\frac{(e^{-iHt}j_xe^{-\beta H}e^{iHt})_\mathrm{1e}}{\mathrm{Tr}_\mathrm{1e}\:e^{-\beta H}}\\
    &=\iota_e(t)+\iota_{ph}(t).
\end{split}
\end{equation}
In the first equality, the subscript ``1e'' in the numerator emphasizes that $\iota(t)$ is to be considered in the aforementioned subspace.
The decomposition in the second equality follows from Eq.~\eqref{eq:current}, where $\iota_e(t)$ [$\iota_{ph}(t)$] is obtained by replacing $j_x$ in the first equality with its electronic (phonon-assisted) contribution.
The HEOM representation $\{\iota_\mathbf{n}^{(n)}(t)\}$ of $\iota(t)$ satisfies~\cite{2025_Jankovic_I}
\begin{equation}
\label{Eq:HEOM-j-j}
\begin{split}
 &\partial_t\langle k|\iota_\mathbf{n}^{(n)}(t)|k+k_\mathbf{n}\rangle=\\&-i(\varepsilon_k-\varepsilon_{k+k_\mathbf{n}}+\mu_\mathbf{n})\langle k|\iota_\mathbf{n}^{(n)}(t)|k+k_\mathbf{n}\rangle\\
 &-\frac{i}{\sqrt{N}}\sum_{qm}\sqrt{1+n_{qm}}\sqrt{c_m}g_{k,-q}^*\langle k-q|\iota_{\mathbf{n}_{qm}^+}^{(n+1)}(t)|k+k_\mathbf{n}\rangle\\
 &+\frac{i}{\sqrt{N}}\sum_{qm}\sqrt{1+n_{qm}}\sqrt{c_m}g_{k+k_\mathbf{n},q}\langle k|\iota_{\mathbf{n}_{qm}^+}^{(n+1)}(t)|k+k_\mathbf{n}+q\rangle\\
 &-\frac{i}{\sqrt{N}}\sum_{qm}\sqrt{n_{qm}}\sqrt{c_m}g_{k,q}^*\langle k+q|\iota_{\mathbf{n}_{qm}^-}^{(n-1)}(t)|k+k_\mathbf{n}\rangle\\
 &+\frac{i}{\sqrt{N}}\sum_{qm}\sqrt{n_{qm}}\frac{c_{\overline{m}}}{\sqrt{c_m}}g_{k+k_\mathbf{n},-q}\langle k|\iota_{\mathbf{n}_{qm}^-}^{(n-1)}(t)|k+k_\mathbf{n}-q\rangle\\
 &-\delta_{n,D}\frac{1}{2}\left(\tau_k^{-1}+\tau_{k+k_\mathbf{n}}^{-1}\right)\langle k|\iota_\mathbf{n}^{(n)}(t)|k+k_\mathbf{n}\rangle.
\end{split}
\end{equation}
The very same equations govern the dynamics of the HEOM representations $\{\iota_{e,\mathbf{n}}^{(n)}(t)\}$ and $\{\iota_{ph,\mathbf{n}}^{(n)}(t)\}$ of $\iota_{e}(t)$ and $\iota_{ph}(t)$, respectively.
In Eq.~\eqref{Eq:HEOM-j-j}, $c_{\overline{m}}=n_\mathrm{ph}+\frac{1-(-1)^m}{2}$.
At each instant $t\geq 0$, the current autocorrelation function reads~\cite{2025_Jankovic_I}
\begin{equation}
 C_{jj}(t)=C_{jj,e}(t)+C_{jj,ph}(t)+C_{jj,x}(t),   
\end{equation}
with the purely electronic contribution
\begin{equation}
\label{Eq:C_jj_e_t}
    C_{jj,e}(t)=\sum_k(\partial_k\varepsilon_k)\langle k|\iota_{e,\mathbf{0}}^{(0)}(t)|k\rangle,
\end{equation}
the phonon-assisted contribution
\begin{equation}
\label{Eq:C_jj_ph_t}
    C_{jj,ph}(t)=\frac{1}{\sqrt{N}}\sum_{kqm}\sqrt{c_m}(\partial_k g_{k,q})\langle k|\iota_{ph,\mathbf{0}_{qm}^+}^{(1)}(t)|k+q\rangle,
\end{equation}
and the cross contribution
\begin{equation}
\label{Eq:C_jj_x_t}
\begin{split}
 &C_{jj,x}(t)=\sum_k(\partial_k\varepsilon_k)\langle k|\iota_{ph,\mathbf{0}}^{(0)}(t)|k\rangle\\
 &+\frac{1}{\sqrt{N}}\sum_{kqm}\sqrt{c_m}(\partial_k g_{k,q})\langle k|\iota_{e,\mathbf{0}_{qm}^+}^{(1)}(t)|k+q\rangle.
\end{split}
\end{equation}
Inserting Eqs.~\eqref{Eq:C_jj_e_t},~\eqref{Eq:C_jj_ph_t}, and \eqref{Eq:C_jj_x_t} into Eq.~\eqref{eq:kubo_form}, we obtain the HEOM full results $\mu_{e,\mathrm{HEOM}}^\mathrm{full}$, $\mu_{ph,\mathrm{HEOM}}^\mathrm{full}$, and $\mu_{x,\mathrm{HEOM}}^\mathrm{full}$ for the three contributions to the total mobility, see Eq.~\eqref{eq:total_mobility_4342}. 

The initial condition for Eq.~\eqref{Eq:HEOM-j-j} is connected with the HEOM representation $\rho_\mathbf{n}^{(n,\mathrm{eq})}$ of the operator $\rho^\mathrm{eq}=(e^{-\beta H})_\mathrm{1e}/\mathrm{Tr}_\mathrm{1e}\:e^{-\beta H}$, which reads
\begin{equation}
\langle k|\rho_\mathbf{n}^{(n,\mathrm{eq})}|k+k_\mathbf{n}\rangle=\frac{\sigma^{(n)}_{\mathbf{n},k}(\beta)}{\sum_k\sigma^{(0)}_{\mathbf{0},k}(\beta)}.
\end{equation}
The connection is provided by the generalized Wick's theorem~\cite{2025_Jankovic_I, 2018_Zhang}, whose application ultimately yields~\cite{2025_Jankovic_I}
\begin{equation}
\label{Eq:iota_t_0_e_heom_rep}
 \begin{split}
  \langle k|\iota_{e,\mathbf{n}}^{(n)}(0)|k+k_\mathbf{n}\rangle=(\partial_k\varepsilon_k)\langle k|\rho_\mathbf{n}^{(n,\mathrm{eq})}|k+k_\mathbf{n}\rangle,
 \end{split}
\end{equation}
\begin{equation}
\label{Eq:iota_t_0_ph_heom_rep}
 \begin{split}
  &\langle k|\iota_{ph,\mathbf{n}}^{(n)}(0)|k+k_\mathbf{n}\rangle=\\
  &\frac{1}{\sqrt{N}}\sum_{qm}\sqrt{1+n_{qm}}\sqrt{c_m}(\partial_k g_{k,-q}^*)\langle k-q|\rho_{\mathbf{n}_{qm}^+}^{(n+1,\mathrm{eq})}|k+k_\mathbf{n}\rangle\\
  &+\frac{1}{\sqrt{N}}\sum_{qm}\sqrt{n_{qm}}\sqrt{c_m}(\partial_k g_{k,q}^*)\langle k+q|\rho_{\mathbf{n}_{qm}^-}^{(n-1,\mathrm{eq})}|k+k_\mathbf{n}\rangle.
 \end{split}
\end{equation}
%

In contrast to the HEOM for single-particle properties [Eq.~\eqref{Eq:HEOM-GF}], the closing term in Eq.~\eqref{Eq:HEOM-j-j} is crucial in avoiding the long-time numerical instabilities characteristic for electron--phonon models with a finite number of phonon modes~\cite{2023_Jankovic,2025_Jankovic_I,2019_Dunn}.
We always choose a sufficiently large $D$ so that the closing term merely stabilizes the long-time dynamics of the zeroth- and first-tier auxiliaries entering 
Eqs.~\eqref{Eq:C_jj_e_t}--\eqref{Eq:C_jj_x_t} without compromising the numerical exactness of the HEOM mobility~\cite{2025_Jankovic_I}.
We have also checked that using other closing terms~\cite{2022_Fay, 2018_Zhang} does not change the mobility obtained by propagating Eq.~\eqref{Eq:HEOM-j-j}~\cite{2025_Jankovic_I,2025_Jankovic_II}.

HEOM computations of the fully numerically exact electron mobility in the one-dimensional Peierls model are viable at moderate to high temperatures ($T/\omega_0\gtrsim 2$) and up to moderate interactions ($\lambda\lesssim 1$)~\cite{2025_Jankovic_I,2025_Jankovic_II}.
We estimated that the relative uncertainty in $\mu$ due to finite-chain and finite-depth effects is typically below 10\%~\cite{2025_Jankovic_I,2025_Jankovic_II}.
The corresponding data are openly available~\cite{Jankovic_Zenodo_Peierls}.
Meanwhile, HEOM computations of the IPA mobility are possible in a wider range of model parameters as these can safely be performed on somewhat shorter chains.

\section{Results for Specific Electron–Phonon Models}
\label{sec:results}

\subsection{Peierls model}
\label{sec:peierls_model-53232}

\subsubsection{Quasiparticle properties}
\label{sec:qp_prop4}

Although our primary interest lies in transport properties, analyzing quasiparticle (QP) properties can also provide valuable insights into the accuracy of different methods. Here, we will compare QP predictions of CE, SCMA, and MA with the predictions of numerically exact benchmark — the generalized Green’s function cluster expansion (GGCE) \cite{GGCE_1, GGCE_2, GGCE_3}.
\begin{figure}[t!]
  \centering
  \includegraphics[width=0.9\linewidth]{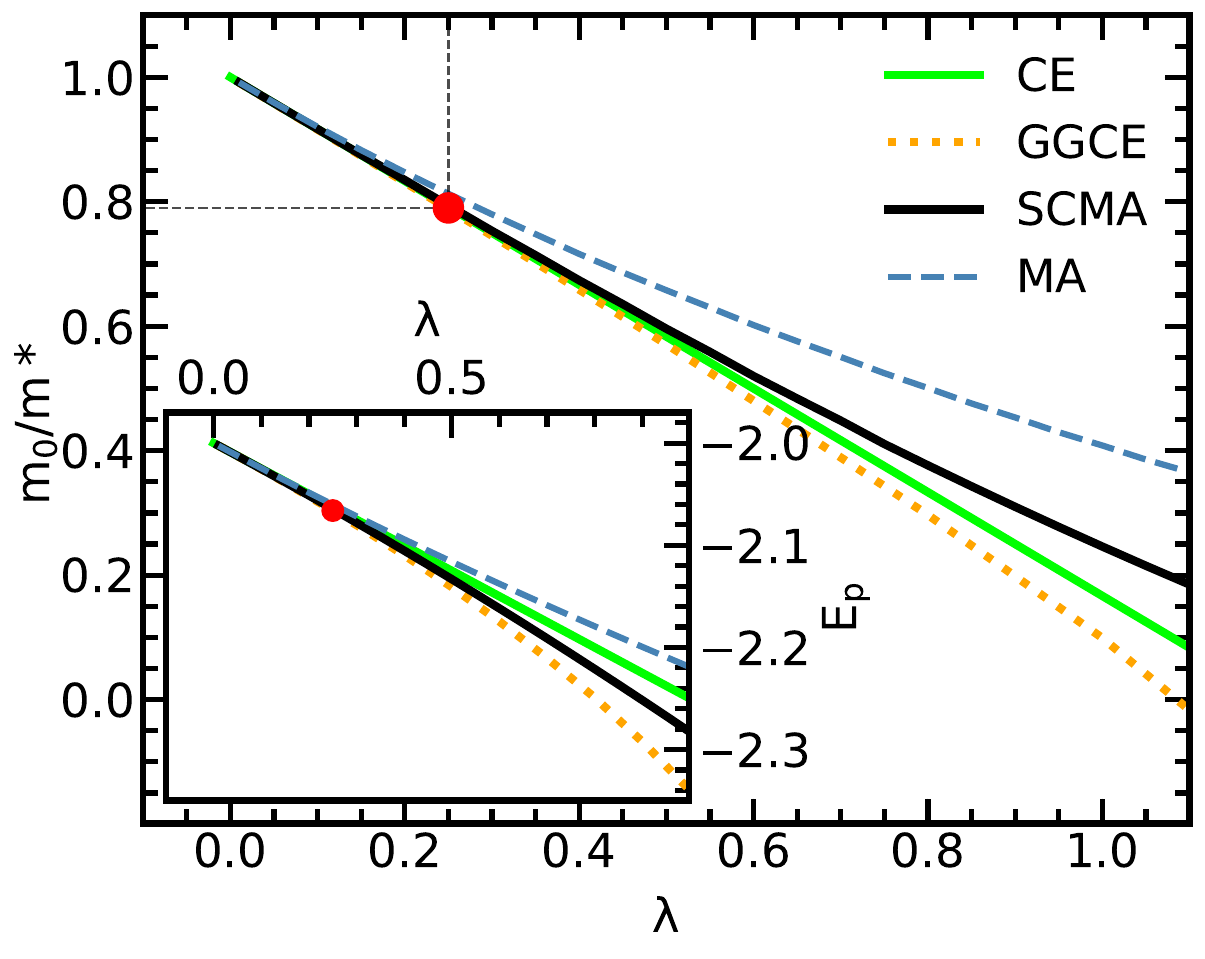}
  \caption{Comparison of the band-mass to renormalized-mass ratio $m_0/m^*$ and the ground-state energy $E_p$ obtained from CE, GGCE, SCMA, and MA for $\omega_0 = 0.5$ at $T = 0$. The red point corresponds to the CE result for $\lambda=0.25$.}
  \label{fig:qp_prop}
\end{figure}

For the GGCE results, we used publicly available code \cite{GGCE_1, GGCE_2, GGCE_3} to compute numerically exact spectral functions at $T = 0$. The ground-state energy is then extracted from the position of the lowest-energy peak at $k = 0$, while the effective mass $m^*$ is obtained by determining the peak positions at several momenta and fitting the resulting dispersion to a parabolic function.

Within MA and SCMA we seek for minimal $\omega$ for each $k$, and we denote it by $\omega=E_{p,k}$, such that the denominator of Eq.~\eqref{eq:dyson_eq} is vanishing
\begin{equation}
    E_{p,k} = \varepsilon_k + \mathrm{Re} \Sigma_k(E_{p,k}); \quad 
    \varepsilon_k = -2J \cos k.
\end{equation}
The ground state energy $E_p$ and the effective mass $m^*$ are then determined from 
\begin{align}
    E_p &= E_{p,k=0}, \\ 
    E_{p,k} &= \mathrm{const}. +\frac{k^2}{2m^*}, \quad \text{for} \;  k\approx 0.
\end{align}
Using this, and the analytic form of the MA and SCMA self-energies (see Eqs.~\eqref{eq:ma_selfen_analytic_form_123}~and~\eqref{eq:functional_form_selfen_scma}), the expression for $m^*$ can be explicitly written as 
\begin{equation}
    m^* = \frac{m_0 \left[ 
    1- \frac{\partial \mathrm{Re} \Sigma_{k=0}(\omega)}{\partial \omega}\Big|_{\omega = E_p}
    \right]}{
    1 + 2m_0 \, \mathrm{Re}\left[ 
    \Sigma_{k=\pi/2}(E_p) - \Sigma_{k=0}(E_p)
    \right]
    },
\end{equation}
where $m_0 = 1/ (2J)$. 
We note that ground state energy can correspond to finite $k$ within the Peierls model \cite{2010_Marchand}, but in the regimes we will be examining it will always correspond to $k=0$. 

Finally, for the CE quasiparticle properties, we rely on the fact these coincide with the predictions of the Rayleigh-Schr\"odinger perturbation theory \cite{2023_Mitric}
\begin{equation}
    E_{p, k} = 
    \varepsilon_{  k} + \mathrm{Re} \Sigma^\mathrm{MA}_{  k} (\varepsilon_k),
\end{equation}
and that $\Sigma^\mathrm{MA}_{  k} (\omega)$ is known analytically; see Eqs.~\eqref{eq:dsada1239454}~and~\eqref{eq:solution_PgjB}. Using these, we obtain
\begin{align}
    E_{p} &= -2J + \frac{\lambda \omega_0}{2J}\left(
    -2J - \omega_0 + \sqrt{\omega_0(4J+\omega_0)}
    \right),  \\
    \frac{m_0}{m*} &= 
    1+ \lambda \frac{4J\omega_0 + \omega_0^2 - (6J+\omega_0)\sqrt{\omega_0(4J+\omega_0)} }{2J(4J+\omega_0)}.
    \label{eq:ef_masa_1d}
\end{align}
%
The negative effective mass \( m^* \) at $k=0$ is predicted at large \( \lambda \) consistent with the fact that the polaron band minimum is no longer located at \( k = 0 \) for strong coupling \cite{2010_Marchand}.

The numerical results for $\omega_0=0.5$ are shown in Fig.~\ref{fig:qp_prop}, while the results for various other regimes are shown in Sec.~A of SM \cite{SuppMat}. As we see, CE and SCMA are both much closer to the numerically exact benchmark (GGCE) than MA. In fact, the CE effective mass is in remarkably good agreement with GGCE even for large $\lambda$, even though CE is in essence a perturbative approach. The same was also true in the case of the Holstein model \cite{2023_Mitric}. 

\begin{figure*}[t!]
    \centering
    \includegraphics[width=\textwidth]{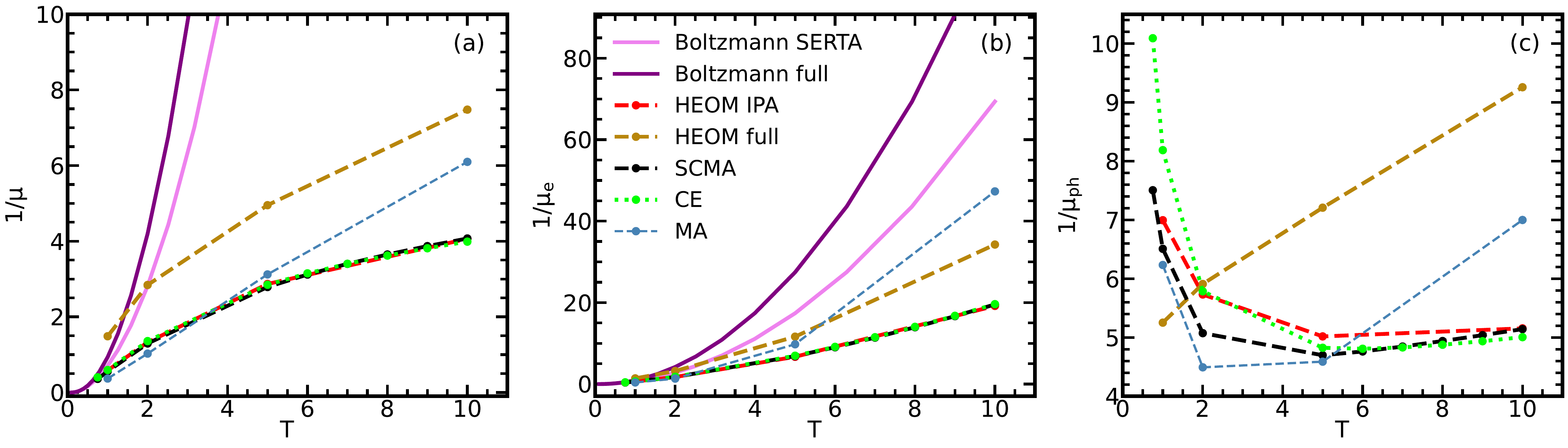}
    \caption{Comparison of different methods for calculating the inverse of (a) the total mobility $1/\mu$ (see Eq.~\eqref{eq:total_mobility_4342}), (b) the electronic contribution to the mobility $1/\mu_e$ (see Eq.~\eqref{muedc}), and (c) the phononic contribution to the mobility $1/\mu_{ph}$ (see Eq.~\eqref{eq:muph}), for $\omega_0 = 0.5$ and $\lambda = 0.25$.}
    \label{fig:selective_regime}
\end{figure*}

Before moving on to the transport properties, let us also calculate the CE effective lifetime $\tau_k$, which within this method is given by \cite{2023_Mitric}
\begin{equation} \label{eq:serta416}
    \tau_{  k}^{-1} = -2 \mathrm{Im}
    \Sigma^\mathrm{MA}_{  k} (\varepsilon_{  k}).
\end{equation}
This is useful from the practical standpoint as it indicates how long should we propagate $C_k(t)$ before the corresponding Green's function attenuates. In the case of Peierls model (see Eqs.~\eqref{eq:dsada1239454}~and~\eqref{eq:solution_PgjB})
\begin{multline} \label{Eq:tau_k}
    \tau_k^{-1} = \frac{4g^2}{J} \Big[ 
    (\nph +1) s_k\left(\frac{\omega_0}{2J} + \cos k \right) + 
    \\
    \nph s_k\left(\frac{\omega_0}{2J} - \cos k \right)
    \Big]
\end{multline}
where $s_k(x)= (1-x^2 + \sin^2 k)\, \theta(1-x^2)/\sqrt{1-x^2}$, with $\theta$ being the Heaviside step function. From this expression, it follows that $\tau_{k}$ diverges for $\frac{\omega_0}{2J} \geq 1 + |\cos k|$ when $T > 0$, whereas at $T = 0$ the divergence occurs when $\frac{\omega_0}{2J} \geq 2 \sin^2 \frac{k}{2}$. Therefore, in principle, for $T> 0$, CE transport properties can be obtained only for $\omega_0 < 2J$ when the corresponding Green's function attenuates for all momenta. An illustration of this is presented in Sec.~A of SM \cite{SuppMat}.

\subsubsection{Numerical challenges, accuracy, and limitations of the CE method for the calculation of transport properties: Illustrative case $\omega_0=0.5$, $\lambda = 0.25$}
\label{subsec:num_challenge}

Figure~\ref{fig:qp_prop} illustrates that the parameter choice $\omega_0=0.5$, $\lambda = 0.25$ corresponds to a regime where the ratio of band to effective mass is approximately $\approx 0.8$. In this regime, all methods yield reasonably accurate QP properties, while for larger $\lambda$, MA begins to deviate more noticeably. Since the interaction strength is neither too weak nor too strong, this represents an ideal case for a detailed comparison of transport properties obtained from different methods. This is the focus of the present subsection. The general trends identified here will be further corroborated in the next subsection, where transport properties are examined over a broader range of parameters.

In the expressions for $\mu_e$, $\mu_{ph}$, and $n_e$ (see Eqs.~\eqref{muedc}, \eqref{eq:muph}, and \eqref{eq:def_ne_4342}, respectively), there is a numerical instability when integrating over large negative frequencies due to the Boltzmann factor $e^{-\beta \omega}$ that becomes exponentially large. This is why in practice we introduce a finite frequency cutoff $\int_{-\infty}^\infty \to \int_{-\Lambda}^\infty$, denoted by $\Lambda$, which is increased until the results fully converge. In addittion, we always use sufficiently large number of $k$ points in Eqs.~\eqref{muedc}, \eqref{eq:muph}, and \eqref{eq:def_ne_4342}, so that the obtained results are representative of the thermodynamic limit.

The results are shown in Fig.~\ref{fig:selective_regime}. Since most points in Fig.~\ref{fig:selective_regime} lie in the higher-temperature range, where the mobility \(\mu\) is small, we chose to plot \(1/\mu (T)\) for better clarity. For completeness, the plot of $\mu(T)$ is provided in the following subsection; see Fig.~\ref{fig:all_mob_1}($a_2$)--($c_2$).
The curves labeled "HEOM full" and "HEOM IPA" represent numerically exact benchmarks for the mobility with [Eqs.~\eqref{Eq:def_iota_t}--\eqref{Eq:iota_t_0_ph_heom_rep}] and without [Eqs.~\eqref{Eq:mu_e_IPA} and~\eqref{Eq:mu_ph_IPA}] vertex corrections, respectively.
As seen from the figure, the discrepancy between the two clearly demonstrates that vertex corrections are not small, although the IPA result remains of the same order of magnitude as the full result.
This difference is not solely due to the fact that $\mu_x$ vanishes within IPA [see Eq.~\eqref{eq:mux}]; it is also evident in both the electronic [Panel~(b)] and phononic [Panel~(c)] contributions.
In particular, while the IPA captures the qualitative temperature dependence of the electronic term in Panel~(b), 
the phonon-assisted contribution in Panel~(c) is not even qualitatively reproduced.
Equation~\eqref{Eq:mu_ph_IPA} reveals that the overall temperature dependence of $\mu_{ph,\mathrm{HEOM}}^\mathrm{IPA}$ is determined by the competition of the temperature effects on the phonon propagator and on the product of particle and hole propagators [see also Fig.~\ref{fig:dijag}(b)].
While the former promote $\mu_{ph,\mathrm{HEOM}}^\mathrm{IPA}$ increasing with temperature due to the enhanced phonon occupation, and dominate at lower temperatures, the latter promote $\mu_{ph,\mathrm{HEOM}}^\mathrm{IPA}$ decreasing with temperature, similarly to $\mu_{e,\mathrm{HEOM}}^\mathrm{IPA}$ [see Eq.~\eqref{Eq:mu_e_IPA} and Fig.~\ref{fig:selective_regime}(b)], and dominate at higher temperatures.
Therefore, $\mu_{ph,\mathrm{HEOM}}^\mathrm{IPA}$ reaches its maximum in the intermediate-temperature range.
In addition to the above-discussed Wick-decoupled contribution [see also Eq.~\eqref{eq:muph_aux}], the full solution $\mu_{ph,\mathrm{HEOM}}^\mathrm{full}$ contains both the genuine particle--hole correlations, which are behind the difference between $\mu_{e,\mathrm{HEOM}}^\mathrm{full}$ and $\mu_{e,\mathrm{HEOM}}^\mathrm{IPA}$ (and also the vertex corrections in the Holstein model~\cite{2024_Jankovic}), and genuine electron--phonon correlations, which are formally embodied in the first- and second-tier HEOM auxiliaries~\cite{2025_Jankovic_I}.
These all stem from the cluster expansion-based decomposition~\cite{Kirabook} of the two-electron--two-phonon correlation function
$$\langle c^\dagger_{\bf k_2 + q_2}(t) c_{\bf k_2 }(t) X_{\bf q_2}(t) c_{\bf k_1 + q_1}^\dagger c_{\bf k_1} X_{\bf q_1} \rangle$$
determining $\mu_{ph}$.
In Sec.~\ref{sec:mob_res_wide_range}, we argue that the electron--phonon correlations are ultimately responsible for turning the nonmonotonic temperature dependence of $\mu_{ph,\mathrm{HEOM}}^\mathrm{IPA}$ into the monotonic decrease of $\mu_{ph,\mathrm{HEOM}}^\mathrm{full}$ with temperature.
%
We finally note that, for similar values of $\omega_0,\lambda,$ and $T$, vertex corrections were found to be less pronounced in the Holstein model \cite{2024_Jankovic}.
In contrast, their stronger impact in the Peierls model can be attributed to the pronounced momentum dependence of the electron--phonon matrix elements $g_{k,q}$ in both $ k $ and $ q $ [see Eq.~\eqref{eq:def_peierls}], as well as to the genuine electron--phonon correlations, which are not present in models without the phonon-assisted current.

The counterparts of the IPA and full results within the Boltzmann approach, labeled “Boltzmann SERTA” and “Boltzmann full,” are in complete disagreement with the HEOM results, indicating that we are far outside the quasiparticle regime where the Boltzmann method can be reliably applied. Nevertheless, it should be noted that Boltzmann approach correctly predicts that the vertex corrections in $\mu$ are negative.

All other methods (MA, SCMA, CE) are calculated within IPA, and thus need to be compared with HEOM IPA.
The above rationalization of the temperature dependence of $\mu_{ph}$ within the IPA is general and not limited to the HEOM method.
Indeed, Fig.~\ref{fig:selective_regime}(c) reveals that MA, SCMA, and CE capture the nonmonotonic temperature dependence of the HEOM IPA benchmark.
However, the MA shows quantitative discrepancies when compared to HEOM IPA, while the functional form of $ \mu(T)$ at high temperatures, where the phonon-assisted contribution dominates, predicts a significantly steeper increase of $ 1/\mu $ with temperature than the benchmark.
To improve upon the predictions of MA, one can use either SCMA or CE. Both of them are in very good agreement with HEOM IPA. However, while SCMA has self-consistency, CE is completely one-shot. This property makes it a computationally efficient and appealing alternative to other methods.
\begin{figure}[t!]
  \centering
  \includegraphics[width=0.91\linewidth]{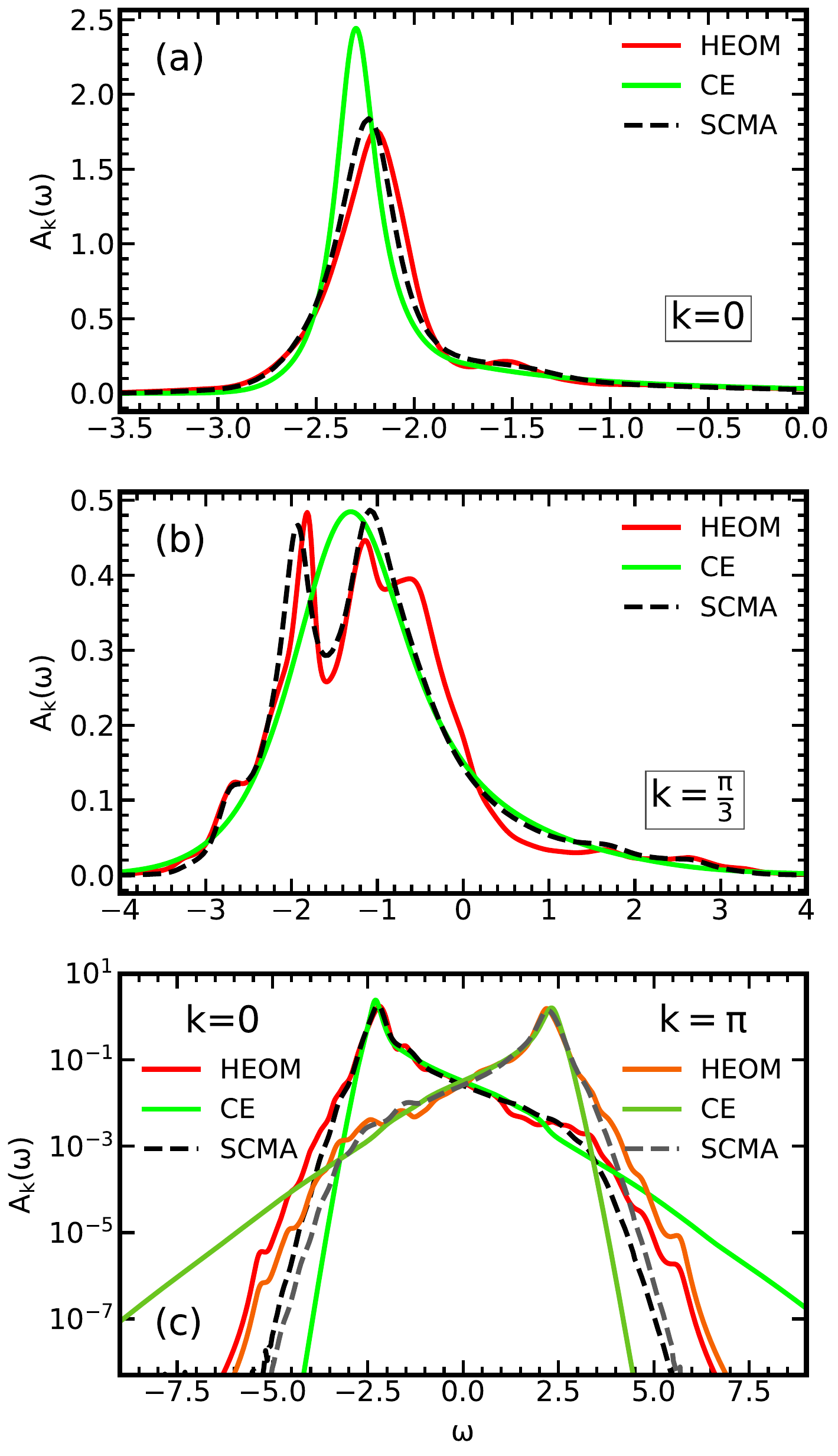}
  \caption{Comparison of HEOM, CE, and SCMA spectral functions for $\omega_0=0.5$, $\lambda=0.25$, $T=1.0$.}
  \label{fig:specf}
\end{figure}

It is worth noting that while CE results can be readily obtained for temperatures higher than those shown in Fig.~\ref{fig:selective_regime}, calculations could not be performed at lower temperatures (\(T < 0.55\)), as we were unable to achieve convergence with respect to the frequency cutoff \(\Lambda\). This limitation is not present for SCMA, which can provide results even at much lower temperatures. To understand the origin of this behavior, we plot in Fig.~\ref{fig:specf} the spectral functions used to calculate \(\mu\). Panels~(a) and~(b) demonstrate that CE captures only the overall shape of the HEOM benchmark, whereas SCMA yields a solution in much better agreement with HEOM, even resolving the multipeak structure in Panel~(b). Nevertheless, the overall shape correctly captured by CE is sufficient to produce accurate mobility results in the temperature range presented in Fig.~\ref{fig:selective_regime}.
\begin{figure*}[t!]
    \centering
    \includegraphics[width=\textwidth]{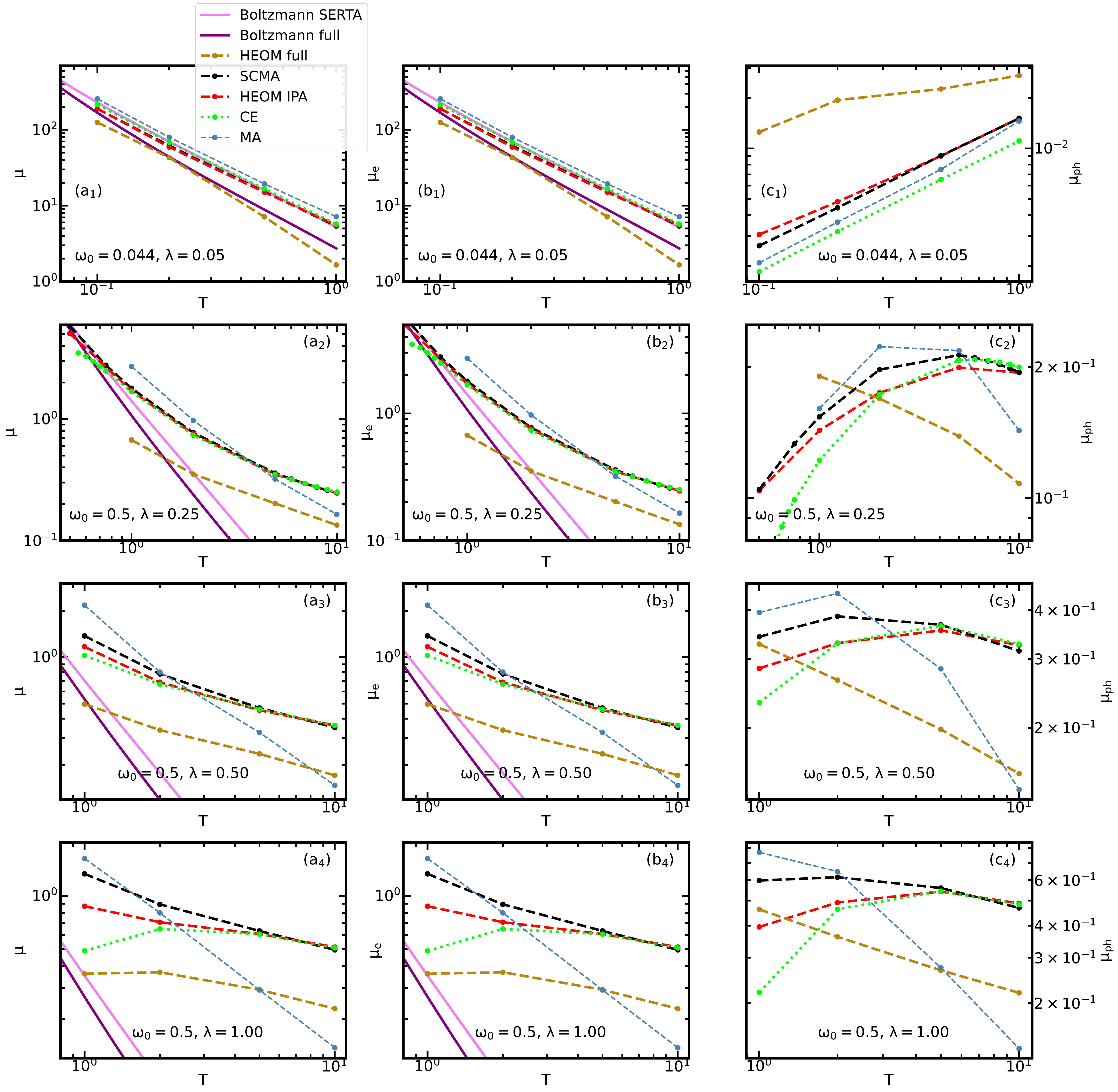}
    \caption{Comparison of different methods for calculating the (a) the total mobility $\mu$ (see Eq.~\eqref{eq:total_mobility_4342}), (b) the electronic contribution to the mobility $\mu_e$ (see Eq.~\eqref{muedc}), and (c) the phononic contribution to the mobility $\mu_{ph}$ (see Eq.~\eqref{eq:muph}).}
    \label{fig:all_mob_1}
\end{figure*}
\begin{figure*}[t!]
    \centering
    \includegraphics[width=\textwidth]{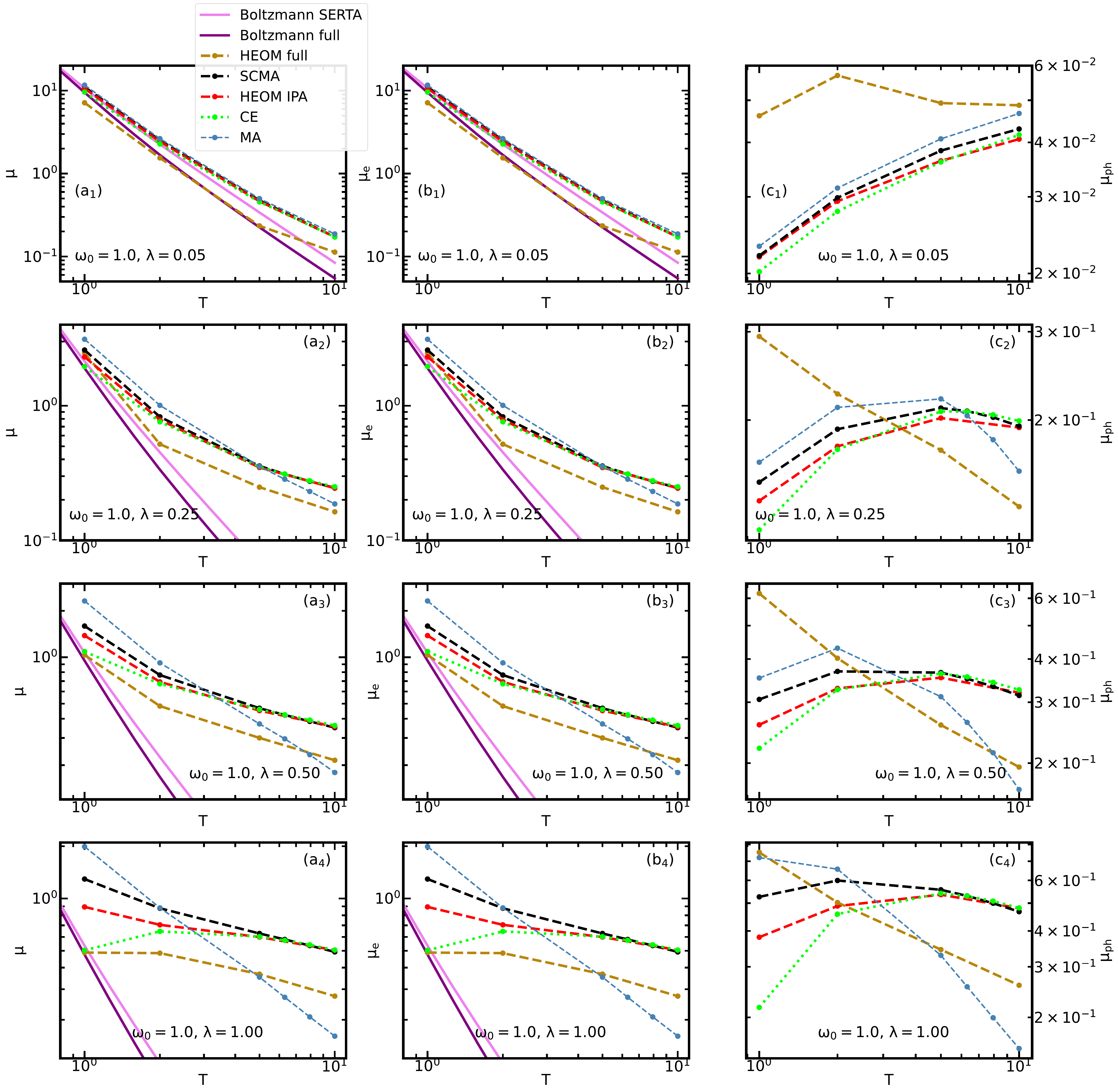}
    \caption{Comparison of different methods for calculating the (a) the total mobility $\mu$ (see Eq.~\eqref{eq:total_mobility_4342}), (b) the electronic contribution to the mobility $\mu_e$ (see Eq.~\eqref{muedc}), and (c) the phononic contribution to the mobility $\mu_{ph}$ (see Eq.~\eqref{eq:muph}), for $\omega_0=1$ in a wide range of electron-phonon coupling strenghts $\lambda$.}
    \label{fig:all_mob_2}
\end{figure*}
%
At low temperatures, however, due to the Boltzmann factor \(e^{-\beta \omega}\) in Eqs.~\eqref{muedc}, \eqref{eq:muph}, and \eqref{eq:def_ne_4342}, the behavior of the spectral functions at large negative frequencies contributes more strongly. This effect is most pronounced in the normalization \(n_e\) [see Eq.~\eqref{eq:def_ne_4342}], which appears in the denominator of Eqs.~\eqref{muedc} and~\eqref{eq:muph}. As can be seen, the factor \(e^{-\beta \nu} A_k(\nu)^2\) in the numerator of Eq.~\eqref{muedc} is less sensitive to the frequency cutoff \(\Lambda\) (for large \(\Lambda\)) than the factor \(e^{-\beta \nu} A_k(\nu)\) in the denominator, since \(A_k(\nu)^2\) decays faster than \(A_k(\nu)\) for \(\nu \to -\infty\).  It is precisely these large negative frequencies that are problematic in CE: as seen from Fig.~\ref{fig:specf}(c), CE exhibits an unphysically large tail towards negative frequencies, which makes convergence with respect to \(\Lambda\) difficult. This artifact of the CE solution will be discussed in more detail in the following section and from a different point of view in Sec.~\ref{sec:spectral_sum_rules}.
\subsubsection{Mobility results in a wide range of parameter regimes}
\label{sec:mob_res_wide_range}

In Figs.~\ref{fig:all_mob_1} and \ref{fig:all_mob_2}, we present the mobility results over a wide range of parameter regimes. We do not discuss each panel individually, as the main conclusions are analogous to those from the previous subsection. Specifically: i) Vertex corrections are clearly visible but do not alter the order of magnitude of the results. In contrast to the Holstein model~\cite{2024_Jankovic}, however, we now also observe non-negligible vertex corrections even at weak interaction strengths [see, for example, Fig.~\ref{fig:all_mob_2}($a_1$)]. This can be attributed to the strong dependence of the matrix elements \( g_{k,q} \) on both \( k \) and \( q \).
Additionally, in contrast to the intermediate- and strong-interaction regimes, the IPA qualitatively captures the temperature dependence of $\mu_{ph}$ in the weak-interaction regime [see, for example, Fig.~\ref{fig:all_mob_1}($c_1$)].
Then, one expects that the genuine electron--phonon correlations are weak, so that the quantitative discrepancies between the HEOM full and HEOM IPA results for both $\mu_{ph}$ and $\mu_e$ originate from the particle--hole correlations that the IPA misses.
Their absence does not alter the correct temperature dependence of $\mu_e$, at least in the ranges of the temperature and interaction strength we examine [see Figs.~\ref{fig:all_mob_1}($b_1-b_4$) and~\ref{fig:all_mob_2}($b_1-b_4$)], and the same then applies to $\mu_{ph}$ [see Fig.~\ref{fig:all_mob_1}($c_1$)].
The above reasoning then strongly suggests that the qualitatively incorrect temperature dependence of $\mu_{ph}$ within the IPA [see Figs.~\ref{fig:all_mob_1}($c_2-c_4$) and~\ref{fig:all_mob_2}($c_2-c_4$)] is due to the neglect of genuine electron--phonon correlations, and not due to the neglect of particle--hole correlations.
%
ii) The CE approach successfully reproduces the correct IPA result well beyond the range of validity of the Boltzmann quasiparticle picture. Unlike CE, which is a one-shot method, SCMA incorporates self-consistency, allowing it to achieve accurate results that also agree with the numerically exact HEOM IPA benchmark. Overall, both CE and SCMA provide a clear improvement over MA, which often deviates from the exact result except in the very weak coupling regime.

At lower temperatures, it becomes increasingly difficult to converge the CE results (this issue does not arise within SCMA). We verified that this problem is related to the long tail of the CE spectral functions extending toward negative frequencies, in close analogy with Fig.~\ref{fig:specf}(c); see the corresponding spectral functions in Sec.~B of SM \cite{SuppMat}. We also observed that this issue is even more pronounced 
at stronger coupling strengths. This behavior is expected, as the spectral functions become broader in this regime, thereby enhancing the contribution of the tail. 

In these parameter regimes, the following issue arises. 
Although it is sometimes possible, with considerable effort, to converge the CE mobility results with respect to the frequency cutoff $\Lambda$ at stronger couplings and lower temperatures, CE typically underestimates the HEOM IPA benchmark in these cases [see, for example, Fig.~\ref{fig:all_mob_1}($a_4$)].  
This can also be understood as a direct consequence of the spectral function tail. As discussed at the end of Sec.~\ref{subsec:num_challenge}, the tail has the largest impact on $n_e$ (see Eq.~\eqref{eq:def_ne_4342}), making it unphysically large. Since $n_e$ appears in the denominator of the expressions for $\mu_e$ and $\mu_{ph}$ [see Eqs.~\eqref{muedc} and \eqref{eq:muph}], these quantities within CE are systematically underestimated compared to the exact solution. 

We note that the issues with the tail of the spectral function that we identify are not the same as the ones identified in Ref.~\cite{2025_Lihm_nonperturbative}. Therein, the authors found a tail with a slow $1/\omega^4$ decay, which originates from artificial Lorentzian broadening of the bare Green's function. We do not introduce any artificial broadening and hence the presence of the tail that we identify is an inherent shortcoming of the CE method.

\subsection{Analytical Rationale for Employing CE}
\label{sec:spectral_sum_rules}

In the previous subsection, we saw that using CE, which can be interpreted as a postprocessing method, improves the predictions of MA in the case of the Peierls model. Similar conclusions were previously drawn for the Holstein model as well~\cite{2023_Mitric}. However, it is desirable to develop a more general analytical understanding of why and under which conditions CE is expected to perform well. Such understanding would not only place the CE prescription on a firmer theoretical footing, but could also provide criteria for anticipating---without relying on benchmarks---the parameter regimes in which we could expect the CE approach to be accurate and not suffer from convergence (with respect to frequency cutoff $\Lambda$) problems.


We propose that an insight into this important question can be obtained by examining the spectral sum rules. The motivation for this approach stems from the observation that CE performs very well at higher temperatures, where only the overall shape of the spectral function plays a role, whereas the long tail extending toward negative frequencies causes difficulties at lower temperatures (see Sec.~\ref{subsec:num_challenge}). Both of these features are, in principle, encoded in the (first few) spectral sum rules
\begin{equation} \label{Eq:sum_rule_def}
    \mathcal{M}_n ({\bf k}) = \int_{-\infty}^\infty A_{\bf k}(\omega) \omega^n d\omega.
\end{equation}
For the Hamiltonian in Eq.~\eqref{eq:1}, in the case we are examining (when  the chemical potential is $\mu_F \to-\infty$), these can be calculated analytically
\begin{equation} \label{eq:sum_rule_practical}
    \mathcal{M}_n ({\bf k}) = \left\langle 
    \underbrace{\left[\dots \left[\left[ c_{\bf k}, H\right],H\right] \dots ,H\right]}_{n\; \mathrm{times}} c_{\bf k}^\dagger
    \right\rangle_{T}.
\end{equation}
The first few read as follows:
\begin{align}
    \mathcal{M}_0 ({\bf k}) &= 1, \\ 
    \mathcal{M}_1 ({\bf k}) &= \varepsilon_{\bf k},  \\
    \mathcal{M}_2 ({\bf k}) &= \varepsilon_{\bf k}^2 + 
    (2\nph+1) \frac{1}{N} \sum_{\bf q} |g_{\bf k,q}|^2, \\
    \mathcal{M}_3 ({\bf k}) &= \varepsilon_{\bf k}^3 
    +\frac{\omega_0}{N} \sum_{\bf q} |g_{\bf k,q}|^2
    \nonumber \\ &+ 
    \frac{(2\nph+1)}{N} \sum_{\bf q} |g_{\bf k,q}|^2
    \left( \varepsilon_{\bf k+q} + 2\varepsilon_{\bf k} \right),
\end{align}
\begin{align}
        \mathcal{M}_4 ({\bf k}) &=
    \varepsilon_{\bf k}^4 
    +\frac{2\omega_0}{N} \sum_{\bf q} (\varepsilon_{\bf k+q} + \varepsilon_{\bf k} )
    |g_{\bf k,q}|^2 \nonumber \\
    +& \frac{2\nph  + 1}{N} 
    \sum_{\bf q} |g_{\bf k,q}|^2 (
    \varepsilon_{\bf k+q}^2 + 3\varepsilon_{\bf k}^2 + \omega_0^2 + 
    2\varepsilon_{\bf k} \varepsilon_{\bf k+q}) \nonumber \\
    +& \frac{(2\nph +1)^2}{N^2} \sum_{\bf q, q_1} \Bigg[
    |g_{\bf k,q}|^2 |g_{\bf k,q_1}|^2 + 
    |g_{\bf k,q}|^2 |g_{\bf q +k,q_1}|^2
    \nonumber \\
    &+  g_{\bf q+k ,-q}\,
     g_{\bf q_1+k ,q} \, g_{\bf q_1 + q + k, -q_1}\, g_{\bf k, q_1}
     \Bigg]. \label{eq:exact_sum_rules}
\end{align}
Continuing this calculation for higher-order \( n \) becomes increasingly cumbersome, as the number of terms generated by repeatedly applying the commutator in Eq.~\eqref{eq:sum_rule_practical} (prior to taking the thermal expectation value \( \langle \dots \rangle_T \)) grows rapidly. In contrast, evaluating \( \mathcal{M}_n(\mathbf{k}) \) within the CE method is relatively easy, even for higher orders of \( n \)~\cite{2023_Mitric}.
\begin{multline}
\mathcal{M}_n^{\mathrm{CE}}({\bf k}) = \mathrm{Re} \left[ 
    i^n \left( \frac{d}{dt} \right)^n e^{C_{\bf k}(t)} 
    \right] \bigg\rvert_{ t=0} \\
    -
    \sum_{p=1}^n {\binom{n}{p}} (-\varepsilon_{\bf k})^p \mathcal{M}_{n-p}^{\mathrm{CE}}({\bf k}). \label{eq:ce_sum_rules_442}
\end{multline}
Therefore, it is straightforward to verify that $\mathcal{M}_n^{\mathrm{CE}}({\bf k}) = \mathcal{M}_n({\bf k})$ for $n < 4$, while 
\begin{multline}
    \mathcal{M}_4^{\mathrm{CE}}({\bf k}) - \mathcal{M}_4({\bf k}) = 
    \frac{(2\nph +1)^2}{N^2} \sum_{\bf q, q_1} \Bigg[
    2|g_{\bf k,q}|^2 |g_{\bf k,q_1}|^2  \\ 
    -|g_{\bf k,q}|^2 |g_{\bf q +k,q_1}|^2
    -  g_{\bf q+k ,-q}\,
     g_{\bf q_1+k ,q} \, g_{\bf q_1 + q + k, -q_1}\, g_{\bf k, q_1}
     \Bigg], \label{eq:ce_sum_rule_4}
\end{multline}
This expression actually vanishes in the case of a \(\mathbf{k}\)-independent coupling constant \( g_{\bf k,q} \). This implies that CE reproduces the sum rules exactly for \( n \leq 4 \) both in the case of  Holstein and (experimentally more relevant) Fr\"ohlich model. In contrast, Eq.~\eqref{eq:ce_sum_rule_4} indicates that \(\mathbf{k}\)-dependent models are more challenging for CE. The most prominent example of such a case is the Peierls model analyzed in this paper. Nevertheless, the difference \( \mathcal{M}_4^{\mathrm{CE}}(\mathbf{k}) - \mathcal{M}_4(\mathbf{k}) \) remains small when the coupling is not too strong, being of order \( \mathcal{O}(g^4) \). 

However, the fact that MA correctly reproduces the spectral sum rules for \( n < 4 \), and that \( \mathcal{M}_4^{\mathrm{CE}}(\mathbf{k}) - \mathcal{M}_4(\mathbf{k}) \) is of order \( \mathcal{O}(g^4) \), is not sufficient to explain why CE should, as observed in Sec.~\ref{sec:mob_res_wide_range}, improve upon the predictions of MA in the Peierls and other \(\mathbf{k}\)-dependent \( g_{\bf k,q} \) models. In fact, MA exhibits the same property: it exactly reproduces the sum rules for \( n < 4 \), while the error for \( n = 4 \) is of order \( \mathcal{O}(g^4) \). The same holds true even for SCMA which, as we have seen, also outperforms MA. These properties of MA and SCMA, as explained in Refs.~\cite{2006_Berciu,2006_Goodvin}, follow directly from the fact that both methods include exactly the lowest-order self-energy diagram, while neglect some $\sim g^4$ diagrams.

The analysis in Refs.~\cite{2006_Berciu,2006_Goodvin} can also be used to highlight the key difference between these approaches: MA neglects all higher-order diagrams, implying that the \( g^n \) contributions for \( n \geq 4 \) to \( \mathcal{M}_n(\mathbf{k}) \) are entirely absent in \( \mathcal{M}^{\mathrm{MA}}_n(\mathbf{k}) \). In contrast, SCMA at least partially captures some of these terms in \( \mathcal{M}^{\mathrm{SCMA}}_n(\mathbf{k}) \). The same holds for CE, as can be seen explicitly in Eqs.~\eqref{eq:exact_sum_rules} and \eqref{eq:ce_sum_rule_4} for \( n = 4 \).
\begin{figure}[t!]
  \centering
  \includegraphics[width=0.9\linewidth]{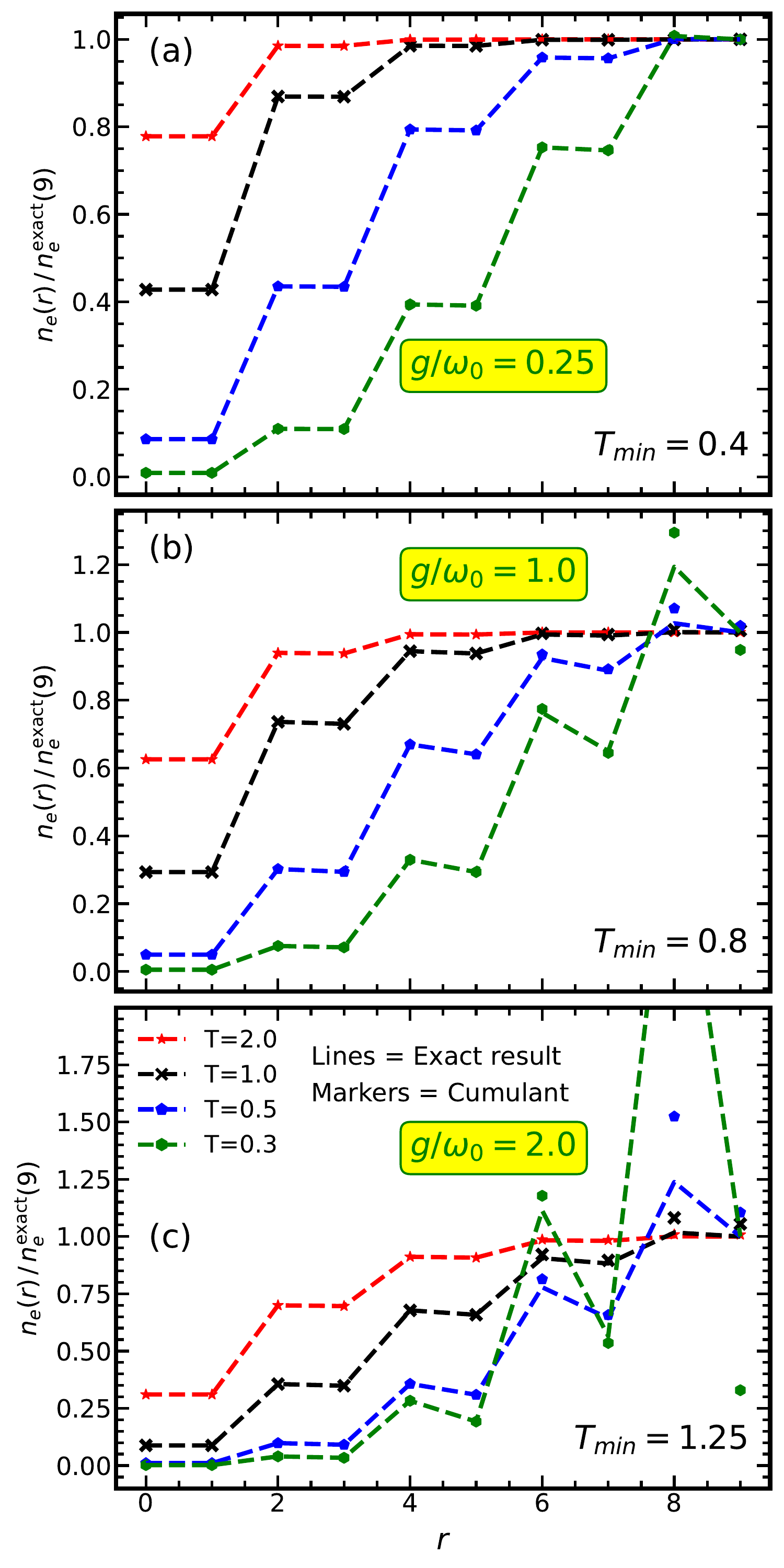}
  \caption{
Comparison of the \( n_e \) (normalized to the exact result for $n_e(r=9)$) predictions [see Eqs.~\eqref{eq:def_ne_4342}, \eqref{eq:ner_def_1321312}, and \eqref{eq:ner_def_1324444}] obtained using the exact and CE spectral sum rules for the Holstein model. In the lower right corner of each panel, the quantity \( T_{\mathrm{min}} \) is shown, representing the lowest temperature at which the CE mobility could be obtained in Ref.~\cite{2023_Mitric} for the corresponding parameter regime of the Holstein model. In all panels $\omega_0=0.5$.
}
  \label{fig:specsum_1}
  \vspace*{-0.1cm}
\end{figure}
\begin{figure}[t!]
  \centering
  \includegraphics[width=0.9\linewidth]{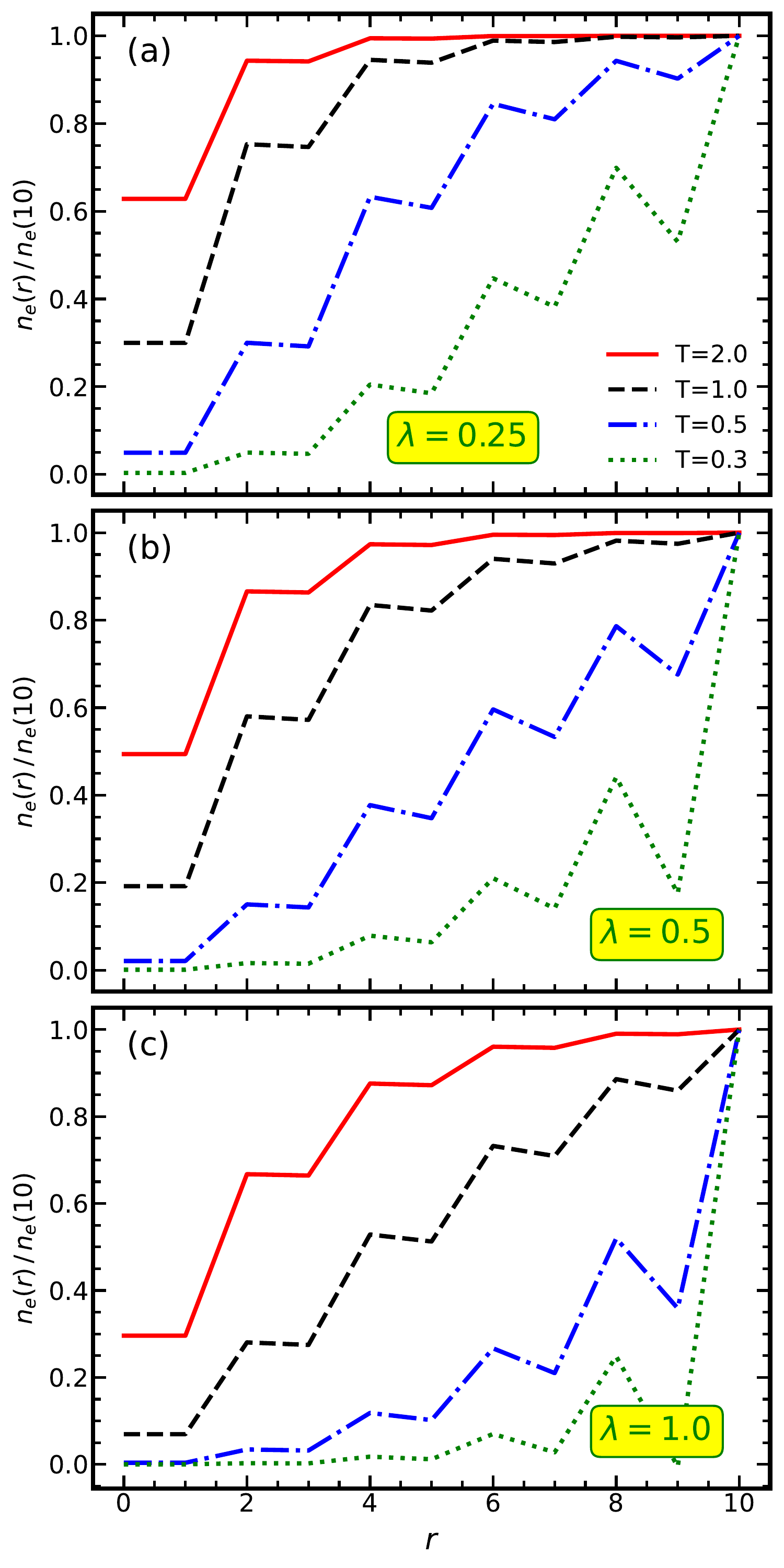}
  \caption{
CE predictions for \( n_e \) (normalized to $n_e(r=10)$) [see Eqs.~\eqref{eq:def_ne_4342}, \eqref{eq:ner_def_1321312}, and \eqref{eq:ner_def_1324444}] obtained for the Peierls model. Here $\omega_0=0.5$.
}
  \label{fig:specsum_2}
\end{figure}
To examine how accurately CE captures higher-order sum rules, we compute both the exact and CE sum rules for orders \( n \leq 9 \) in the Holstein model. This model is chosen because, although the calculation is still quite involved, it is technically the most straightforward in this case. Since the resulting analytical expressions are lengthy and cumbersome, they are presented in Sec.~C of SM \cite{SuppMat}. Here, instead, we focus on numerical results illustrating how the sum rules affect the quantity \( n_e \), defined in Eq.~\eqref{eq:def_ne_4342}. This quantity is particularly relevant because \( n_e \) can be explicitly expressed in terms of \( \mathcal{M}_n(\mathbf{k}) \) 
\begin{align}
    n_e(r) &= \sum_{n=0}^r \frac{(-\beta)^n}{n!} \sum_k \mathcal{M}_n(k), \label{eq:ner_def_1321312}\\
    n_e &= \lim_{n\to\infty} n_e(r), \label{eq:ner_def_1324444}
\end{align}
and, as discussed in Sec.~\ref{subsec:num_challenge}, it appears as the normalization factor in the mobility expressions [see Eqs.~\eqref{muedc} and \eqref{eq:muph}] that has proven to be most sensitive to the presence of unphysical tail in the spectral function.

The results for three different interaction strengths are shown in Fig.~\ref{fig:specsum_1}. We observe that, although the higher-order sum rules are not exact, their contributions to \( n_e \) remain in good agreement with the exact result, not only at very weak couplings but also at stronger coupling strengths. As the temperature is lowered or the interaction strength is increased, an increasingly larger number of \( r \) terms in Eq.~\eqref{eq:ner_def_1324444} is required before \( n_e(r) \) starts to converge. In this regime, the difference \( \mathcal{M}_n^{\mathrm{CE}}(\mathbf{k}) - \mathcal{M}_n(\mathbf{k}) \), which grows with increasing \( n \), becomes more pronounced in the resulting \( n_e \), leading to a visible discrepancy between the CE prediction and the exact result. 

This observation can be used to address the motivating question of this subsection: to formulate a rough criterion for identifying the regimes in which CE is expected to both converge (with respect to the frequency cutoff \( \Lambda \)) and yield accurate results. These regimes correspond to cases where the number \( r \) of spectral sum rules required for \( n_e(r) \) to converge is not very large (than say \( r \sim 10 \)). From another perspective, if \( r \gg 10 \) terms were needed, the problematic negative-frequency tail of the CE spectral function (see Sec.~\ref{subsec:num_challenge}) will have a significant impact on the result, leading to inaccuracies. As numerical support for this criterion, we again refer to Fig.~\ref{fig:specsum_1}, where the lower-right corner of each panel shows the minimal temperature (that we read off from Fig.~11 of Ref.~\cite{2023_Mitric}) for which CE mobility results in the Holstein model could be obtained in the corresponding parameter regime. This temperature is nearly the same as the temperature for which $r\sim 10$ terms are no longer sufficient to obtain $n_e$.

The numerical criterion that we formulated is quite convenient as we do not need the exact results. We only need CE spectral sum rules, and these are quite easily calculated, as shown in Eq.~\eqref{eq:ce_sum_rules_442}. For example, in Fig.~\ref{fig:specsum_2} we show the results for the Peierls model (the corresponding analytical expressions are given in Sec.~C of SM \cite{SuppMat}). From these results, we can estimate that the minimal temperature at which CE converges and provides reasonably accurate results is \( T_{\mathrm{min}} \approx 0.5 \) for \( \lambda = 0.25 \) (see Panel~(a)), \( T_{\mathrm{min}} \approx 1.0 \) for \( \lambda = 0.5 \) (see Panel~(b)), and \( T_{\mathrm{min}} \approx 2.0 \) for \( \lambda = 1.0 \) (see Panel~(c)). These estimates are consistent with the mobility results shown in Fig.~\ref{fig:all_mob_1}(\( a_2 \))--(\( a_4 \)).

\begin{figure}[t!]
  \centering
  \includegraphics[width=0.9\linewidth]{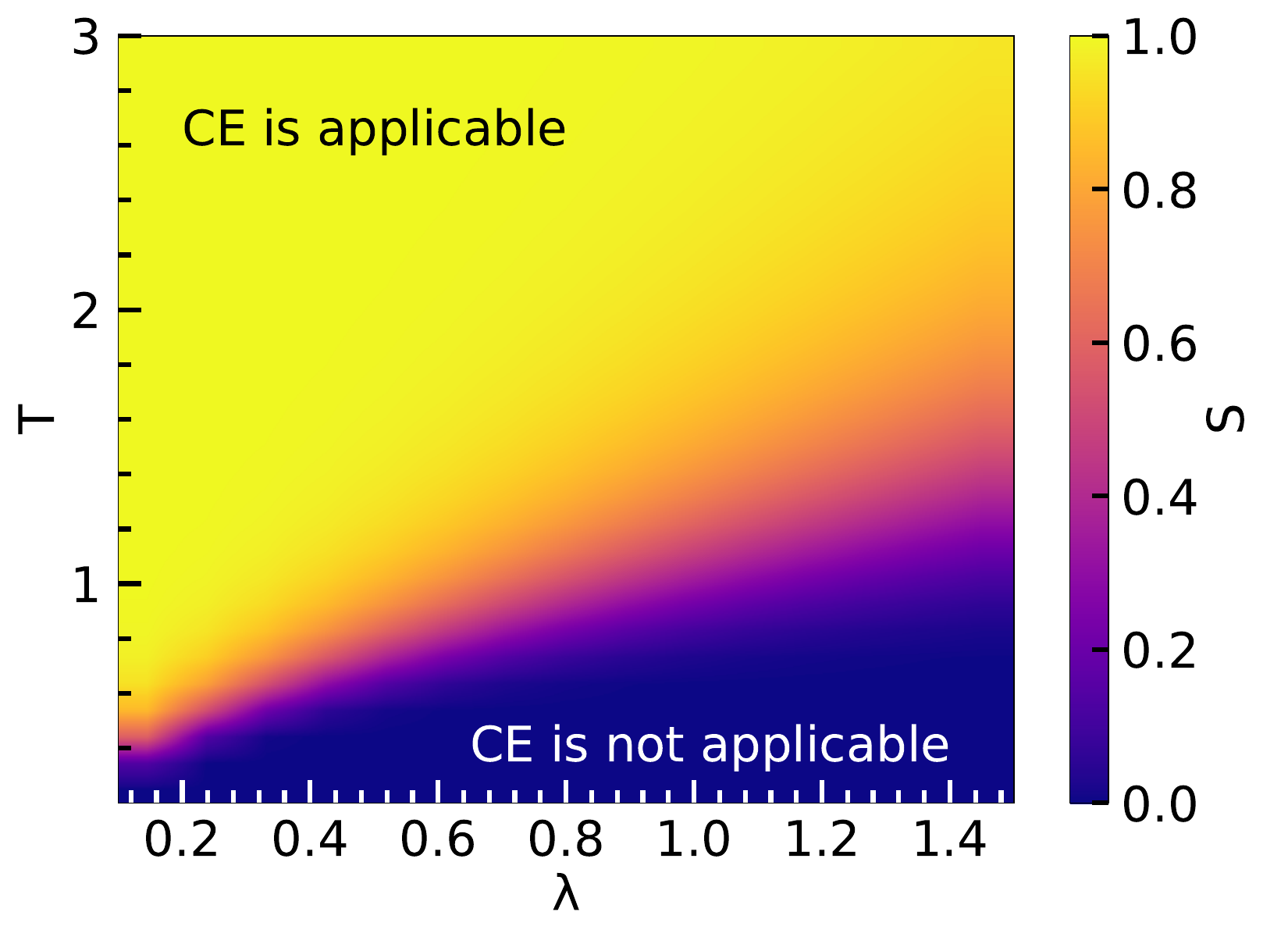}
  \caption{
Illustration of the regions where CE is expected to be accurate in the Peierls model for $\omega_0=0.5$. $S$ is given by Eq.~\eqref{eq:defining_S}, with $r_\mathrm{max}=12$ and $\eta = 0.2$.
}
  \label{fig:ce_heatplot}
\end{figure}

Lastly, these considerations allow us to construct Fig.~\ref{fig:ce_heatplot}, which illustrates regions in the \((\lambda,T)\) parameter space where the CE approach is expected to be reliable. To this end, we introduce a scalar quantity \(S\) that quantifies how close the truncated result \(n_e(r_{\mathrm{max}})\) is to its asymptotic value \(n_e(r\to\infty)\),
\begin{equation}
\label{eq:defining_S}
S(r_{\mathrm{max}})=
\exp\!\left[
-\frac{\Delta n_e(r_{\mathrm{max}})
+\Delta^2 n_e(r_{\mathrm{max}})}
{\eta\,|n_e(r_{\mathrm{max}})|}
\right].
\end{equation}
Here,
\(\Delta n_e(r_{\mathrm{max}})=|n_e(r_{\mathrm{max}})-n_e(r_{\mathrm{max}}-1)|\), 
\(\Delta^2 n_e(r_{\mathrm{max}})=|n_e(r_{\mathrm{max}})-2n_e(r_{\mathrm{max}}-1)+n_e(r_{\mathrm{max}}-2)|\), while $\eta$ is the parameter defining relative convergence tolerance which we take to be $\eta=0.2$ to best illustrate our results.
We emphasize that the inclusion of both \(\Delta n_e\) and \(\Delta^2 n_e\) in Eq.~\eqref{eq:defining_S} is essential. While \(\Delta n_e(r_{\mathrm{max}})\) alone might appear sufficient as a convergence criterion, this can be misleading in practice. In particular, for not too strong couplings one often finds
\(|n_e(r)-n_e(r-1)| \ll |n_e(r)-n_e(r-2)|\),
reflecting the fact that, for odd \(r\), certain contributions to \(\mathcal{M}_r(k)\) are odd in \(k\) and therefore vanish upon summation in Eq.~\eqref{eq:ner_def_1321312}. The inclusion of \(\Delta^2 n_e\) accounts for this effect and thereby provides a more reliable convergence criterion.

\subsection{Fr\"ohlich model}
\label{sec:frohlich_model-53232}
\subsubsection{General remarks}

The CE method in the context of the Fr\"ohlich model was previously examined in Ref.~\cite{2018_Nery}, but only at zero temperature. In the following, however, we turn our attention to the finite-temperature transport properties of the Fr\"ohlich model. Before proceeding, let us note that the results of Ref.~\cite{2018_Nery}, among other things, demonstrate that the quasiparticle properties obtained using CE are significantly closer to the numerically exact benchmark (QMC \cite{2000_Mishchenko} in their case) than those predicted by the one-shot Migdal approximation. The fact that this observation is fully consistent with the conclusions of Sec.~\ref{sec:qp_prop4} for the Peierls model and of Ref.~\cite{2023_Mitric} for the Holstein model suggests that some general, largely model-independent conclusions can be drawn regarding the advantages and limitations of the CE method. Consequently, as we will see, the conclusions reached for the Peierls model also hold for the Fr\"ohlich model. For this reason, we will not provide a detailed analysis as in Sec.~\ref{sec:peierls_model-53232}, but will instead summarize only the main results.

The expression for the mobility, within IPA, is entirely determined by Eq.~\eqref{muedc}, as there is no ${\bf k}$-dependence to electron-phonon matrix elements $g_{\bf k,q}$, implying that the phonon-assisted part in the current operator vanishes; see Eq.~\eqref{eq:current}. Taking that in to account, as well as the spherical symmetry of the problem, we see that the expression for the mobility, in this case, reduces to 
\begin{equation} \label{eq:frohlich_mobility_expression}
    \mu = \frac{\beta\pi}{3} \frac{\int_0^\infty dk \, k^4 
    \int_{-\infty}^\infty d\nu \; e^{-\beta\nu} A_k(\nu)^2}{
    \int_0^\infty dk \, k^2
    \int_{-\infty}^\infty d\nu \; e^{-\beta\nu} A_k(\nu)}{
    },
\end{equation}
where the factor $1/3$ appeared as a consequence of the dimensionality of the problem. 


To apply this formula, the CE and SCMA spectral functions will be calculated using the procedure presented in Secs.~\ref{subsec:frohlich_ce}~and~\ref{subsec_general_migldal_scma}, respectively. For the SCMA, however, we note that the expression given by Eq.~\eqref{eq:selfenSCMA} can be cast into a more convenient form using the explicit form of the electron-phonon matrix elements for the Fr\"ohlich model (see Eq.~\eqref{eq:def_frohlich}), as well as the spherical symmetry of the problem
\begin{multline}
    \Sigma_{\bf k}^\mathrm{SCMA}(\omega) = \frac{|\mathcal{M}_0|^2}{(2\pi)^2}
    \int_0^\infty dq \, \Big[ (\nph +1) G_q(\omega - \omega_0) 
    \\ + \nph G_q(\omega + \omega_0) \Big] 
    \frac{q}{2k} \ln \left( \frac{q+k}{q-k} \right)^2.
\end{multline}

\vspace*{-0.5cm}

\subsubsection{Numerical results}
\begin{figure}[t!]
  \centering
  \includegraphics[width=0.9\linewidth]{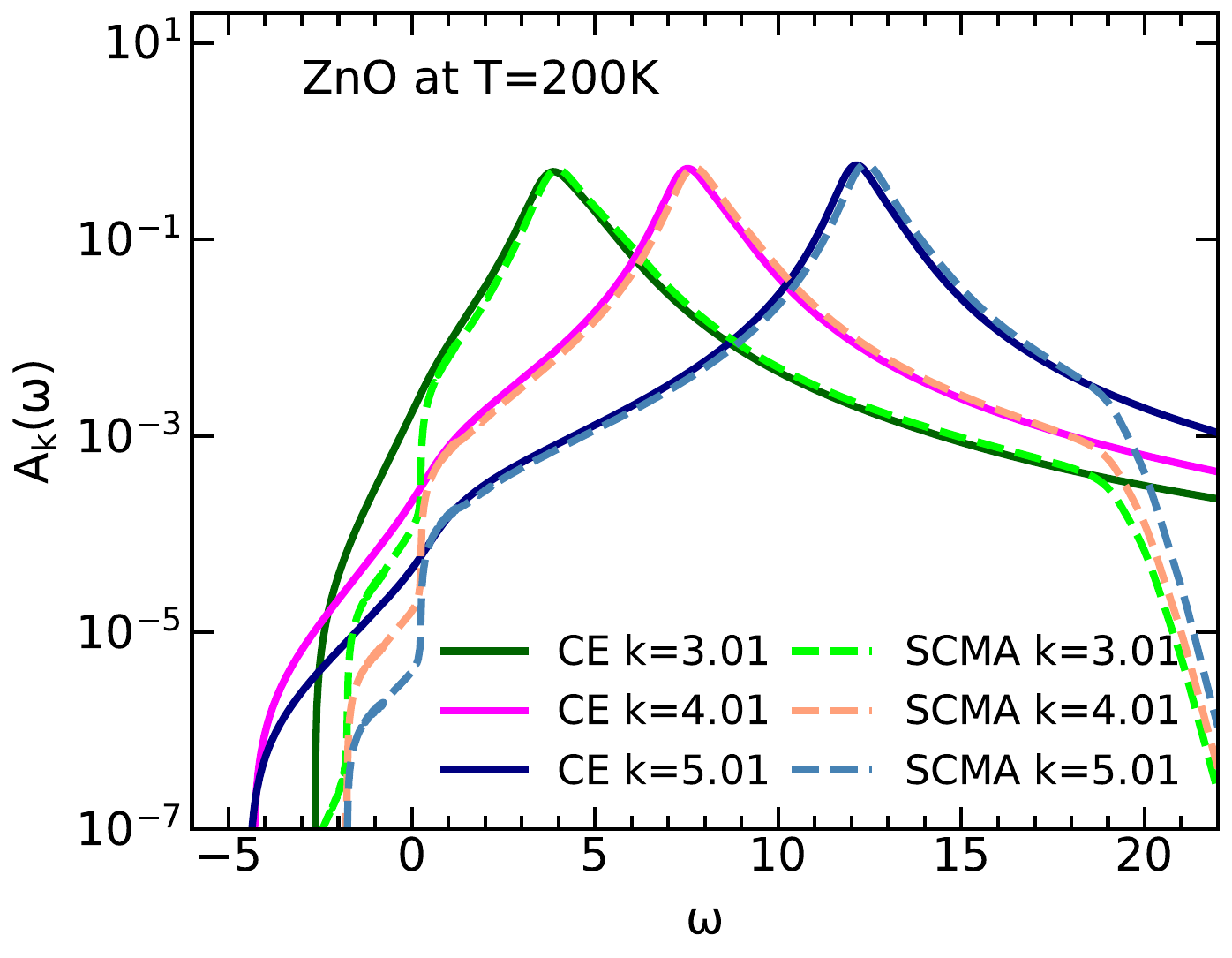}
  \caption{
CE and SCMA spectral functions $A_k(\omega)$, for different momenta $k$, of ZnO at a temperature of $T = 200K$.}
  \label{fig:spec_frolih}
\end{figure}
\begin{figure*}[t!]
    \centering
    \includegraphics[width=\textwidth]{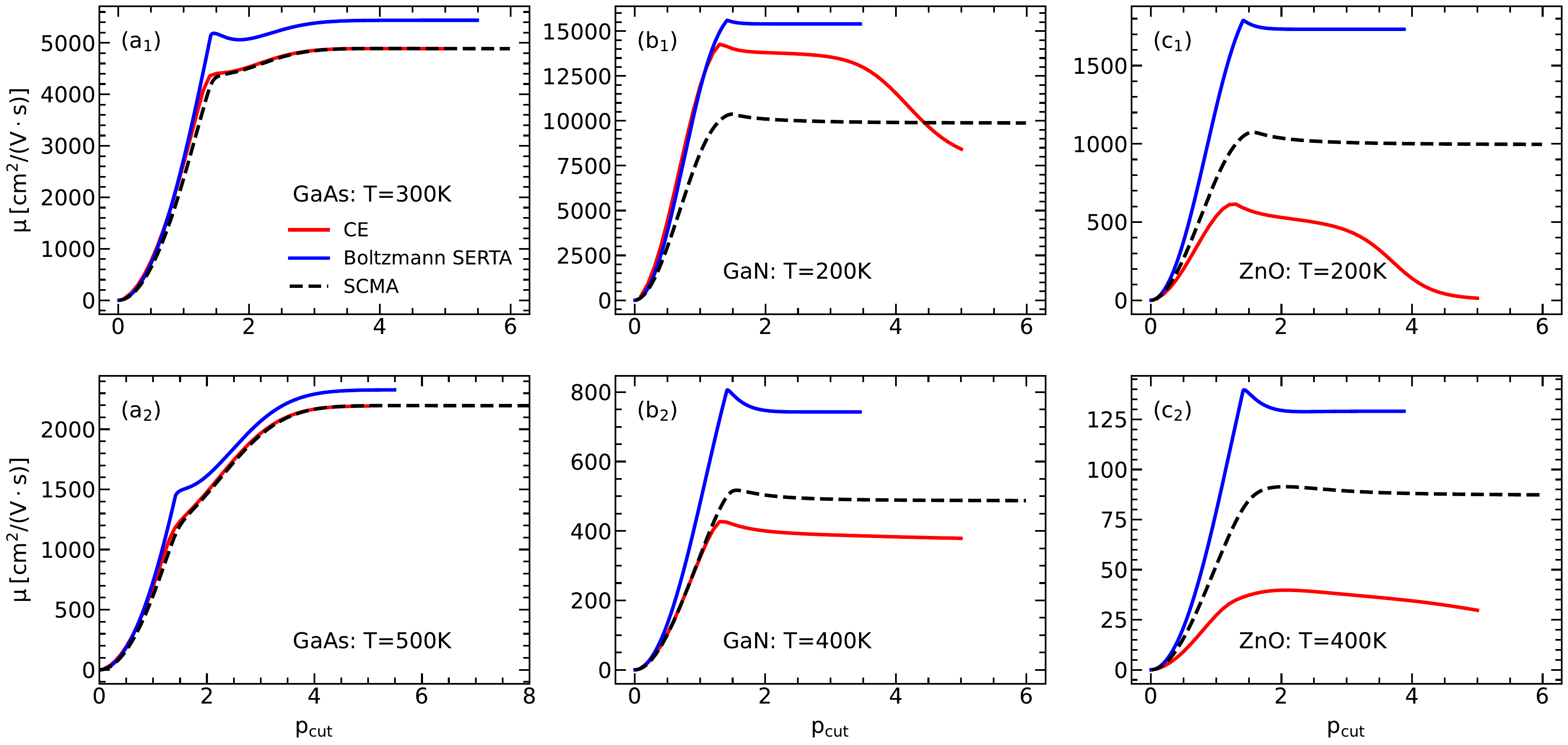}
    \caption{The dependence of mobility on the momentum cutoff $p_{cut}$ for GaAs, GaN, and ZnO at two different temperatures. 
Results are shown for both the Boltzmann SERTA, SCMA and the CE methods.
}
    \label{fig:pcut_conv}
\end{figure*}
\begin{figure}[t!]
  \centering
  \includegraphics[width=0.9\linewidth]{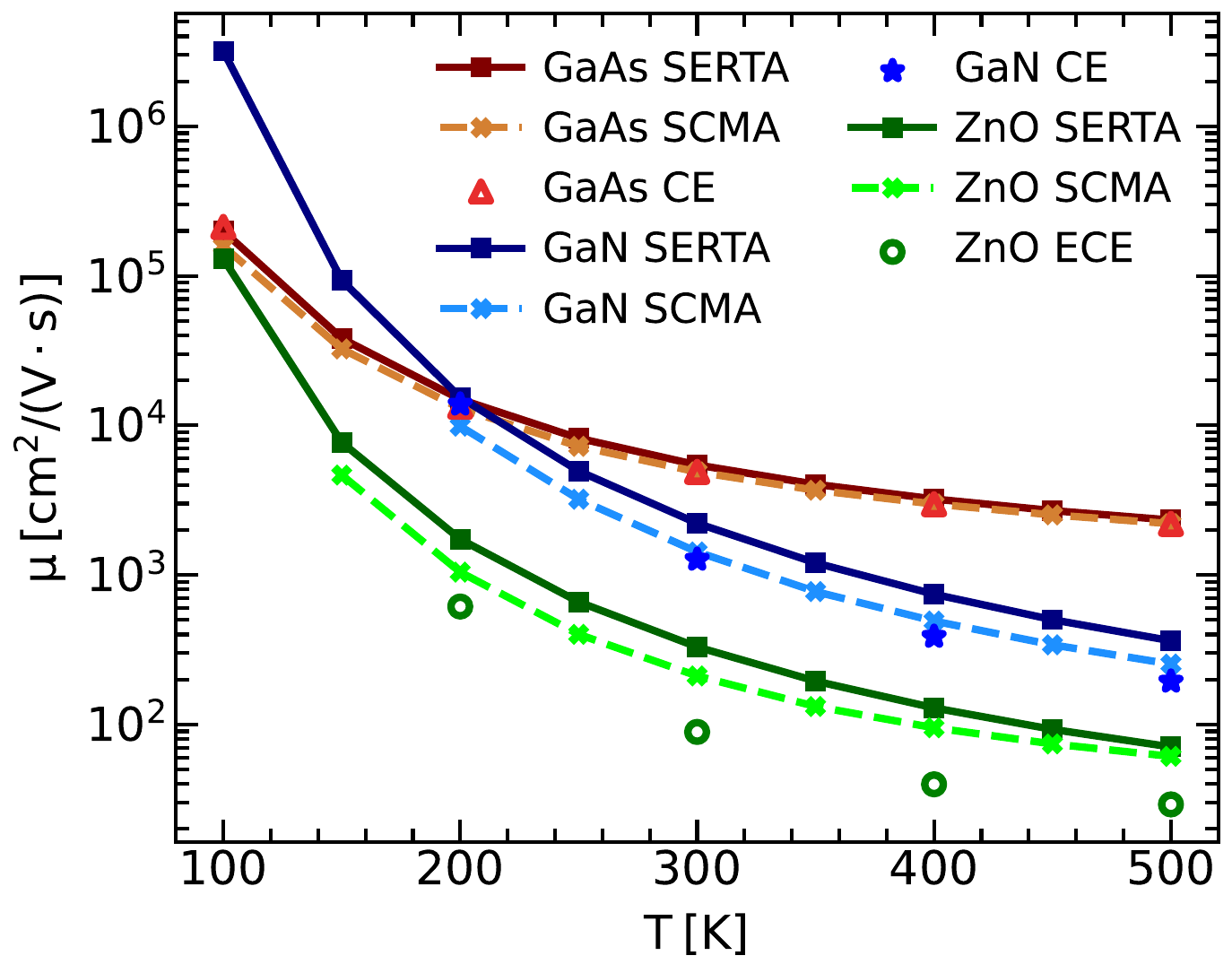}
  \caption{
Temperature dependence of mobility, calculated using Boltzmann SERTA, SCMA and CE approaches, for GaAs, GaN, and ZnO.
}
  \label{fig:mob_frolih}
\end{figure}

The calculations were performed using parameters representative of the conduction bands in three polar semiconductors: GaAs ($\alpha = 0.072,\; \omega_0 = 36\,\mathrm{meV},\; m_0/m_e = 0.067$), GaN ($\alpha = 0.407,\; \omega_0 = 91.2\,\mathrm{meV},\; m_0/m_e = 0.20$), and ZnO ($\alpha = 1.022,\; \omega_0 = 72.3\,\mathrm{meV},\; m_0/m_e = 0.275$), where $m_0/m_e$ denotes the ratio of the electron band mass to the free-electron mass. 
The effective mass parameters for GaAs and GaN were taken from Ref.~\cite{Vurgaftman_2001} while other parameters for GaAs and GaN (energy of longitudinal optical phonon, static and high-frequency dielectric constant) were taken from Ref.~\cite{IoffeNSM}. The parameters for ZnO were taken from Ref. \cite{Madelung}.
In these materials, the ratio of the fully renormalized electron mass $m^\ast$ (including all interaction effects) to the band mass satisfies $m_0/m^\ast \gtrsim 0.8$~\cite{2000_Mishchenko}. In light of this, and based on the results presented in Sec.~\ref{sec:peierls_model-53232} and Refs.~\cite{2023_Mitric, 2022_Mitric} for the Peierls and Holstein models, we expect the self-consistent Migdal approximation to provide accurate mobility estimates over the parameter range considered in this work. This point is important, as entirely real-axis numerically exact mobility data for the Fr\"ohlich model are not available.
We note that, although quantum Monte Carlo (QMC) is a highly reliable method for calculations on the imaginary frequency axis \cite{2019_Misenko_Frolih}, the use of analytical continuation can severely compromise the accuracy of the resulting dc mobility (see, e.g., Fig.~15 of Ref.~\cite{2025_Lihm} and Sec.~II~C of the Supplemental Material of Ref.~\cite{2024_QC_Mitric}). This effect is clearly illustrated in the Holstein model at weak–intermediate coupling, $\lambda = 1/2$: Fig.~1 of Ref.~\cite{2025_Mitric} shows that the hierarchical equations of motion (HEOM) method is in excellent agreement with quantum typicality—another numerically exact approach that does not rely on analytical continuation—whereas Fig.~S5 of the Supplemental Material of Ref.~\cite{2024_Jankovic} demonstrates that QMC underestimates the mobility in the same regime by nearly an order of magnitude. Regardless, the QMC results of Ref.~\cite{2019_Misenko_Frolih} are not suitable as benchmarks here, as they were reported only for $\alpha=2.5$ and $6$, well outside the weak-to-intermediate-coupling regimes relevant for assessing the accuracy of the CE approach.


As shown in Fig.~\ref{fig:spec_frolih} for ZnO, the CE spectral functions in the Fr\"ohlich model exhibit long tails toward negative frequencies at larger momenta, similarly to the Peierls and Holstein models, in contrast to the SCMA results (for a more detailed and linear-scale comparison between CE and SCMA spectral functions, see Sec.~D of SM \cite{SuppMat}). 
These tails may be unphysically wide and may hinder the evaluation of mobility. 
Moreover, unlike in the Holstein and Peierls models—where the momentum $k$ appearing in the mobility expressions is restricted by the finite electronic bandwidth—the Fr\"ohlich model requires momentum sums (i.e., integrals) over an unbounded continuum, see Eq.~\eqref{eq:frohlich_mobility_expression}. Consequently, in addition to the frequency cutoff, we must introduce a momentum cutoff $p_{\mathrm{cut}}$ and verify convergence with respect to it.

The dependence of the mobility $\mu$ on the momentum cutoff $p_{\mathrm{cut}}$ is shown in Fig.~\ref{fig:pcut_conv}. For GaAs, where the electron--phonon interaction is weakest, the CE mobility converges nicely as $p_{\mathrm{cut}}$ increases. In contrast, for GaN and ZnO, where the coupling is stronger, the CE mobility generally fails to converge with increasing $p_{\mathrm{cut}}$. This issue is particularly pronounced at lower temperatures, as seen in the $200\,\mathrm{K}$ results for these two materials. In such cases, $\mu(p_{\mathrm{cut}})$ typically exhibits a maximum at an intermediate cutoff, but does not saturate thereafter. By contrast, the Boltzmann SERTA and SCMA mobilities display well-defined convergence with respect to $p_{\mathrm{cut}}$.

Although, strictly speaking, the CE mobility is not well defined when $\mu(p_{\mathrm{cut}})$ does not saturate for large $p_{\mathrm{cut}}$, one may argue that the maximal value of $\mu(p_{\mathrm{cut}})$ provides a reasonable estimate of the mobility within the CE methodology. This interpretation is motivated by the expectation that momenta above a certain characteristic scale $\bar{p}$ contribute only weakly to transport. To estimate $\bar{p}$, we note that the maximum of $\mu(p_{\mathrm{cut}})$ in the CE curve typically appears close to the range of $p_{\mathrm{cut}}$ where the SERTA and SCMA mobilities begin to saturate; see Fig.~\ref{fig:pcut_conv}. This suggests that the physically relevant momentum contributions are largely contained before this maximum. Based on this reasoning, we take $\max_{p_{\mathrm{cut}}}\mu(p_{\mathrm{cut}})$ as a qualitative estimate of the CE mobility in cases where genuine saturation is not observed. To distinguish such estimates from the cases where the CE mobility properly converges, we label them as ``ECE'' (estimated CE) results. 

After all these considerations, Fig.~\ref{fig:mob_frolih} presents the temperature dependence of the mobility calculated using the Boltzmann SERTA, SCMA, and CE approaches for GaAs, GaN, and ZnO. As expected, for GaAs, the predictions of all three methods are in very good agreement, reflecting the weak electron--phonon coupling in this material. However, Panels~\ref{fig:pcut_conv}$(a_1)$--$(a_2)$ show that CE and SCMA are actually in nearly perfect agreement, while the SERTA results deviate slightly.

For GaN and ZnO, where the electron--phonon coupling is stronger, the discrepancies between the methods are more pronounced. In GaN, CE provides results closer to the SCMA benchmark than the Boltzmann SERTA. In ZnO, SERTA and SCMA appear very close, while CE appears to deviate more significantly. We emphasize that, in this case, the CE results did not fully converge with respect to $p_{\mathrm{cut}}$, so the plotted values correspond to ECE estimates. Moreover, the logarithmic scale used in Fig.~\ref{fig:mob_frolih} can exaggerate the apparent differences. In particular,  panels~\ref{fig:pcut_conv}$(c_1)$--$(c_2)$ reveal that the discrepancy between SCMA and SERTA is actually comparable to that between SCMA and CE. Notably, SERTA tends to overestimate the SCMA benchmark, whereas CE—consistent with our previous results for the Holstein~\cite{2023_Mitric} and Peierls models (see Sec.~\ref{sec:mob_res_wide_range})—underestimates it. Finally, we note that, for these materials, the temperatures considered in Fig.~\ref{fig:mob_frolih}, expressed in Fr\"ohlich units, are relatively low ($T < \omega_0$) compared to those used for the Peierls model in Sec.~\ref{sec:mob_res_wide_range}. Based on our experience with the Peierls model, such low-temperature regimes are more challenging for the CE method, but more favorable for SERTA, which explains why the Boltzmann SERTA results perform comparatively better here than in the Peierls case.

\section{Discussion and Conclusions} \label{Sec:discussion}

The results presented in this paper provide a unified perspective on the applicability of the CE method for the accurate calculation of mobility in electron–phonon systems. 
To examine this thoroughly, we have applied the method to the Peierls and Fr\"ohlich models and, together with the results of our earlier work on the Holstein model~\cite{2023_Mitric},  questioned of whether general, model-independent conclusions can be drawn. 
Although these three models represent fundamentally different scenarios—with the electron–phonon coupling $g_{\mathbf{k},\mathbf{q}}$ having neither $\mathbf{k}$ nor $\mathbf{q}$ dependence in the Holstein model, only $\mathbf{q}$ dependence in the Fr\"ohlich model, and both $\mathbf{k}$ and $\mathbf{q}$ dependence in the Peierls model—we find that the CE method provides accurate mobility results even beyond the regimes where the quasiparticle picture breaks down and beyond the range of applicability of the one-shot Migdal approximation. These conclusions were drawn by comparing Boltzmann, MA, and CE results at intermediate to high temperatures with state-of-the-art benchmarks. It should be emphasized that, although CE spectral functions in this temperature range did not resolve the full multipeak structure that may be present in the exact solution, this had a negligible impact on the mobility. At these elevated temperatures, the mobility is primarily determined by the overall shape of the spectral function, which CE captured accurately.

At low temperatures, however, CE convergence (with respect to the frequency cutoff) issues begin to appear. In addition, for somewhat stronger electron–phonon interactions, as the temperature is reduced, but still before the convergence problems set in, the CE mobility tends to converge to a value that typically underestimates the correct result. All of these low temperature problems, as we deduced, originate due to the fact that CE incorrectly predicts long spectral function tails toward negative frequency, at larger momenta.  

In our work, we also provided an analytic argument indicating why one can expect the CE to perform well. In particular, while the CE, by construction, coincides with the MA (i.e., with the lowest-order Feynman diagram) in the limit of very weak electron--phonon coupling strengths $g$, we additionally showed that the CE generally reproduces the exact sum rules for $n < 4$, whereas the discrepancy between the CE and exact results for $n = 4$ is of order $\mathcal{O}(g^4)$. Although the MA shares this general property, the CE offers two key advantages: 
(i) for the Holstein, Fr\"ohlich, and other models with ${\bf k}$-independent matrix elements $g_{\mathbf{k}, \mathbf{q}}$, the CE sum rule for $n = 4$ exactly coincides with the exact result; 
(ii) for higher-order sum rules ($n \ge 4$), even in models with momentum-dependent couplings $g_{\mathbf{k}, \mathbf{q}}$, the CE---unlike the MA---at least partially captures contributions of orders $\mathcal{O}(g^m)$ with $m \ge 4$.

We also formulated a rule of thumb for estimating whether the CE for mobility calculation is expected to be accurate within a given parameter regime. Specifically, we argued that one should examine the quantity $n_e = \sum_{n=0}^\infty \frac{(-\beta)^n}{n!} \sum_k \mathcal{M}_n(k)$, which appears as the normalization in the mobility expressions, and determine how many terms (in $n$) are needed for convergence within the CE. If convergence requires many terms, the discrepancies between the exact and CE sum rules---which increase with $n$---will render the CE estimates of $n_e$, and consequently of the mobility, inaccurate. This typically occurs for strong coupling or low temperatures, consistent with the fact that in this regime the long tail of the spectral function strongly affects higher-order sum rules. In contrast, if convergence is reached after only a few terms, the spectral-function tail plays a minor role. As this happens for weaker interactions or higher temperatures, the mobility typically depends mainly on the overall spectral function shape, which is well captured by the first few sum rules. This rule of thumb is particularly appealing, since the CE sum rules are, as we have shown, straightforward to evaluate.

Finally, since the CE approach is built upon the IPA (and introduces additional approximations within this framework), for a comprehensive assessment of its reliability for mobility calculations, it was essential to analyze the magnitude of vertex corrections—that is, the difference between the fully numerically exact result and the exact result obtained within the IPA (using no other approximations than IPA). For the Peierls model, we observed that the IPA result is qualitatively similar to the exact mobility result. Quantitatively, the two are of the same order of magnitude, but the discrepancy remains non-negligible, even for relatively weak coupling strengths. In contrast, our earlier findings for the Holstein model demonstrated that vertex corrections are rather small even at moderate coupling strengths $g$, and vanish completely in the limit $g \to 0$ \cite{2024_Jankovic}. This behavior in the Holstien model can be attributed to the local (momentum-independent) nature of the electron-phonon coupling, which implies that the ladder diagrams in the current-current correlation function vanish \cite{2023_Mitric_phd}. These ladder diagrams are important, as they are known to give vertex correction's leading order contribution in the weak coupling limit \cite{1966_Mahan_mobility}.

With this observation in mind, we note that Lihm and Ponc\'e recently applied the self-consistent Migdal approximation to a general electron-phonon Hamiltonian, where they did not restrict themselves to the IPA but instead summed all ladder diagram contributions \cite{2025_Lihm}. Focusing on ladder diagrams is particularly advantageous—not only because they provide the leading-order contribution in the weak-coupling limit, but also because this choice corresponds to a charge-conserving approximation and, from a technical standpoint, allows for a fully self-consistent formulation of the equations. As a result, they were able to (partially) capture the effects of vertex corrections, which in practice led to highly accurate results. This approach is crucial for achieving quantitatively precise mobility values, as our findings clearly demonstrate that vertex corrections in a general electron-phonon system cannot \emph{a priori} be assumed negligible, even at relatively weak electron-phonon coupling strengths.

\acknowledgments

This research was supported by the Science Fund of the
Republic of Serbia, Grant No. 5468, Polaron Mobility in
Model Systems and Real Materials–PolMoReMa. The authors also 
acknowledge funding provided by the Institute of Physics
Belgrade through a grant from the Ministry of Science,
Technological Development, and Innovation of the Republic 
of Serbia. Numerical computations were performed on
the PARADOX-IV supercomputing facility at the Scientific
Computing Laboratory, National Center of Excellence for the
Study of Complex Systems, Institute of Physics Belgrade.

\section*{Authors' contributions}
P. M. worked on the CE, MA, and SCMA methods, performed analytical and numerical calculations of spectral sum rules, and wrote the manuscript. V. J.  worked on the HEOM. D. T. worked on the quasiparticle properties using the GGCE method. N. V. worked on Boltzmann transport and partly on the Fröhlich model. P. M., V. J., and N. V. discussed the results and finalized the manuscript. N. V. secured the funding.

\section*{DATA AVAILABILITY}

The HEOM full transport data are available in Ref. \cite{Jankovic_Zenodo_Peierls}, while all other data supporting the findings of this work are available in Ref. \cite{zenodo_cumulant_2025}.



%

\clearpage
\pagebreak
\newpage

\onecolumngrid
\begin{center}
  \textbf{\large Applicability of the cumulant expansion method for the calculation of transport
properties in electron-phonon systems}\\[.2cm]
  Petar Mitri\'c,$^{1}$ Veljko Jankovi\'c,$^{1}$ Nenad Vukmirovi\'c,$^{1}$ and Darko Tanaskovi\'c$^1$\\[.1cm]
  {\itshape ${}^1$Institute of Physics Belgrade,
University of Belgrade, Pregrevica 118, 11080 Belgrade, Serbia}
  \\[1cm]
\end{center}
\onecolumngrid

\setcounter{equation}{0}
\setcounter{figure}{0}
\setcounter{table}{0}
\makeatletter
\renewcommand{\theequation}{S\arabic{equation}}
\renewcommand{\thefigure}{S\arabic{figure}}
\renewcommand{\bibnumfmt}[1]{[S#1]}
\renewcommand{\citenumfont}[1]{S#1}
\renewcommand{\thetable}{S\arabic{table}}
\renewcommand{\thesection}{\Alph{section}}
\setcounter{section}{0}
\setcounter{subsection}{0}
\renewcommand{\thesubsection}{\arabic{subsection}}
\allowdisplaybreaks

Here, we supplement the results of the main text by providing additional numerical and analytical results.

\section{Quasiparticle properties for the Peierls model}

\begin{figure}[ht]
    \centering
    \begin{minipage}{0.45\textwidth}
        \centering
        \includegraphics[width=\linewidth]{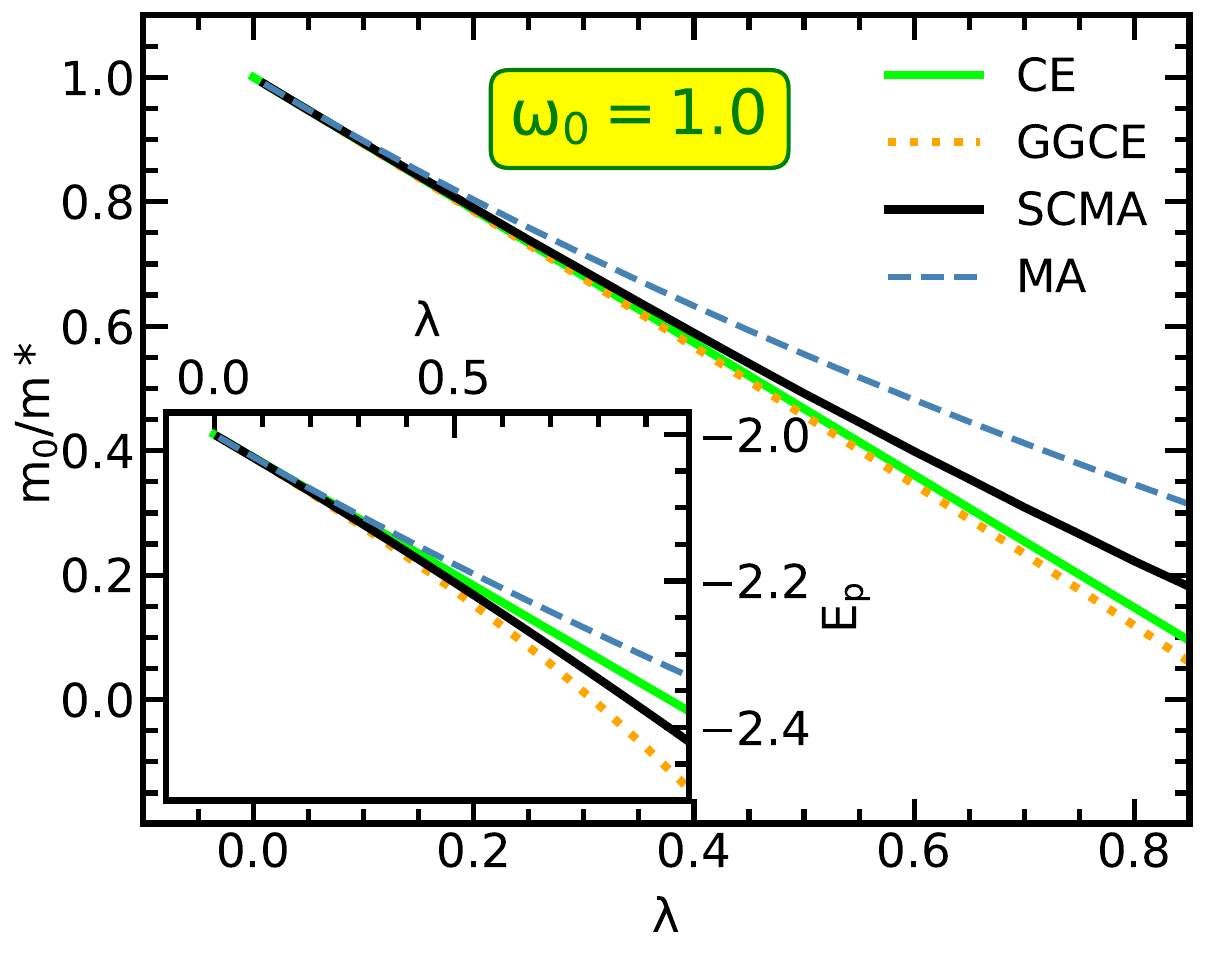}
        \caption{\centering Renormalized mass $m^*$ and ground-state energy $E_p$ for $\omega_0 = 1.0$ and $T = 0$.}
        \label{fig:image1}
    \end{minipage}\hfill
    \begin{minipage}{0.45\textwidth}
        \centering
        \includegraphics[width=\linewidth]{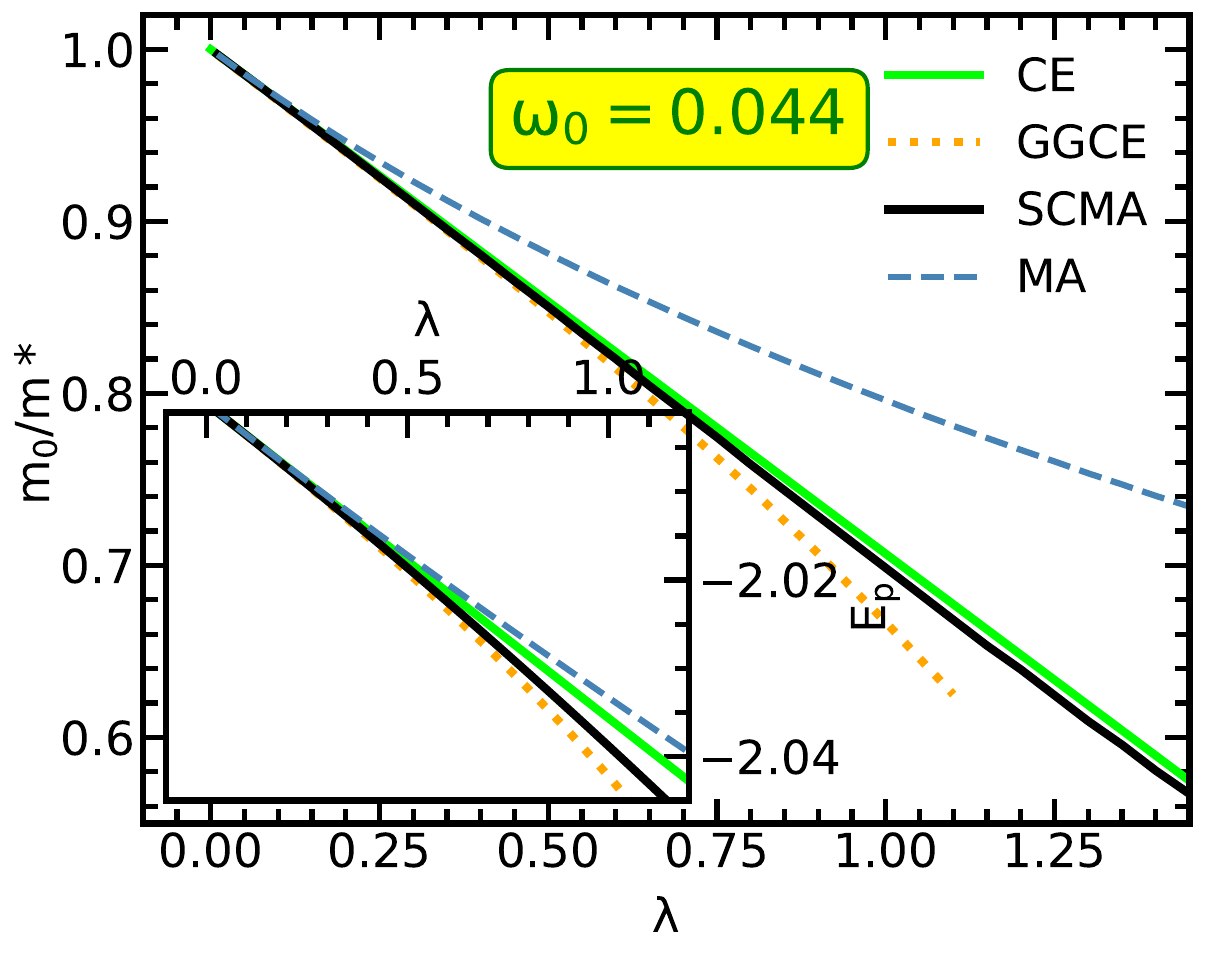}
        \caption{\centering Renormalized mass $m^*$ and ground-state energy $E_p$ for $\omega_0 = 0.044$ and $T = 0$.}
        \label{fig:image2}
    \end{minipage}
    
    \vspace{0.5cm} 
    
    \begin{minipage}{0.45\textwidth}
        \centering
        \includegraphics[width=\linewidth]{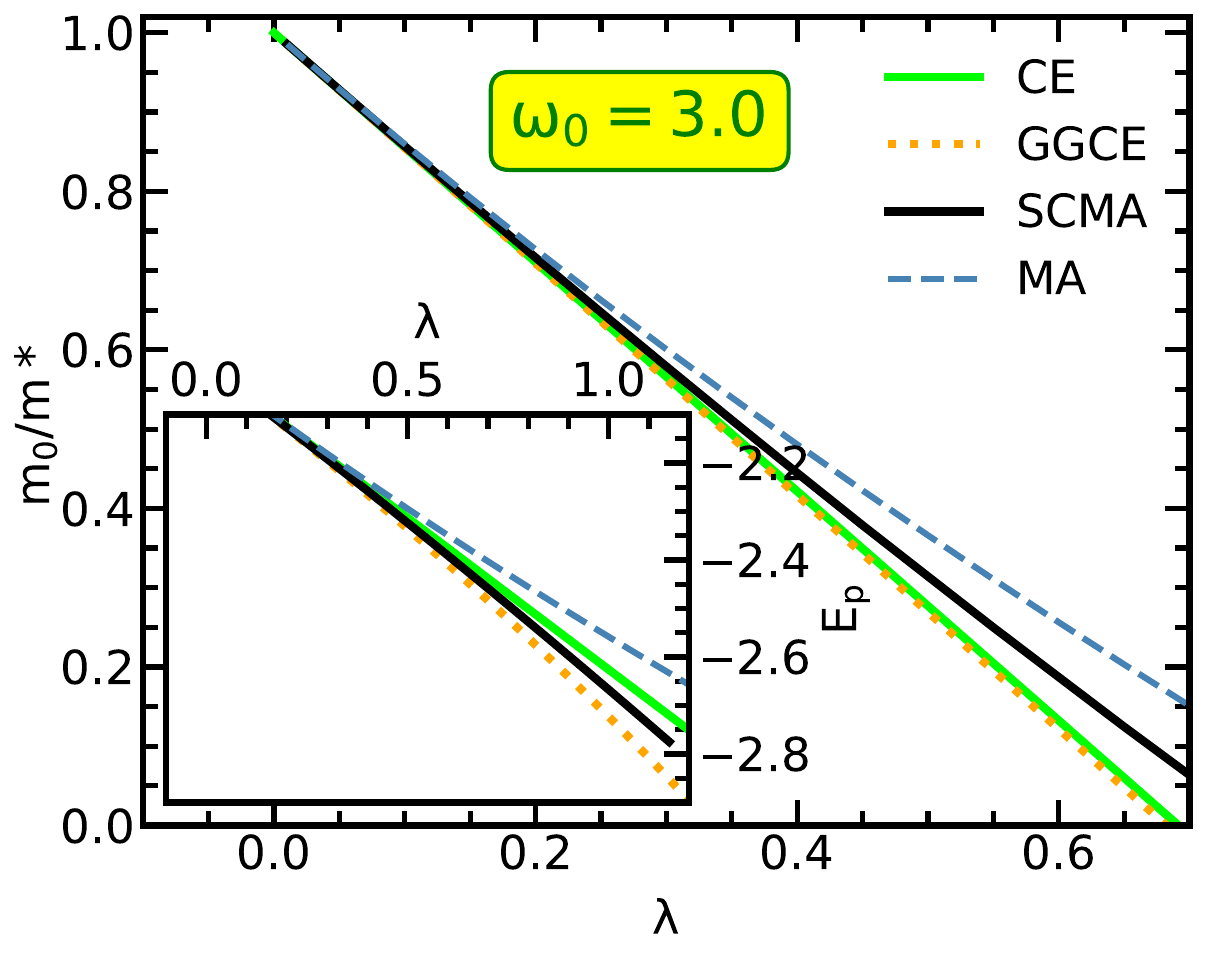}
        \caption{\centering Renormalized mass $m^*$ and ground-state energy $E_p$ for $\omega_0 = 3.0$ and $T = 0$.}
        \label{fig:image3}
    \end{minipage}\hfill
    \begin{minipage}{0.53\textwidth}
        \centering
        \hspace*{1cm}
        \includegraphics[width=\linewidth]{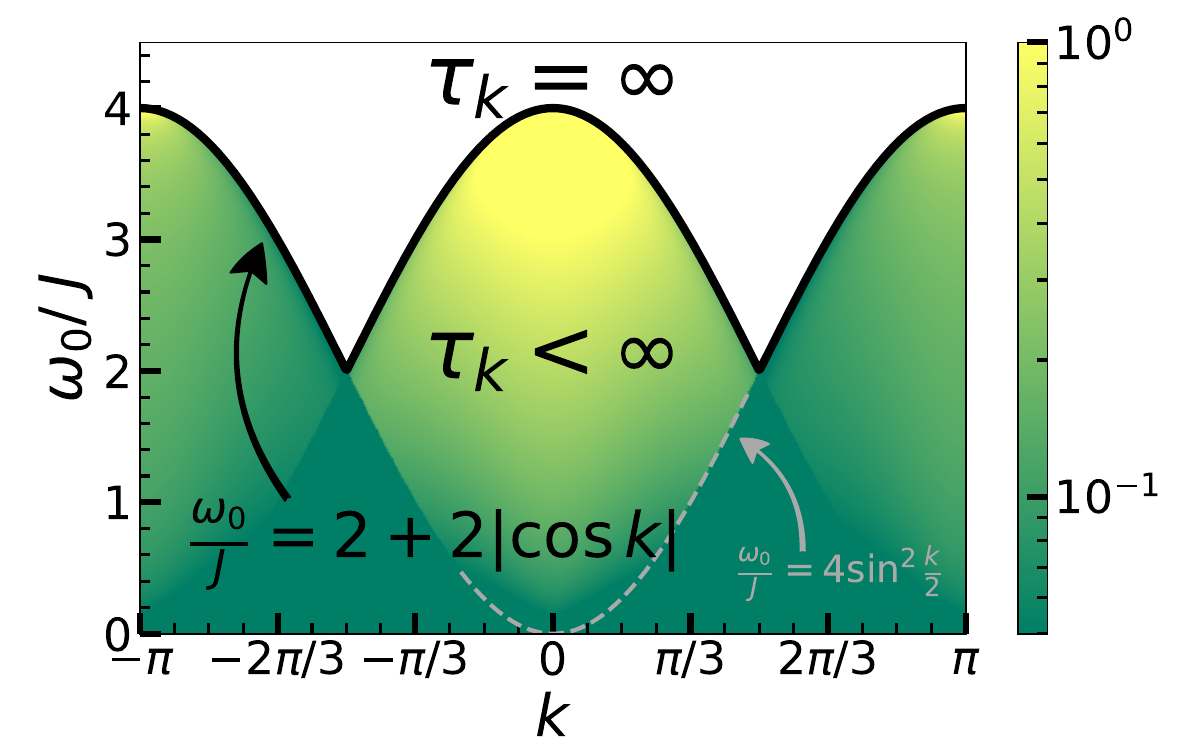}
        \caption{\centering Lifetime $\tau_{  k}$ within the CE method for $J=1$, $T=2$ and $g=1$.}
        \label{fig:image4}
    \end{minipage}
\end{figure}

In Sec.~IV~A~1 of the main text, we showed the cumulant expansion (CE), self-consistent Migdal approximation (SCMA), Migdal approximation (MA), and generalized Green’s function cluster expansion (GGCE) benchmark predictions for the band-mass to renormalized-mass ratio $m_0 / m^*$ (note that $m_0 = 1/(2J)$) and the ground-state energy $E_p$ for $\omega_0=0.5$ and $T=0$. Here, in  Figs.~\ref{fig:image1}--~\ref{fig:image3} we additionally show results for $\omega_0=1.0$, $\omega_0=0.044$, and $\omega_0=3.0$. Also, in Fig.~\ref{fig:image4} we illustrate CE lifetime for $J=g=1$, $T=2$; see Eq.~(63) from the main text. 


\section{Spectral functions in the Peierls model}

\begin{figure*}[t!]
  \centering
  \includegraphics[width=1.0\linewidth]{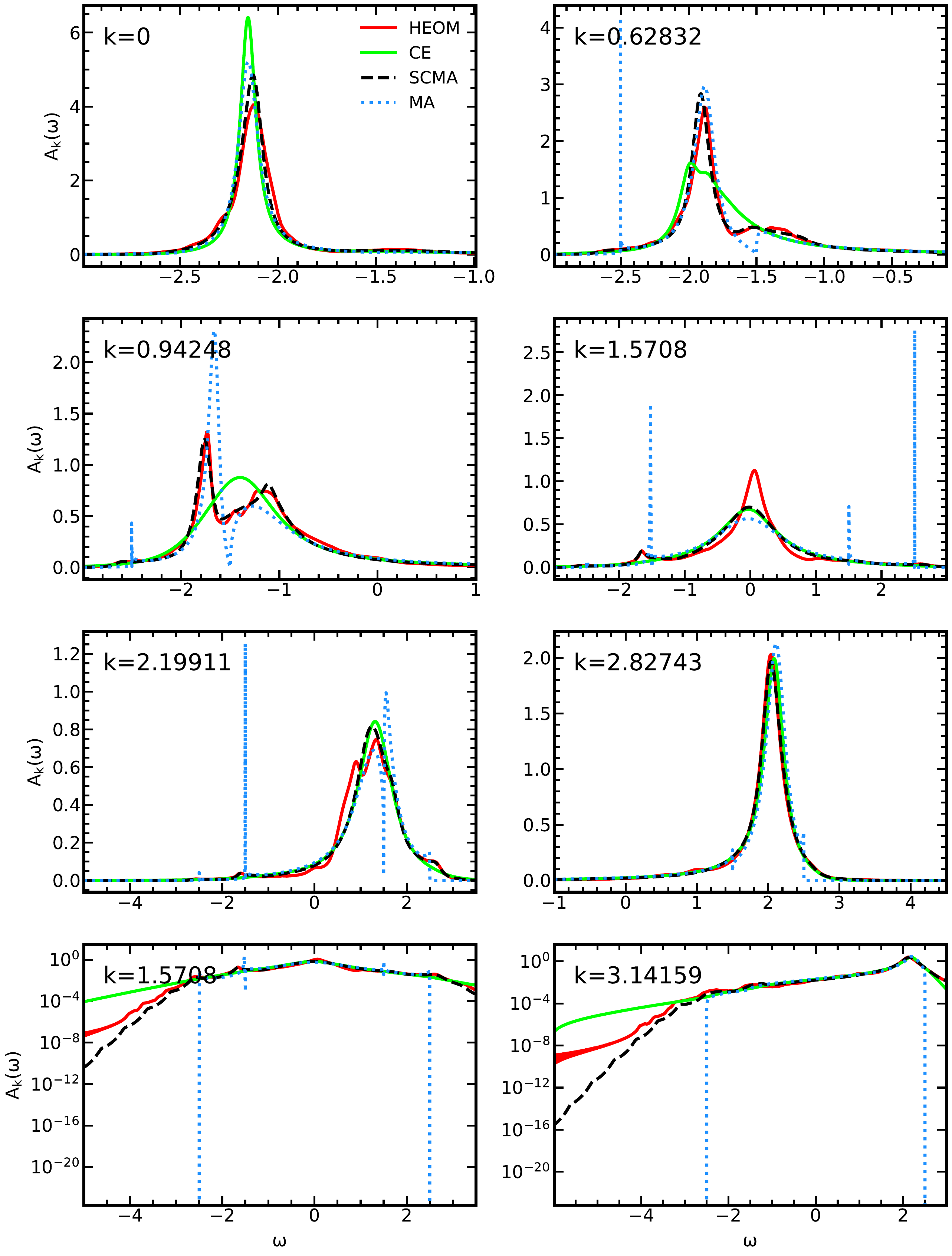}
  \caption{Comparison of the spectral functions for $J=1$, $\omega_0=0.5$, $\lambda=0.25$, $T=0.5$.}
  \label{fig:5}
\end{figure*}

\begin{figure*}[t!]
  \centering
  \includegraphics[width=1.0\linewidth]{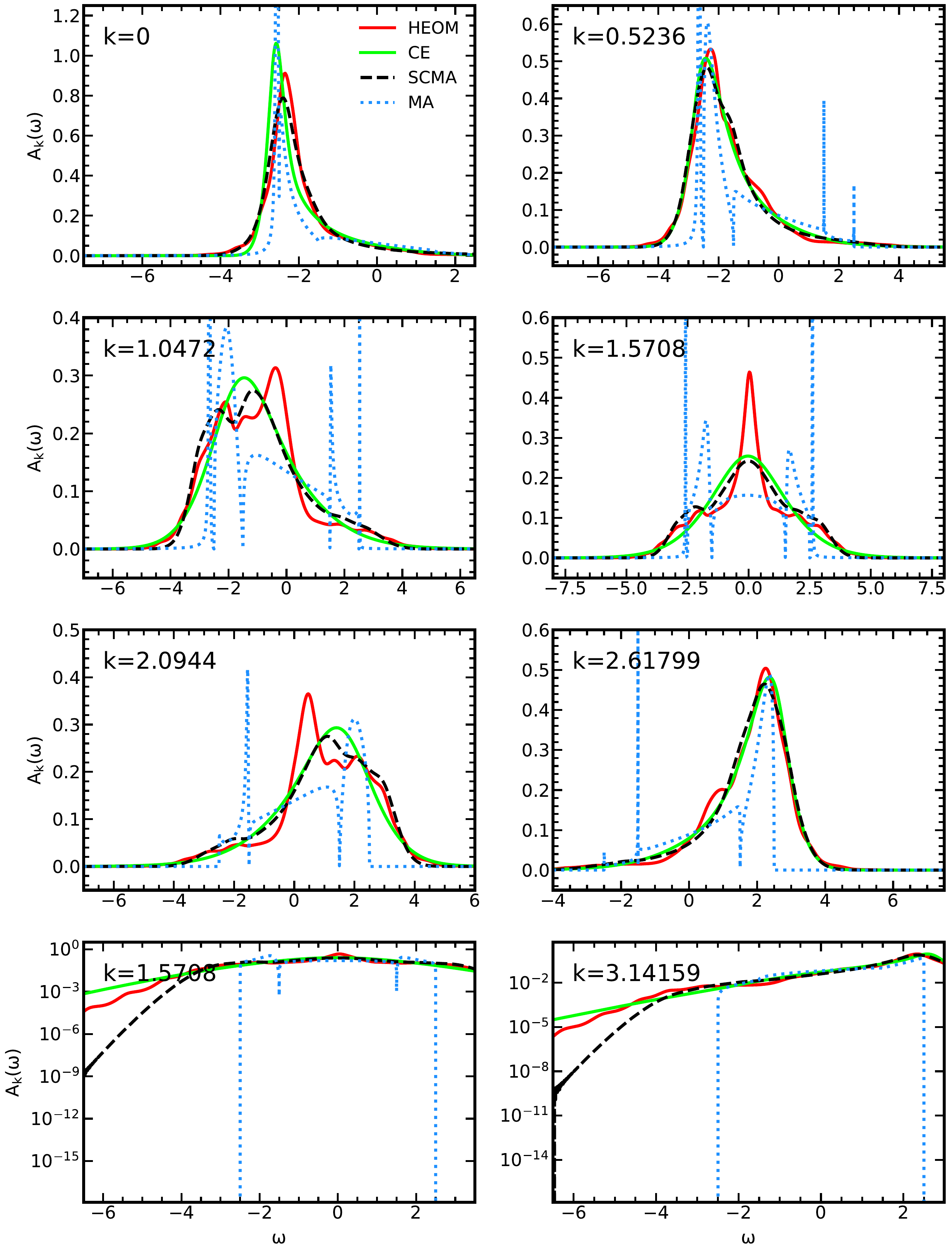}
  \caption{Comparison of the spectral functions for $J=1$, $\omega_0=0.5$, $\lambda=0.25$, $T=2.0$.}
  \label{fig:6}
\end{figure*}

\begin{figure*}[t!]
  \centering
  \includegraphics[width=1.0\linewidth]{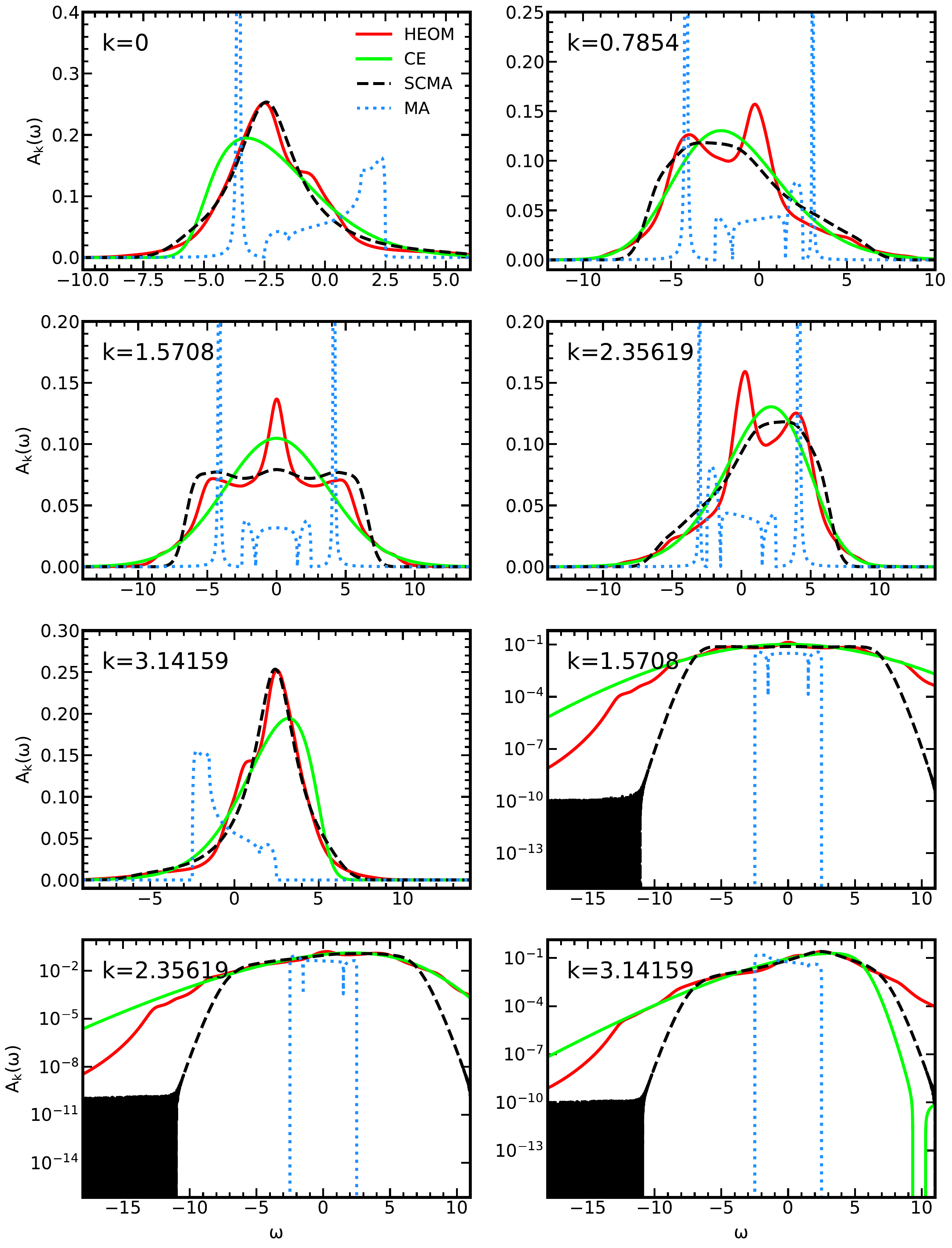}
  \caption{Comparison of the spectral functions for $J=1$, $\omega_0=0.5$, $\lambda=0.25$, $T=10.0$.}
  \label{fig:7}
\end{figure*}

\begin{figure*}[t!]
  \centering
  \includegraphics[width=1.0\linewidth]{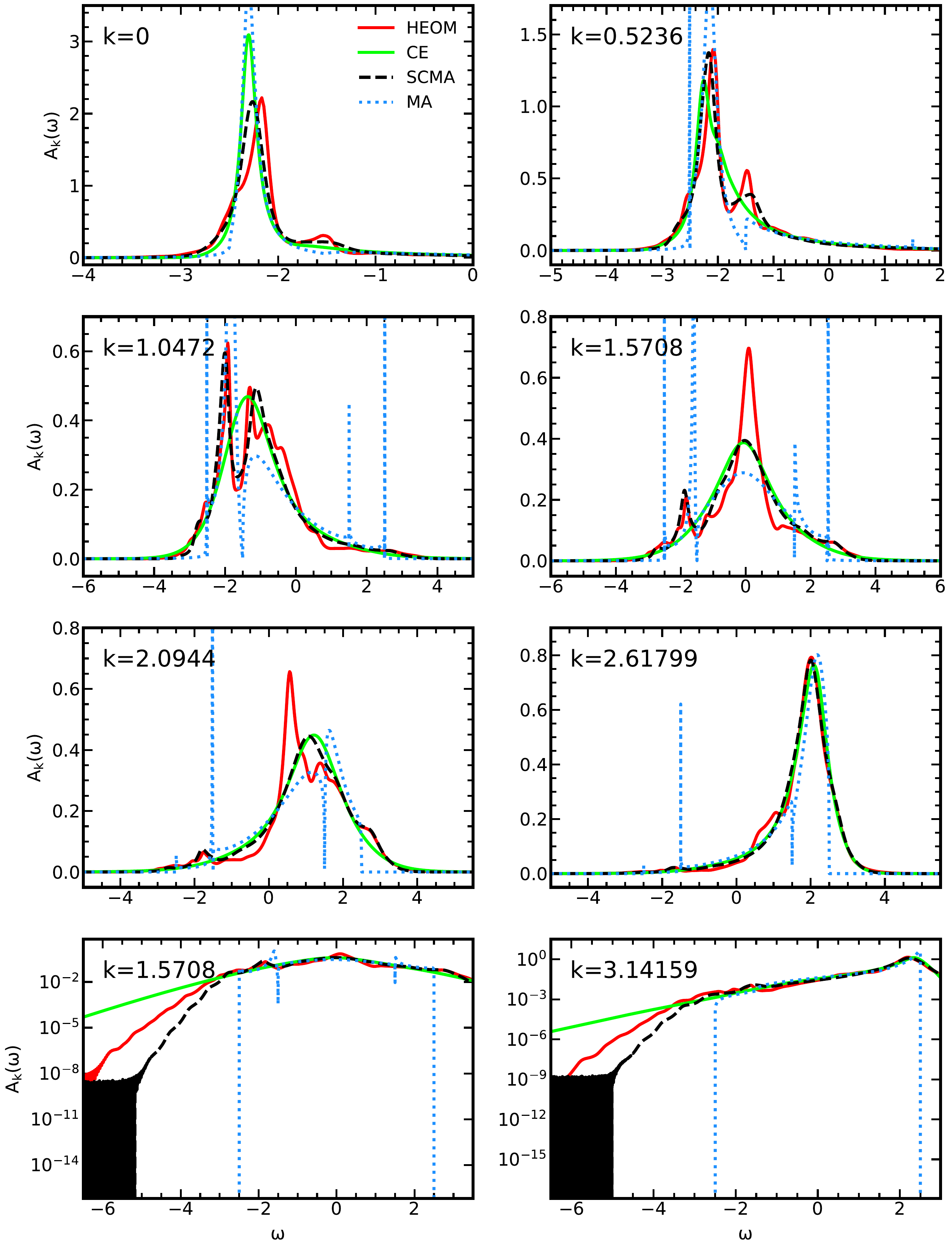}
  \caption{Comparison of the spectral functions for $J=1$, $\omega_0=0.5$, $\lambda=0.5$, $T=0.5$.}
  \label{fig:8}
\end{figure*}

\begin{figure*}[t!]
  \centering
  \includegraphics[width=1.0\linewidth]{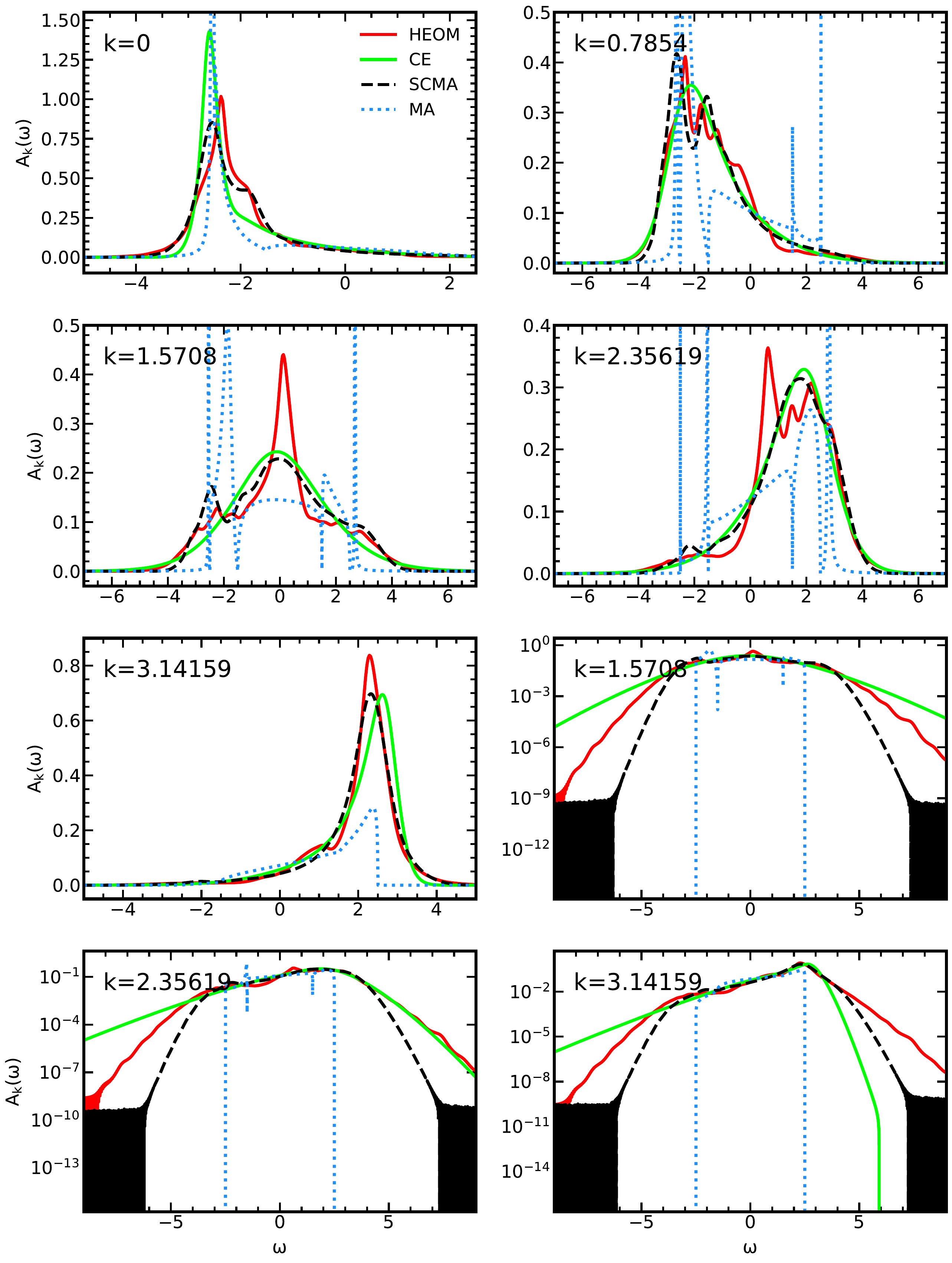}
  \caption{Comparison of the spectral functions for $J=1$, $\omega_0=0.5$, $\lambda=1.0$, $T=0.5$.}
  \label{fig:8A}
\end{figure*}

\begin{figure*}[t!]
  \centering
  \includegraphics[width=1.0\linewidth]{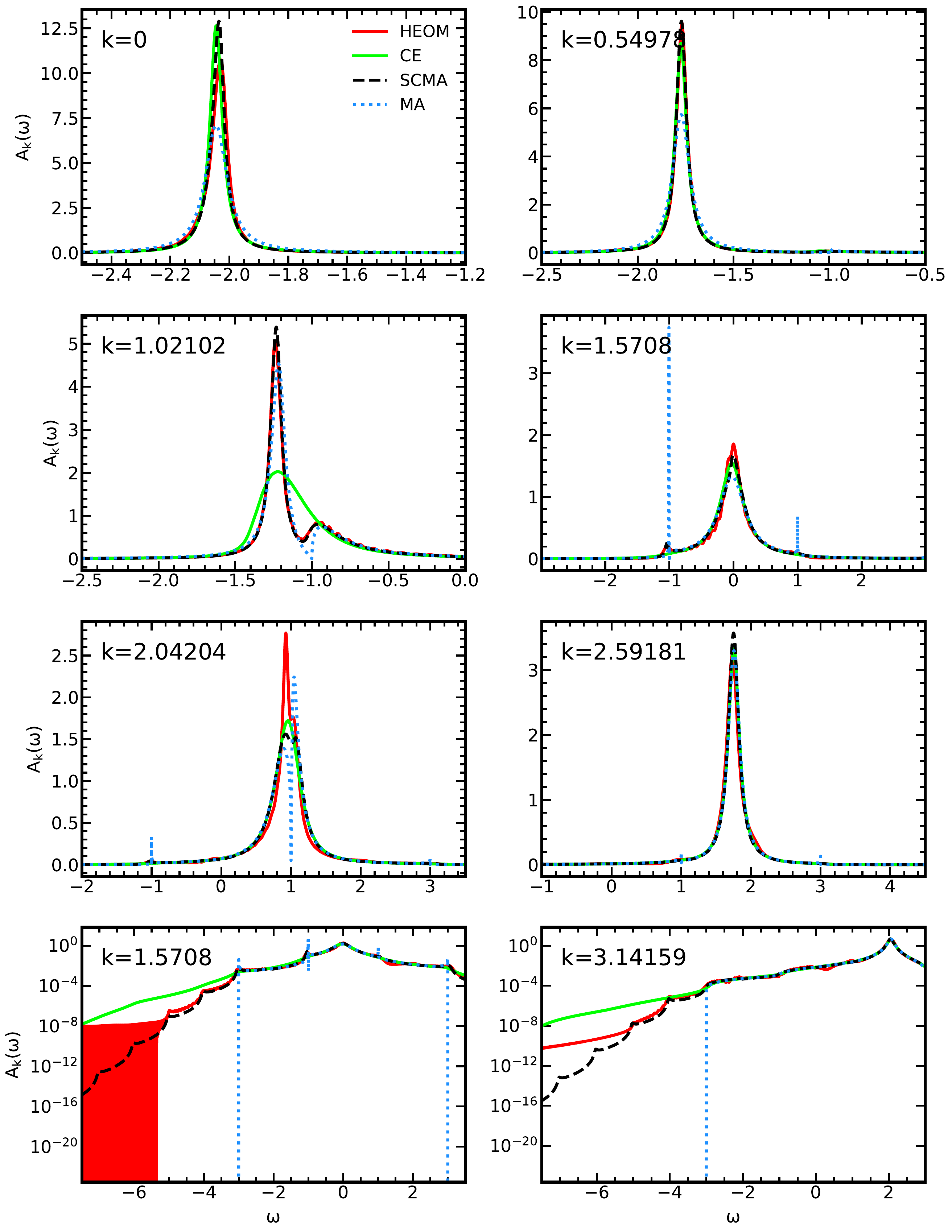}
  \caption{Comparison of the spectral functions for $J=1$, $\omega_0=1.0$, $\lambda=0.05$, $T=1.0$.}
  \label{fig:8B}
\end{figure*}

\begin{figure*}[t!]
  \centering
  \includegraphics[width=1.0\linewidth]{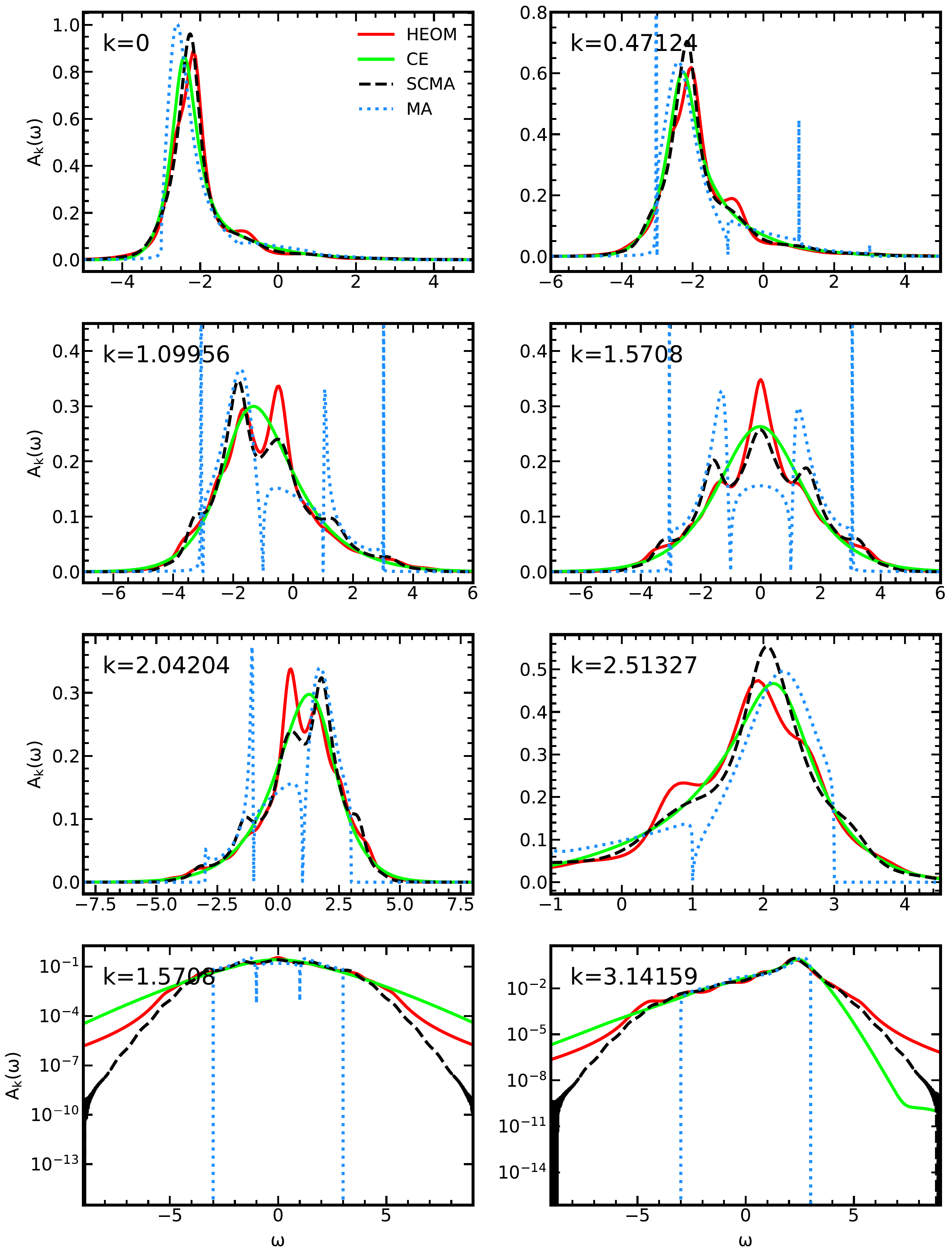}
  \caption{Comparison of the spectral functions for $J=1$, $\omega_0=1.0$, $\lambda=0.05$, $T=10.0$.}
  \label{fig:8C}
\end{figure*}

\begin{figure*}[t!]
  \centering
  \includegraphics[width=1.0\linewidth]{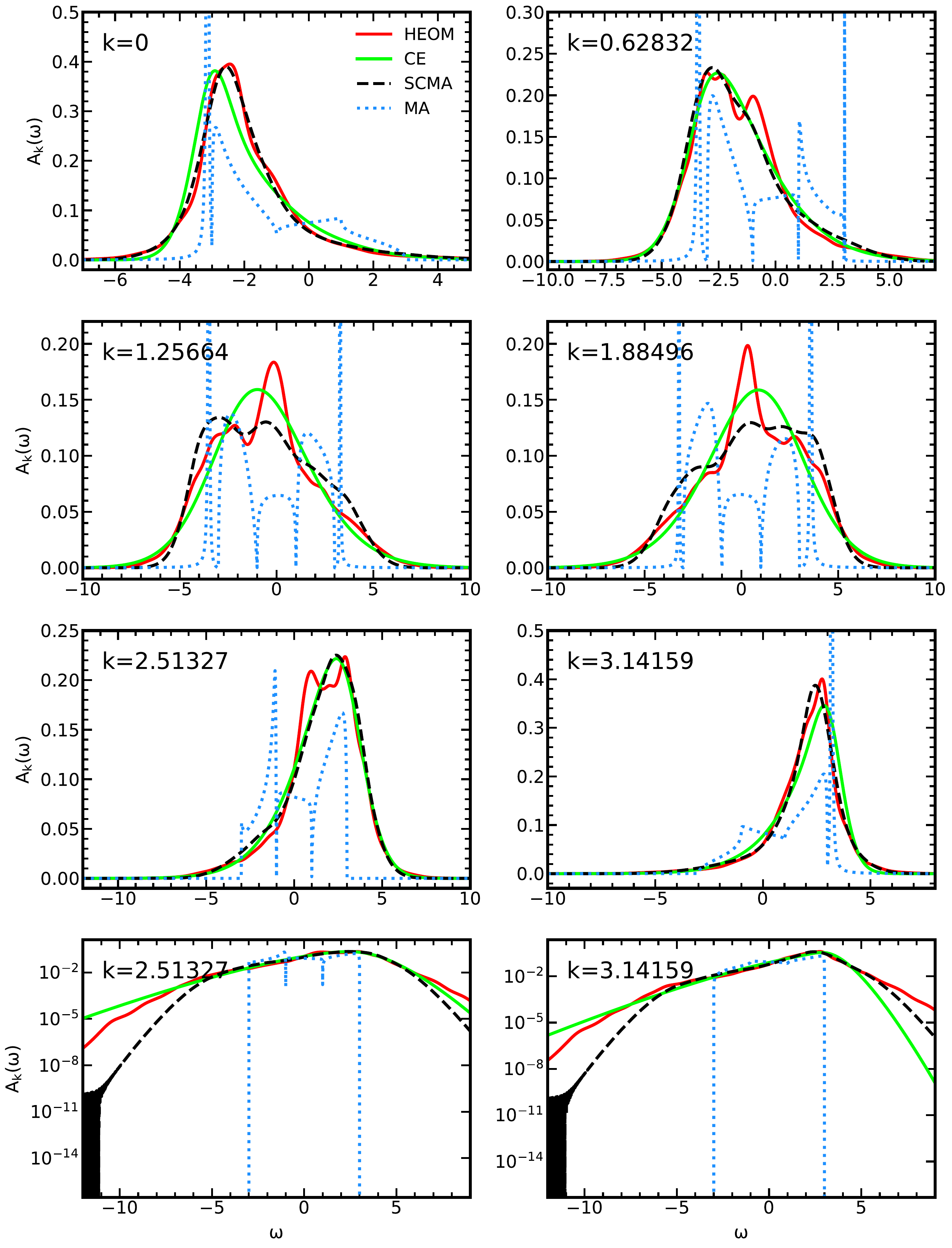}
  \caption{Comparison of the spectral functions for $J=1$, $\omega_0=1.0$, $\lambda=0.25$, $T=5.0$.}
  \label{fig:9}
\end{figure*}

In Fig.~5 of the main text, we presented single-particle spectral functions $A_k(\omega)$ obtained using CE, SCMA, and the numerically exact hierarchical equations of motion (HEOM) method, for $\omega_0 = 0.5$, $\lambda = 0.25$, and $T = 1.0$. We concluded that CE captures the overall shape of the exact solution but fails to resolve the multipeak structure, if such is present in the exact result. In contrast, SCMA typically yields significantly better agreement with HEOM. We also noted that CE exhibits an unphysically long tail extending towards negative frequencies, which is most apparent on a logarithmic scale. To demonstrate that these observations are not limited to this particular choice of parameters, but are in fact generic, in Figs.~\ref{fig:5}--~\ref{fig:9} we provide additional spectral-function results covering a wide range of regimes.

%


\clearpage
\newpage
\section{Spectral sum rules}
In Sec.~IV~B of the main text, we presented the exact expressions for the sum rules $\mathcal{M}_n({\bf k})$, as well as their counterparts obtained within the CE method, for $n \leq 4$, considering a general electron–phonon Hamiltonian with a single electronic band and a single optical phonon mode. Here, in Sec.~\ref{sec:holstein_sum_rules}, we provide both the exact and CE sum rules for the Holstein model up to $n \leq 9$. For the Peierls model (Sec.~\ref{sec:peierls_sum_rules}), we present only the CE sum rules up to $n \leq 10$.

We note that the exact and CE sum rules were calculated using Eqs.~(82)~and~(88) from the main text, respectivelly.

\subsection{Holstein model}
\label{sec:holstein_sum_rules}

The exact sum rules for the Holstein model read as follows:

\begin{align}
\mathcal{M}_0(\mathbf{k}) &= 1 \\
\mathcal{M}_1(\mathbf{k}) &= \varepsilon_{\mathbf{k}} \\
\mathcal{M}_2(\mathbf{k}) &= g^2(2n_{\mathrm{ph}}+ 1) + \varepsilon_{\mathbf{k}}^2 \\
\mathcal{M}_3(\mathbf{k}) &= 2g^2(2n_{\mathrm{ph}}+ 1)\varepsilon_{\mathbf{k}} + g^2\omega_0 + \varepsilon_{\mathbf{k}}^3 \\
\mathcal{M}_4(\mathbf{k}) &= 3g^4(2n_{\mathrm{ph}}+ 1)^2 + g^2(2n_{\mathrm{ph}}+ 1)(3\varepsilon_{\mathbf{k}}^2 + 2t_0^2 + \omega_0^2) + 2g^2\omega_0\varepsilon_{\mathbf{k}} + \varepsilon_{\mathbf{k}}^4 \\
\mathcal{M}_5(\mathbf{k}) &= 7g^4(2n_{\mathrm{ph}}+ 1)^2\varepsilon_{\mathbf{k}} + g^2(3\omega_0\varepsilon_{\mathbf{k}}^2 + 6t_0^2\omega_0 + \omega_0^3) + (2n_{\mathrm{ph}}+ 1)(10g^4\omega_0 + g^2(4t_0^2\varepsilon_{\mathbf{k}} + 2\omega_0^2\varepsilon_{\mathbf{k}} + 4\varepsilon_{\mathbf{k}}^3)) + \varepsilon_{\mathbf{k}}^5 \\
\mathcal{M}_6(\mathbf{k}) &= 15g^6(2n_{\mathrm{ph}}+ 1)^3 + g^4(2n_{\mathrm{ph}}+ 1)^2(12\varepsilon_{\mathbf{k}}^2 + 18t_0^2 + 15\omega_0^2) + 10g^4\omega_0^2 + g^2(12t_0^2\omega_0\varepsilon_{\mathbf{k}} + 4\omega_0\varepsilon_{\mathbf{k}}^3 + 2\omega_0^3\varepsilon_{\mathbf{k}}) \nonumber \\ &+ (2n_{\mathrm{ph}}+ 1)(22g^4\omega_0\varepsilon_{\mathbf{k}} + g^2(6t_0^2\varepsilon_{\mathbf{k}}^2 + 3\omega_0^2\varepsilon_{\mathbf{k}}^2 + 5\varepsilon_{\mathbf{k}}^4 + 12t_0^2\omega_0^2 + 6t_0^4 + \omega_0^4)) + \varepsilon_{\mathbf{k}}^6 \\
\mathcal{M}_7(\mathbf{k}) &= 36g^6(2n_{\mathrm{ph}}+ 1)^3\varepsilon_{\mathbf{k}} + 21g^4\omega_0^2\varepsilon_{\mathbf{k}} + g^2(18t_0^2\omega_0\varepsilon_{\mathbf{k}}^2 + 3\omega_0^3\varepsilon_{\mathbf{k}}^2 + 5\omega_0\varepsilon_{\mathbf{k}}^4 + 20t_0^2\omega_0^3 + 30t_0^4\omega_0 + \omega_0^5) \nonumber \\ &+ (2n_{\mathrm{ph}}+ 1)^2(105g^6\omega_0 + g^4(41t_0^2\varepsilon_{\mathbf{k}} + 32\omega_0^2\varepsilon_{\mathbf{k}} + 18\varepsilon_{\mathbf{k}}^3)) + (2n_{\mathrm{ph}}+ 1)g^4(36\omega_0\varepsilon_{\mathbf{k}}^2 + 108t_0^2\omega_0 + 56\omega_0^3) \nonumber \\ &+ (2n_{\mathrm{ph}}+ 1)g^2(2\varepsilon_{\mathbf{k}}(12t_0^2\omega_0^2 + 6t_0^4 + \omega_0^4) + 8t_0^2\varepsilon_{\mathbf{k}}^3 + 4\omega_0^2\varepsilon_{\mathbf{k}}^3 + 6\varepsilon_{\mathbf{k}}^5) + \varepsilon_{\mathbf{k}}^7 \\
\mathcal{M}_8(\mathbf{k}) &= 105g^8(2n_{\mathrm{ph}}+ 1)^4 + g^6(2n_{\mathrm{ph}}+ 1)^3(64\varepsilon_{\mathbf{k}}^2 + 160t_0^2 + 210\omega_0^2) + g^4(33\omega_0^2\varepsilon_{\mathbf{k}}^2 + 158t_0^2\omega_0^2 + 56\omega_0^4) \nonumber \\ &+ g^2(24t_0^2\omega_0\varepsilon_{\mathbf{k}}^3 + 40t_0^2\omega_0^3\varepsilon_{\mathbf{k}} + 60t_0^4\omega_0\varepsilon_{\mathbf{k}} + 6\omega_0\varepsilon_{\mathbf{k}}^5 + 4\omega_0^3\varepsilon_{\mathbf{k}}^3 + 2\omega_0^5\varepsilon_{\mathbf{k}}) \nonumber \\ &+ (2n_{\mathrm{ph}}+ 1)^2(236g^6\omega_0\varepsilon_{\mathbf{k}} + g^4(68t_0^2\varepsilon_{\mathbf{k}}^2 + 51\omega_0^2\varepsilon_{\mathbf{k}}^2 + 25\varepsilon_{\mathbf{k}}^4 + 258t_0^2\omega_0^2 + 94t_0^4 + 63\omega_0^4)) \nonumber \\ &+ 280(2n_{\mathrm{ph}}+ 1)g^6\omega_0^2 + (2n_{\mathrm{ph}}+ 1)g^4(240t_0^2\omega_0\varepsilon_{\mathbf{k}} + 52\omega_0\varepsilon_{\mathbf{k}}^3 + 116\omega_0^3\varepsilon_{\mathbf{k}}) \nonumber \\ &+ 3(2n_{\mathrm{ph}}+ 1)g^2\varepsilon_{\mathbf{k}}^2(12t_0^2\omega_0^2 + 6t_0^4 + \omega_0^4) \nonumber \\ &+ (2n_{\mathrm{ph}}+ 1)g^2(10t_0^2\varepsilon_{\mathbf{k}}^4 + 5\omega_0^2\varepsilon_{\mathbf{k}}^4 + 7\varepsilon_{\mathbf{k}}^6 + 90t_0^4\omega_0^2 + 30t_0^2\omega_0^4 + 20t_0^6 + \omega_0^6) + \varepsilon_{\mathbf{k}}^8 \\
\mathcal{M}_9(\mathbf{k}) &= 249g^8(2n_{\mathrm{ph}}+ 1)^4\varepsilon_{\mathbf{k}} + 280g^6\omega_0^3 + g^4(346t_0^2\omega_0^2\varepsilon_{\mathbf{k}} + 114\omega_0^4\varepsilon_{\mathbf{k}} + 46\omega_0^2\varepsilon_{\mathbf{k}}^3) \nonumber \\ &+ g^2(60t_0^2\omega_0^3\varepsilon_{\mathbf{k}}^2 + 30t_0^2\omega_0\varepsilon_{\mathbf{k}}^4 + 90t_0^4\omega_0\varepsilon_{\mathbf{k}}^2 + 3\omega_0^5\varepsilon_{\mathbf{k}}^2 + 5\omega_0^3\varepsilon_{\mathbf{k}}^4 + 7\omega_0\varepsilon_{\mathbf{k}}^6) \nonumber \\ &+ g^2(42t_0^2\omega_0^5 + 210t_0^4\omega_0^3 + 140t_0^6\omega_0 + \omega_0^7) + (2n_{\mathrm{ph}}+ 1)^3(1260g^8\omega_0 + g^6(384t_0^2\varepsilon_{\mathbf{k}} + 456\omega_0^2\varepsilon_{\mathbf{k}} + 100\varepsilon_{\mathbf{k}}^3)) \nonumber \\ &+ (2n_{\mathrm{ph}}+ 1)^2g^6(396\omega_0\varepsilon_{\mathbf{k}}^2 + 1632t_0^2\omega_0 + 1638\omega_0^3) \nonumber \\ &+ (2n_{\mathrm{ph}}+ 1)^2g^4(564t_0^2\omega_0^2\varepsilon_{\mathbf{k}} + 99t_0^2\varepsilon_{\mathbf{k}}^3 + 213t_0^4\varepsilon_{\mathbf{k}} + 72\omega_0^2\varepsilon_{\mathbf{k}}^3 + 129\omega_0^4\varepsilon_{\mathbf{k}} + 33\varepsilon_{\mathbf{k}}^5) \nonumber \\ &+ 600(2n_{\mathrm{ph}}+ 1)g^6\omega_0^2\varepsilon_{\mathbf{k}} + (2n_{\mathrm{ph}}+ 1)g^4(388t_0^2\omega_0\varepsilon_{\mathbf{k}}^2 + 70\omega_0\varepsilon_{\mathbf{k}}^4 + 180\omega_0^3\varepsilon_{\mathbf{k}}^2 + 2\omega_0(660t_0^2\omega_0^2 + 404t_0^4 + 123\omega_0^4)) \nonumber \\ &+ (2n_{\mathrm{ph}}+ 1)g^2(4\varepsilon_{\mathbf{k}}^3(12t_0^2\omega_0^2 + 6t_0^4 + \omega_0^4) + 2\varepsilon_{\mathbf{k}}(90t_0^4\omega_0^2 + 30t_0^2\omega_0^4 + 20t_0^6 + \omega_0^6)) \nonumber \\ &+ (2n_{\mathrm{ph}}+ 1)g^2(12t_0^2\varepsilon_{\mathbf{k}}^5 + 6\omega_0^2\varepsilon_{\mathbf{k}}^5 + 8\varepsilon_{\mathbf{k}}^7) + \varepsilon_{\mathbf{k}}^9
\end{align}

The CE sum rules $\mathcal{M}_n^\mathrm{CE}({\bf k})$ coincide with the exact result for $n\leq 4$. For $n > 4$, the difference $\mathcal{M}_n^\mathrm{CE}({\bf k}) - \mathcal{M}_n({\bf k})$ is given by:
\clearpage \newpage
\begin{align}
    \mathcal{M}_5^\mathrm{CE}({\bf k}) - \mathcal{M}_5({\bf k}) &= 
    -2 g^4 (2n_{\mathrm{ph}}+ 1)^2 \varepsilon_{\mathbf{k}}  \\
    \mathcal{M}_6^\mathrm{CE}({\bf k}) - \mathcal{M}_6({\bf k}) &= 
    -2 g^4 (2n_{\mathrm{ph}}+ 1) \left(-6 (2n_{\mathrm{ph}}+ 1) + 6 \omega_0 \varepsilon_{\mathbf{k}} + (2n_{\mathrm{ph}}+ 1) \varepsilon_{\mathbf{k}}^2\right) \\
    \mathcal{M}_7^\mathrm{CE}({\bf k}) - \mathcal{M}_7({\bf k}) &= 
    -g^4 \Big[-88 (2n_{\mathrm{ph}}+ 1) \omega_0 + \varepsilon_{\mathbf{k}} \Big( 9 (2n_{\mathrm{ph}}+ 1)^2 (3 + 4 g^2 (2n_{\mathrm{ph}}+ 1)) \nonumber \\ &+ (21 + 25 (2n_{\mathrm{ph}}+ 1)^2) \omega_0^2 + 4 (2n_{\mathrm{ph}}+ 1) \varepsilon_{\mathbf{k}} (2 \omega_0 + (2n_{\mathrm{ph}}+ 1) \varepsilon_{\mathbf{k}}) \Big) \Big] \\
    \mathcal{M}_8^\mathrm{CE}({\bf k}) - \mathcal{M}_8({\bf k}) &= 
    g^4 \Big[ 178 \omega_0^2 + 2 (2n_{\mathrm{ph}}+ 1)^2 (107 + 130 g^2 (2n_{\mathrm{ph}}+ 1) + 109 \omega_0^2) \nonumber \\ &+ \varepsilon_{\mathbf{k}} \Big(-8 (2n_{\mathrm{ph}}+ 1) \omega_0 (37 + 47 g^2 (2n_{\mathrm{ph}}+ 1) + 18 \omega_0^2) \nonumber \\ &- \varepsilon_{\mathbf{k}} \Big(-2 (2n_{\mathrm{ph}}+ 1)^2 (8 + 3 g^2 (2n_{\mathrm{ph}}+ 1)) \nonumber \\ &+ (5 + 9 (2n_{\mathrm{ph}}+ 1)^2) \omega_0^2 + 4 (2n_{\mathrm{ph}}+ 1) \varepsilon_{\mathbf{k}} (6 \omega_0 + (2n_{\mathrm{ph}}+ 1) \varepsilon_{\mathbf{k}})\Big)\Big) \Big] \\
    \mathcal{M}_9^\mathrm{CE}({\bf k}) - \mathcal{M}_9({\bf k}) &=
    -g^4 \Big[-4 (2n_{\mathrm{ph}}+ 1) \omega_0 (572 + 789 g^2 (2n_{\mathrm{ph}}+ 1) + 354 \omega_0^2) \nonumber \\ &+ \varepsilon_{\mathbf{k}} \Big(3 (2n_{\mathrm{ph}}+ 1)^2 (179 + 4 g^2 (2n_{\mathrm{ph}}+ 1) (116 + 47 g^2 (2n_{\mathrm{ph}}+ 1))) \nonumber \\ &+ 10 \left(85 + 48 (2n_{\mathrm{ph}}+ 1) (2 (2n_{\mathrm{ph}}+ 1) + g^2 (3 + 2 (2n_{\mathrm{ph}}+ 1)^2))\right) \omega_0^2 + 6 (33 + 34 (2n_{\mathrm{ph}}+ 1)^2) \omega_0^4 \nonumber \\ &+ \varepsilon_{\mathbf{k}} \Big(-2 (2n_{\mathrm{ph}}+ 1) \omega_0 (130 + 159 g^2 (2n_{\mathrm{ph}}+ 1) + 6 \omega_0^2) \nonumber \\ &+ \varepsilon_{\mathbf{k}} \Big(46 \omega_0^2 + (2n_{\mathrm{ph}}+ 1)^2 (63 + 128 g^2 (2n_{\mathrm{ph}}+ 1) + 54 \omega_0^2) \nonumber \\ &+ 2 (2n_{\mathrm{ph}}+ 1) \varepsilon_{\mathbf{k}} (8 \omega_0 + 3 (2n_{\mathrm{ph}}+ 1) \varepsilon_{\mathbf{k}})\Big)\Big)\Big)\Big]
\end{align}

\subsection{Peierls model}
\label{sec:peierls_sum_rules}

For the Peierls model, the CE sum rules read as follows:
\begin{align}
    \mathcal{M}_0^\mathrm{CE}({\bf k}) &= 1 \\
\mathcal{M}_1^\mathrm{CE}({\bf k}) &=  
    \varepsilon_{\mathbf{k}} \\
    \mathcal{M}_2^\mathrm{CE}({\bf k}) &= 
    g^2 (4 + 8 n_\mathrm{ph}) + \varepsilon_{\mathbf{k}}^2 - 2 g^2 (1 + 2 n_\mathrm{ph}) \cos 2k \\
    \mathcal{M}_3^\mathrm{CE}({\bf k}) &= 
    \varepsilon_{\mathbf{k}}^3 + 4 g^2 (\omega_0 + 2 \varepsilon_{\mathbf{k}} + 4 n_\mathrm{ph} \varepsilon_{\mathbf{k}}) - 2 g^2 (\omega_0 + 2 \varepsilon_{\mathbf{k}} + 4 n_\mathrm{ph} \varepsilon_{\mathbf{k}}) \cos 2k \\
    \mathcal{M}_4^\mathrm{CE}({\bf k}) &= 
    54 g^4 (1 + 2 n_\mathrm{ph})^2 + \varepsilon_{\mathbf{k}}^4 + 2 g^2 \bigl( (1 + 2 n_\mathrm{ph}) (3 + 2 \omega_0^2) + 4 \omega_0 \varepsilon_{\mathbf{k}} + 6 (1 + 2 n_\mathrm{ph}) \varepsilon_{\mathbf{k}}^2 \bigr) \nonumber \\
& - 2 g^2 \bigl( 24 (g + 2 g n_\mathrm{ph})^2 + (1 + 2 n_\mathrm{ph}) (2 + \omega_0^2) + 2 \omega_0 \varepsilon_{\mathbf{k}} + 3 (1 + 2 n_\mathrm{ph}) \varepsilon_{\mathbf{k}}^2 \bigr) \cos 2k \nonumber \\
& + 6 g^4 (1 + 2 n_\mathrm{ph})^2 \cos 4k \\
\mathcal{M}_5^\mathrm{CE}({\bf k}) &= 
\varepsilon_{\mathbf{k}}^5 + 90 g^4 (1 + 2 n_\mathrm{ph}) (2 \omega_0 + \varepsilon_{\mathbf{k}} + 2 n_\mathrm{ph} \varepsilon_{\mathbf{k}}) \nonumber \\
& + 2 g^2 \bigl( 2 \omega_0^3 + 4 \omega_0^2 (\varepsilon_{\mathbf{k}} + 2 n_\mathrm{ph} \varepsilon_{\mathbf{k}}) + 2 (1 + 2 n_\mathrm{ph}) \varepsilon_{\mathbf{k}} (3 + 4 \varepsilon_{\mathbf{k}}^2) + \omega_0 (9 + 6 \varepsilon_{\mathbf{k}}^2) \bigr) \nonumber \\
& - 2 g^2 \bigl( \omega_0 (6 + 80 g^2 (1 + 2 n_\mathrm{ph}) + \omega_0^2) + 2 (1 + 2 n_\mathrm{ph}) (2 + 20 g^2 (1 + 2 n_\mathrm{ph}) + \omega_0^2) \varepsilon_{\mathbf{k}} \nonumber \\
& \quad + 3 \omega_0 \varepsilon_{\mathbf{k}}^2 + 4 (1 + 2 n_\mathrm{ph}) \varepsilon_{\mathbf{k}}^3 \bigr) \cos 2k \nonumber \\
& + 10 g^4 (1 + 2 n_\mathrm{ph}) (2 \omega_0 + \varepsilon_{\mathbf{k}} + 2 n_\mathrm{ph} \varepsilon_{\mathbf{k}}) \cos 4k \\
\mathcal{M}_6^\mathrm{CE}({\bf k}) &= 
1320 g^6 (1 + 2 n_\mathrm{ph})^3 + \varepsilon_{\mathbf{k}}^6 \nonumber \\
& + 2 g^2 \bigl( 2 (1 + 2 n_\mathrm{ph}) (4 + 9 \omega_0^2 + \omega_0^4) + 2 \omega_0 (9 + 2 \omega_0^2) \varepsilon_{\mathbf{k}} + 3 (1 + 2 n_\mathrm{ph}) (3 + 2 \omega_0^2) \varepsilon_{\mathbf{k}}^2 \nonumber \\
& \quad + 8 \omega_0 \varepsilon_{\mathbf{k}}^3 + 10 (1 + 2 n_\mathrm{ph}) \varepsilon_{\mathbf{k}}^4 \bigr) \nonumber \\
& + 30 g^4 \bigl( 14 + 15 \omega_0^2 + 4 n_\mathrm{ph}(1 + n_\mathrm{ph}) (14 + 9 \omega_0^2) + 6 (1 + 2 n_\mathrm{ph}) \omega_0 \varepsilon_{\mathbf{k}} + 6 (\varepsilon_{\mathbf{k}} + 2 n_\mathrm{ph} \varepsilon_{\mathbf{k}})^2 \bigr) \nonumber \\
& - 2 g^2 \bigl( 765 g^4 (1 + 2 n_\mathrm{ph})^3 + (1 + 2 n_\mathrm{ph}) (6 + 12 \omega_0^2 + \omega_0^4) + 2 \omega_0 (6 + \omega_0^2) \varepsilon_{\mathbf{k}} \nonumber \\
& \quad + 3 (1 + 2 n_\mathrm{ph}) (2 + \omega_0^2) \varepsilon_{\mathbf{k}}^2 + 4 \omega_0 \varepsilon_{\mathbf{k}}^3 + 5 (1 + 2 n_\mathrm{ph}) \varepsilon_{\mathbf{k}}^4 \nonumber \\
& \quad + 10 g^2 \bigl( 21 + 20 \omega_0^2 + 12 n_\mathrm{ph}(1 + n_\mathrm{ph}) (7 + 4 \omega_0^2) + 8 (1 + 2 n_\mathrm{ph}) \omega_0 \varepsilon_{\mathbf{k}} + 8 (\varepsilon_{\mathbf{k}} + 2 n_\mathrm{ph} \varepsilon_{\mathbf{k}})^2 \bigr) \bigr) \cos 2k \nonumber \\
& + 10 g^4 \bigl( 6 + 36 g^2 (1 + 2 n_\mathrm{ph})^3 + 5 \omega_0^2 + 2 \bigl( 6 n_\mathrm{ph}(1 + n_\mathrm{ph}) (2 + \omega_0^2) + (1 + 2 n_\mathrm{ph}) \omega_0 \varepsilon_{\mathbf{k}} + (\varepsilon_{\mathbf{k}} + 2 n_\mathrm{ph} \varepsilon_{\mathbf{k}})^2 \bigr) \bigr) \cos 4k \nonumber \\
& - 30 g^6 (1 + 2 n_\mathrm{ph})^3 \cos 6k \\
\mathcal{M}_7^\mathrm{CE}({\bf k}) &= 
9240 g^6 (1 + 2 n_\mathrm{ph})^2 \omega_0 + \varepsilon_{\mathbf{k}}^7 \nonumber \\
& + 14 g^4 (1 + 2 n_\mathrm{ph}) \bigl( 72 \omega_0^3 + 9 \omega_0^2 (\varepsilon_{\mathbf{k}} + 2 n_\mathrm{ph} \varepsilon_{\mathbf{k}}) + 2 (1 + 2 n_\mathrm{ph}) \varepsilon_{\mathbf{k}} (7 + 9 \varepsilon_{\mathbf{k}}^2) + 4 \omega_0 (49 + 9 \varepsilon_{\mathbf{k}}^2) \bigr) \nonumber \\
& + 2 g^2 \bigl( 2 \omega_0^5 + 4 \omega_0^4 (\varepsilon_{\mathbf{k}} + 2 n_\mathrm{ph} \varepsilon_{\mathbf{k}}) + 6 \omega_0^3 (5 + \varepsilon_{\mathbf{k}}^2) + 4 (1 + 2 n_\mathrm{ph}) \omega_0^2 \varepsilon_{\mathbf{k}} (9 + 2 \varepsilon_{\mathbf{k}}^2) \nonumber \\
& \quad + \omega_0 (40 + 27 \varepsilon_{\mathbf{k}}^2 + 10 \varepsilon_{\mathbf{k}}^4) + 4 (1 + 2 n_\mathrm{ph}) \varepsilon_{\mathbf{k}} (4 + 3 (\varepsilon_{\mathbf{k}}^2 + \varepsilon_{\mathbf{k}}^4)) \bigr) \nonumber \\
& - 2 g^2 \bigl( \omega_0 (30 + 5355 g^4 (1 + 2 n_\mathrm{ph})^2 + 20 \omega_0^2 + \omega_0^4 + 28 g^2 (1 + 2 n_\mathrm{ph}) (49 + 16 \omega_0^2)) \nonumber \\
& \quad + 2 (1 + 2 n_\mathrm{ph}) (6 + 12 \omega_0^2 + \omega_0^4 + 7 g^2 (1 + 2 n_\mathrm{ph}) (7 + 4 \omega_0^2)) \varepsilon_{\mathbf{k}} \nonumber \\
& \quad + \omega_0 (224 g^2 (1 + 2 n_\mathrm{ph}) + 3 (6 + \omega_0^2)) \varepsilon_{\mathbf{k}}^2 \nonumber \\
& \quad + 4 (1 + 2 n_\mathrm{ph}) (2 + 28 g^2 (1 + 2 n_\mathrm{ph}) + \omega_0^2) \varepsilon_{\mathbf{k}}^3 + 5 \omega_0 \varepsilon_{\mathbf{k}}^4 + 6 (1 + 2 n_\mathrm{ph}) \varepsilon_{\mathbf{k}}^5 \bigr) \cos 2k \nonumber \\
& + 14 g^4 (1 + 2 n_\mathrm{ph}) \bigl( (8 \omega_0^3 + \omega_0^2 (\varepsilon_{\mathbf{k}} + 2 n_\mathrm{ph} \varepsilon_{\mathbf{k}}) + 2 (1 + 2 n_\mathrm{ph}) \varepsilon_{\mathbf{k}} (1 + \varepsilon_{\mathbf{k}}^2) \nonumber \\
& \quad + 4 \omega_0 (7 + 45 g^2 (1 + 2 n_\mathrm{ph}) + \varepsilon_{\mathbf{k}}^2)) \cos 4k - 15 g^2 (1 + 2 n_\mathrm{ph}) \omega_0 \cos 6k \bigr) 
\end{align}

\begin{align}
\mathcal{M}_8^\mathrm{CE}({\bf k}) &= 
47670 g^8 (1 + 2 n_\mathrm{ph})^4 + \varepsilon_{\mathbf{k}}^8 \nonumber \\
& + 14 g^4 \bigl( 262 + 1048 n_\mathrm{ph} + 1048 n_\mathrm{ph}^2 + 812 \omega_0^2 + 1904 n_\mathrm{ph} \omega_0^2 + 1904 n_\mathrm{ph}^2 \omega_0^2 + 153 \omega_0^4 \nonumber \\
& \quad + 324 n_\mathrm{ph} \omega_0^4 + 324 n_\mathrm{ph}^2 \omega_0^4 - 4 (1 + 2 n_\mathrm{ph}) \omega_0 (14 + 9 \omega_0^2) \varepsilon_{\mathbf{k}} \nonumber \\
& \quad + 6 (14 + 15 \omega_0^2 + 4 n_\mathrm{ph}(1 + n_\mathrm{ph}) (14 + 9 \omega_0^2)) \varepsilon_{\mathbf{k}}^2 + 36 (1 + 2 n_\mathrm{ph}) \omega_0 \varepsilon_{\mathbf{k}}^3 + 27 (1 + 2 n_\mathrm{ph})^2 \varepsilon_{\mathbf{k}}^4 \bigr) \nonumber \\
& + 2 g^2 \bigl( (1 + 2 n_\mathrm{ph}) (25 + 120 \omega_0^2 + 45 \omega_0^4 + 2 \omega_0^6) + 4 \omega_0 (20 + 15 \omega_0^2 + \omega_0^4) \varepsilon_{\mathbf{k}} \nonumber \\
& \quad + 6 (1 + 2 n_\mathrm{ph}) (4 + 9 \omega_0^2 + \omega_0^4) \varepsilon_{\mathbf{k}}^2 + 4 \omega_0 (9 + 2 \omega_0^2) \varepsilon_{\mathbf{k}}^3 \nonumber \\
& \quad + 5 (1 + 2 n_\mathrm{ph}) (3 + 2 \omega_0^2) \varepsilon_{\mathbf{k}}^4 + 12 \omega_0 \varepsilon_{\mathbf{k}}^5 + 14 (1 + 2 n_\mathrm{ph}) \varepsilon_{\mathbf{k}}^6 \bigr) \nonumber \\
& + 280 g^6 (1 + 2 n_\mathrm{ph}) \bigl( 7 (15 + 22 \omega_0^2) + 12 n_\mathrm{ph}(1 + n_\mathrm{ph}) (35 + 22 \omega_0^2) \nonumber \\
& \quad - 44 (1 + 2 n_\mathrm{ph}) \omega_0 \varepsilon_{\mathbf{k}} + 22 (\varepsilon_{\mathbf{k}} + 2 n_\mathrm{ph} \varepsilon_{\mathbf{k}})^2 \bigr) \nonumber \\
& - 2 g^2 \bigl( 31920 g^6 (1 + 2 n_\mathrm{ph})^4 + (1 + 2 n_\mathrm{ph}) (20 + 90 \omega_0^2 + 30 \omega_0^4 + \omega_0^6) \nonumber \\
& \quad + 2 \omega_0 (30 + 20 \omega_0^2 + \omega_0^4) \varepsilon_{\mathbf{k}} + 3 (1 + 2 n_\mathrm{ph}) (6 + 12 \omega_0^2 + \omega_0^4) \varepsilon_{\mathbf{k}}^2 \nonumber \\
& \quad + 4 \omega_0 (6 + \omega_0^2) \varepsilon_{\mathbf{k}}^3 + 5 (1 + 2 n_\mathrm{ph}) (2 + \omega_0^2) \varepsilon_{\mathbf{k}}^4 + 6 \omega_0 \varepsilon_{\mathbf{k}}^5 + 7 (1 + 2 n_\mathrm{ph}) \varepsilon_{\mathbf{k}}^6 \nonumber \\
& \quad + 28 g^2 \bigl( 70 + 280 n_\mathrm{ph}(1 + n_\mathrm{ph}) + 203 \omega_0^2 + 476 n_\mathrm{ph}(1 + n_\mathrm{ph}) \omega_0^2 + 2 (17 + 36 n_\mathrm{ph}(1 + n_\mathrm{ph})) \omega_0^4 \nonumber \\
& \quad - 2 (1 + 2 n_\mathrm{ph}) \omega_0 (7 + 4 \omega_0^2) \varepsilon_{\mathbf{k}} + (21 + 20 \omega_0^2 + 12 n_\mathrm{ph}(1 + n_\mathrm{ph}) (7 + 4 \omega_0^2)) \varepsilon_{\mathbf{k}}^2 \nonumber \\
& \quad + 8 (1 + 2 n_\mathrm{ph}) \omega_0 \varepsilon_{\mathbf{k}}^3 + 6 (1 + 2 n_\mathrm{ph})^2 \varepsilon_{\mathbf{k}}^4 \bigr) \nonumber \\
& \quad + 210 g^4 (1 + 2 n_\mathrm{ph}) \bigl( 86 + 119 \omega_0^2 + 4 n_\mathrm{ph}(1 + n_\mathrm{ph}) (86 + 51 \omega_0^2) \nonumber \\
& \quad - 34 (1 + 2 n_\mathrm{ph}) \omega_0 \varepsilon_{\mathbf{k}} + 17 (\varepsilon_{\mathbf{k}} + 2 n_\mathrm{ph} \varepsilon_{\mathbf{k}})^2 \bigr) \bigr) \cos 2k \nonumber \\
& + 14 g^4 \bigl( \bigl( 44 + 176 n_\mathrm{ph} + 176 n_\mathrm{ph}^2 + 1500 (g + 2 g n_\mathrm{ph})^4 + 116 \omega_0^2 + 272 n_\mathrm{ph} \omega_0^2 + 272 n_\mathrm{ph}^2 \omega_0^2 \nonumber \\
& \quad + 17 \omega_0^4 + 36 n_\mathrm{ph} \omega_0^4 + 36 n_\mathrm{ph}^2 \omega_0^4 - 4 (1 + 2 n_\mathrm{ph}) \omega_0 (2 + \omega_0^2) \varepsilon_{\mathbf{k}} \nonumber \\
& \quad + 2 (6 + 5 \omega_0^2 + 12 n_\mathrm{ph}(1 + n_\mathrm{ph}) (2 + \omega_0^2)) \varepsilon_{\mathbf{k}}^2 + 4 (1 + 2 n_\mathrm{ph}) \omega_0 \varepsilon_{\mathbf{k}}^3 + 3 (1 + 2 n_\mathrm{ph})^2 \varepsilon_{\mathbf{k}}^4 \nonumber \\
& \quad + 60 g^2 (1 + 2 n_\mathrm{ph}) \bigl( 11 + 14 \omega_0^2 + 4 n_\mathrm{ph}(1 + n_\mathrm{ph}) (11 + 6 \omega_0^2) - 4 (1 + 2 n_\mathrm{ph}) \omega_0 \varepsilon_{\mathbf{k}} \nonumber \\
& \quad + 2 (\varepsilon_{\mathbf{k}} + 2 n_\mathrm{ph} \varepsilon_{\mathbf{k}})^2 \bigr) \bigr) \cos 4k \nonumber \\
& - 10 g^2 (1 + 2 n_\mathrm{ph}) \bigl( 6 + 24 g^2 (1 + 2 n_\mathrm{ph})^3 + 7 \omega_0^2 + 12 n_\mathrm{ph}(1 + n_\mathrm{ph}) (2 + \omega_0^2) \nonumber \\
& \quad - 2 (1 + 2 n_\mathrm{ph}) \omega_0 \varepsilon_{\mathbf{k}} + (\varepsilon_{\mathbf{k}} + 2 n_\mathrm{ph} \varepsilon_{\mathbf{k}})^2 \bigr) \cos 6k + 15 (g + 2 g n_\mathrm{ph})^4 \cos 8k \bigr)
\end{align}

\begin{align} 
  \mathcal{M}_9^\mathrm{CE}({\bf k}) &= 
\varepsilon_{\mathbf{k}}^9 - 143010 g^8 (1 + 2 n_\mathrm{ph})^3 (-4 \omega_0 + \varepsilon_{\mathbf{k}} + 2 n_\mathrm{ph} \varepsilon_{\mathbf{k}}) \nonumber \\
& + 56 g^6 \bigl( 22 (137 + 468 n_\mathrm{ph}(1 + n_\mathrm{ph})) \omega_0^3 - 1056 (2 + n_\mathrm{ph}(7 + 9 n_\mathrm{ph} + 6 n_\mathrm{ph}^2)) \omega_0^2 \varepsilon_{\mathbf{k}} \nonumber \\
& \quad - 4 (1 + 2 n_\mathrm{ph})^3 \varepsilon_{\mathbf{k}} (315 + 11 \varepsilon_{\mathbf{k}}^2) + 3 (1 + 2 n_\mathrm{ph})^2 \omega_0 (1995 + 374 \varepsilon_{\mathbf{k}}^2) \bigr) \nonumber \\
& + 2 g^2 \bigl( 2 \omega_0^7 + 4 \omega_0^6 (\varepsilon_{\mathbf{k}} + 2 n_\mathrm{ph} \varepsilon_{\mathbf{k}}) + 2 (1 + 2 n_\mathrm{ph}) \omega_0^4 \varepsilon_{\mathbf{k}} (45 + 4 \varepsilon_{\mathbf{k}}^2) + \omega_0^5 (63 + 6 \varepsilon_{\mathbf{k}}^2) \nonumber \\
& \quad + 12 (1 + 2 n_\mathrm{ph}) \omega_0^2 \varepsilon_{\mathbf{k}} (20 + 6 \varepsilon_{\mathbf{k}}^2 + \varepsilon_{\mathbf{k}}^4) + 10 \omega_0^3 (28 + 9 \varepsilon_{\mathbf{k}}^2 + \varepsilon_{\mathbf{k}}^4) \nonumber \\
& \quad + 2 (1 + 2 n_\mathrm{ph}) \varepsilon_{\mathbf{k}} (25 + 16 \varepsilon_{\mathbf{k}}^2 + 9 \varepsilon_{\mathbf{k}}^4 + 8 \varepsilon_{\mathbf{k}}^6) \nonumber \\
& \quad + \omega_0 (175 + 120 \varepsilon_{\mathbf{k}}^2 + 45 \varepsilon_{\mathbf{k}}^4 + 14 \varepsilon_{\mathbf{k}}^6) \bigr) \nonumber \\
& + 6 g^4 \bigl( 738 (1 + 2 n_\mathrm{ph}) \omega_0^5 - 9 (53 + 100 n_\mathrm{ph}(1 + n_\mathrm{ph})) \omega_0^4 \varepsilon_{\mathbf{k}} \nonumber \\
& \quad + 48 (1 + 2 n_\mathrm{ph}) \omega_0^3 (133 + 12 \varepsilon_{\mathbf{k}}^2) + (1 + 2 n_\mathrm{ph})^2 \varepsilon_{\mathbf{k}} (-614 + 84 \varepsilon_{\mathbf{k}}^2 + 81 \varepsilon_{\mathbf{k}}^4) \nonumber \\
& \quad + 2 (1 + 2 n_\mathrm{ph}) \omega_0 (3058 + 756 \varepsilon_{\mathbf{k}}^2 + 81 \varepsilon_{\mathbf{k}}^4) \nonumber \\
& \quad + 6 \omega_0^2 \varepsilon_{\mathbf{k}} (-350 - 616 n_\mathrm{ph}(1 + n_\mathrm{ph}) + 9 (\varepsilon_{\mathbf{k}} + 2 n_\mathrm{ph} \varepsilon_{\mathbf{k}})^2) \bigr) \nonumber \\
& - 2 g^2 \bigl( \omega_0^7 + 2 \omega_0^6 (\varepsilon_{\mathbf{k}} + 2 n_\mathrm{ph} \varepsilon_{\mathbf{k}}) + 3 \omega_0^5 (14 + 656 g^2 (1 + 2 n_\mathrm{ph}) + \varepsilon_{\mathbf{k}}^2) \nonumber \\
& \quad + 4 \omega_0^4 \varepsilon_{\mathbf{k}} (-6 g^2 (53 + 100 n_\mathrm{ph}(1 + n_\mathrm{ph})) + (1 + 2 n_\mathrm{ph}) (15 + \varepsilon_{\mathbf{k}}^2)) \nonumber \\
& \quad + \omega_0^3 (714 g^4 (137 + 468 n_\mathrm{ph}(1 + n_\mathrm{ph})) + 48 g^2 (1 + 2 n_\mathrm{ph}) (399 + 32 \varepsilon_{\mathbf{k}}^2) + 5 (42 + 12 \varepsilon_{\mathbf{k}}^2 + \varepsilon_{\mathbf{k}}^4)) \nonumber \\
& \quad + 4 (1 + 2 n_\mathrm{ph}) \varepsilon_{\mathbf{k}} (10 - 23940 g^6 (1 + 2 n_\mathrm{ph})^3 + 6 \varepsilon_{\mathbf{k}}^2 + 3 \varepsilon_{\mathbf{k}}^4 + 2 \varepsilon_{\mathbf{k}}^6 \nonumber \\
& \quad - 21 g^4 (1 + 2 n_\mathrm{ph})^2 (516 + 17 \varepsilon_{\mathbf{k}}^2) + 3 g^2 (1 + 2 n_\mathrm{ph}) (-166 + 21 \varepsilon_{\mathbf{k}}^2 + 18 \varepsilon_{\mathbf{k}}^4)) \nonumber \\
& \quad + \omega_0 (140 + 383040 g^6 (1 + 2 n_\mathrm{ph})^3 + 90 \varepsilon_{\mathbf{k}}^2 + 30 \varepsilon_{\mathbf{k}}^4 + 7 \varepsilon_{\mathbf{k}}^6 \nonumber \\
& \quad + 126 g^4 (1 + 2 n_\mathrm{ph})^2 (1634 + 289 \varepsilon_{\mathbf{k}}^2) + 24 g^2 (1 + 2 n_\mathrm{ph}) (818 + 189 \varepsilon_{\mathbf{k}}^2 + 18 \varepsilon_{\mathbf{k}}^4)) \nonumber \\
& \quad + 6 \omega_0^2 \varepsilon_{\mathbf{k}} (-5712 g^4 (2 + n_\mathrm{ph}(7 + 9 n_\mathrm{ph} + 6 n_\mathrm{ph}^2)) + (1 + 2 n_\mathrm{ph}) (30 + 8 \varepsilon_{\mathbf{k}}^2 + \varepsilon_{\mathbf{k}}^4) \nonumber \\
& \quad + 6 g^2 (-7 (25 + 44 n_\mathrm{ph}(1 + n_\mathrm{ph})) + 4 (\varepsilon_{\mathbf{k}} + 2 n_\mathrm{ph} \varepsilon_{\mathbf{k}})^2)) \bigr) \cos 2k \nonumber \\
& + 6 g^4 \bigl( 82 (1 + 2 n_\mathrm{ph}) \omega_0^5 - (53 + 100 n_\mathrm{ph}(1 + n_\mathrm{ph})) \omega_0^4 \varepsilon_{\mathbf{k}} \nonumber \\
& \quad + 6 \omega_0^2 \varepsilon_{\mathbf{k}} (-50 - 88 n_\mathrm{ph}(1 + n_\mathrm{ph}) - 448 g^2 (2 + n_\mathrm{ph}(7 + 9 n_\mathrm{ph} + 6 n_\mathrm{ph}^2)) + (\varepsilon_{\mathbf{k}} + 2 n_\mathrm{ph} \varepsilon_{\mathbf{k}})^2) \nonumber \\
& \quad - (1 + 2 n_\mathrm{ph})^2 \varepsilon_{\mathbf{k}} (10500 g^4 (1 + 2 n_\mathrm{ph})^2 - 9 \varepsilon_{\mathbf{k}}^4 - 12 (-9 + \varepsilon_{\mathbf{k}}^2) + 112 g^2 (1 + 2 n_\mathrm{ph}) (33 + \varepsilon_{\mathbf{k}}^2)) \nonumber \\
& \quad + 8 \omega_0^3 (7 g^2 (137 + 468 n_\mathrm{ph}(1 + n_\mathrm{ph})) + 2 (1 + 2 n_\mathrm{ph}) (57 + 4 \varepsilon_{\mathbf{k}}^2)) \nonumber \\
& \quad + 6 (1 + 2 n_\mathrm{ph}) \omega_0 (172 + 7000 g^4 (1 + 2 n_\mathrm{ph})^2 + 36 \varepsilon_{\mathbf{k}}^2 + 3 \varepsilon_{\mathbf{k}}^4 + 14 g^2 (1 + 2 n_\mathrm{ph}) (209 + 34 \varepsilon_{\mathbf{k}}^2)) \bigr) \cos 4k \nonumber \\
& - 28 g^6 \bigl( (137 + 468 n_\mathrm{ph}(1 + n_\mathrm{ph})) \omega_0^3 - 48 (2 + n_\mathrm{ph}(7 + 9 n_\mathrm{ph} + 6 n_\mathrm{ph}^2)) \omega_0^2 \varepsilon_{\mathbf{k}} \nonumber \\
& \quad - 2 (1 + 2 n_\mathrm{ph})^3 \varepsilon_{\mathbf{k}} (36 + 180 g^2 (1 + 2 n_\mathrm{ph}) + \varepsilon_{\mathbf{k}}^2) \nonumber \\
& \quad + 3 (1 + 2 n_\mathrm{ph})^2 \omega_0 (114 + 480 g^2 (1 + 2 n_\mathrm{ph}) + 17 \varepsilon_{\mathbf{k}}^2) \bigr) \cos 6k \nonumber \\
& - 630 g^8 (1 + 2 n_\mathrm{ph})^3 (-4 \omega_0 + \varepsilon_{\mathbf{k}} + 2 n_\mathrm{ph} \varepsilon_{\mathbf{k}}) \cos 8k
\end{align}

\clearpage \newpage
\begin{align}
    \mathcal{M}^\mathrm{CE}_{10}({\bf k}) &= 
    2290680 g^{10} (1 + 2 n_\mathrm{ph})^5 + \varepsilon_{\mathbf{k}}^{10} \nonumber \\
& + 280 g^6 \bigl( 3 (1 + 2 n_\mathrm{ph}) (708 (1 + 2 n_\mathrm{ph})^2 + 35 (79 + 132 n_\mathrm{ph}(1 + n_\mathrm{ph})) \omega_0^2 \nonumber \\
& \quad + 11 (65 + 84 n_\mathrm{ph}(1 + n_\mathrm{ph})) \omega_0^4) - 44 \omega_0 (105 + 58 \omega_0^2 + 12 n_\mathrm{ph}(1 + n_\mathrm{ph}) (35 + 16 \omega_0^2)) \varepsilon_{\mathbf{k}} \nonumber \\
& \quad + 6 (1 + 2 n_\mathrm{ph}) (175 + 286 \omega_0^2 + 20 n_\mathrm{ph}(1 + n_\mathrm{ph}) (35 + 22 \omega_0^2)) \varepsilon_{\mathbf{k}}^2 \nonumber \\
& \quad - 396 (1 + 2 n_\mathrm{ph})^2 \omega_0 \varepsilon_{\mathbf{k}}^3 + 77 (1 + 2 n_\mathrm{ph})^3 \varepsilon_{\mathbf{k}}^4 \bigr) \nonumber \\
& + 6 g^4 \bigl( 6000 + 24000 n_\mathrm{ph} + 24000 n_\mathrm{ph}^2 + 36366 \omega_0^2 + 81800 n_\mathrm{ph} \omega_0^2 + 81800 n_\mathrm{ph}^2 \omega_0^2 \nonumber \\
& \quad + 19138 \omega_0^4 + 40040 n_\mathrm{ph} \omega_0^4 + 40040 n_\mathrm{ph}^2 \omega_0^4 + 1503 \omega_0^6 + 3060 n_\mathrm{ph} \omega_0^6 + 3060 n_\mathrm{ph}^2 \omega_0^6 \nonumber \\
& \quad - 2 (1 + 2 n_\mathrm{ph}) \omega_0 (5786 + 6356 \omega_0^2 + 819 \omega_0^4) \varepsilon_{\mathbf{k}} \nonumber \\
& \quad + 2 (1358 (1 + 2 n_\mathrm{ph})^2 + 84 (51 + 112 n_\mathrm{ph}(1 + n_\mathrm{ph})) \omega_0^2 + 9 (95 + 196 n_\mathrm{ph}(1 + n_\mathrm{ph})) \omega_0^4) \varepsilon_{\mathbf{k}}^2 \nonumber \\
& \quad - 24 (1 + 2 n_\mathrm{ph}) \omega_0 (28 + 15 \omega_0^2) \varepsilon_{\mathbf{k}}^3 \nonumber \\
& \quad + 27 (14 + 15 \omega_0^2 + 4 n_\mathrm{ph}(1 + n_\mathrm{ph}) (14 + 9 \omega_0^2)) \varepsilon_{\mathbf{k}}^4 + 162 (1 + 2 n_\mathrm{ph}) \omega_0 \varepsilon_{\mathbf{k}}^5 \nonumber \\
& \quad + 108 (1 + 2 n_\mathrm{ph})^2 \varepsilon_{\mathbf{k}}^6 \bigr) \nonumber \\
& + 2 g^2 \bigl( 2 (1 + 2 n_\mathrm{ph}) (42 + 350 \omega_0^2 + 280 \omega_0^4 + 42 \omega_0^6 + \omega_0^8) \nonumber \\
& \quad + 2 \omega_0 (175 + 280 \omega_0^2 + 63 \omega_0^4 + 2 \omega_0^6) \varepsilon_{\mathbf{k}} \nonumber \\
& \quad + 3 (1 + 2 n_\mathrm{ph}) (25 + 120 \omega_0^2 + 45 \omega_0^4 + 2 \omega_0^6) \varepsilon_{\mathbf{k}}^2 \nonumber \\
& \quad + 8 \omega_0 (20 + 15 \omega_0^2 + \omega_0^4) \varepsilon_{\mathbf{k}}^3 + 10 (1 + 2 n_\mathrm{ph}) (4 + 9 \omega_0^2 + \omega_0^4) \varepsilon_{\mathbf{k}}^4 \nonumber \\
& \quad + 6 \omega_0 (9 + 2 \omega_0^2) \varepsilon_{\mathbf{k}}^5 + 7 (1 + 2 n_\mathrm{ph}) (3 + 2 \omega_0^2) \varepsilon_{\mathbf{k}}^6 + 16 \omega_0 \varepsilon_{\mathbf{k}}^7 \nonumber \\
& \quad + 18 (1 + 2 n_\mathrm{ph}) \varepsilon_{\mathbf{k}}^8 \bigr) \nonumber \\
& + 3150 g^8 (1 + 2 n_\mathrm{ph})^2 \bigl( 732 (1 + 2 n_\mathrm{ph})^2 + 454 (3 + 4 n_\mathrm{ph}(1 + n_\mathrm{ph})) \omega_0^2 \nonumber \\
& \quad - 908 (1 + 2 n_\mathrm{ph}) \omega_0 \varepsilon_{\mathbf{k}} + 227 (\varepsilon_{\mathbf{k}} + 2 n_\mathrm{ph} \varepsilon_{\mathbf{k}})^2 \bigr) \nonumber \\
& - 2 g^2 \bigl( 1672650 g^8 (1 + 2 n_\mathrm{ph})^5 + (1 + 2 n_\mathrm{ph}) (70 + 560 \omega_0^2 + 420 \omega_0^4 + 56 \omega_0^6 + \omega_0^8) \nonumber \\
& \quad + 2 \omega_0 (140 + 210 \omega_0^2 + 42 \omega_0^4 + \omega_0^6) \varepsilon_{\mathbf{k}} \nonumber \\
& \quad + 3 (1 + 2 n_\mathrm{ph}) (20 + 90 \omega_0^2 + 30 \omega_0^4 + \omega_0^6) \varepsilon_{\mathbf{k}}^2 \nonumber \\
& \quad + 4 \omega_0 (30 + 20 \omega_0^2 + \omega_0^4) \varepsilon_{\mathbf{k}}^3 + 5 (1 + 2 n_\mathrm{ph}) (6 + 12 \omega_0^2 + \omega_0^4) \varepsilon_{\mathbf{k}}^4 \nonumber \\
& \quad + 6 \omega_0 (6 + \omega_0^2) \varepsilon_{\mathbf{k}}^5 + 7 (1 + 2 n_\mathrm{ph}) (2 + \omega_0^2) \varepsilon_{\mathbf{k}}^6 + 8 \omega_0 \varepsilon_{\mathbf{k}}^7 \nonumber \\
& \quad + 9 (1 + 2 n_\mathrm{ph}) \varepsilon_{\mathbf{k}}^8 \nonumber \\
& \quad + 105 g^4 \bigl( (1 + 2 n_\mathrm{ph}) (3612 (1 + 2 n_\mathrm{ph})^2 + 172 (79 + 132 n_\mathrm{ph}(1 + n_\mathrm{ph})) \omega_0^2 \nonumber \\
& \quad + 51 (65 + 84 n_\mathrm{ph}(1 + n_\mathrm{ph})) \omega_0^4) - 8 \omega_0 (946 (1 + 2 n_\mathrm{ph})^2 \nonumber \\
& \quad + 17 (29 + 96 n_\mathrm{ph}(1 + n_\mathrm{ph})) \omega_0^2) \varepsilon_{\mathbf{k}} \nonumber \\
& \quad + 4 (1 + 2 n_\mathrm{ph}) (430 + 663 \omega_0^2 + 20 n_\mathrm{ph}(1 + n_\mathrm{ph}) (86 + 51 \omega_0^2)) \varepsilon_{\mathbf{k}}^2 \nonumber \\
& \quad - 612 (1 + 2 n_\mathrm{ph})^2 \omega_0 \varepsilon_{\mathbf{k}}^3 + 119 (1 + 2 n_\mathrm{ph})^3 \varepsilon_{\mathbf{k}}^4 \bigr) \nonumber \\
& \quad + 6 g^2 \bigl( 3355 + 13420 n_\mathrm{ph} + 13420 n_\mathrm{ph}^2 + 19476 \omega_0^2 + 43760 n_\mathrm{ph} \omega_0^2 + 43760 n_\mathrm{ph}^2 \omega_0^2 \nonumber \\
& \quad + 9569 \omega_0^4 + 20020 n_\mathrm{ph} \omega_0^4 + 20020 n_\mathrm{ph}^2 \omega_0^4 + 668 \omega_0^6 + 1360 n_\mathrm{ph} \omega_0^6 + 1360 n_\mathrm{ph}^2 \omega_0^6 \nonumber \\
& \quad - 4 (1 + 2 n_\mathrm{ph}) \omega_0 (1558 + 1589 \omega_0^2 + 182 \omega_0^4) \varepsilon_{\mathbf{k}} \nonumber \\
& \quad + 4 (364 (1 + 2 n_\mathrm{ph})^2 + 21 (51 + 112 n_\mathrm{ph}(1 + n_\mathrm{ph})) \omega_0^2 + 2 (95 + 196 n_\mathrm{ph}(1 + n_\mathrm{ph})) \omega_0^4) \varepsilon_{\mathbf{k}}^2 \nonumber \\
& \quad - 16 (1 + 2 n_\mathrm{ph}) \omega_0 (21 + 10 \omega_0^2) \varepsilon_{\mathbf{k}}^3 \nonumber \\
& \quad + 9 (21 + 20 \omega_0^2 + 12 n_\mathrm{ph}(1 + n_\mathrm{ph}) (7 + 4 \omega_0^2)) \varepsilon_{\mathbf{k}}^4 + 72 (1 + 2 n_\mathrm{ph}) \omega_0 \varepsilon_{\mathbf{k}}^5 \nonumber \\
& \quad + 48 (1 + 2 n_\mathrm{ph})^2 \varepsilon_{\mathbf{k}}^6 \bigr) \nonumber \\
& \quad + 6300 g^6 (1 + 2 n_\mathrm{ph})^2 \bigl( 253 + 456 \omega_0^2 + 4 n_\mathrm{ph}(1 + n_\mathrm{ph}) (253 + 152 \omega_0^2) \nonumber \\
& \quad - 304 (1 + 2 n_\mathrm{ph}) \omega_0 \varepsilon_{\mathbf{k}} + 76 (\varepsilon_{\mathbf{k}} + 2 n_\mathrm{ph} \varepsilon_{\mathbf{k}})^2 \bigr) \bigr) \cos 2k \nonumber \\
& + 6 g^4 \bigl( 1140 + 4560 n_\mathrm{ph} + 4560 n_\mathrm{ph}^2 + 226800 g^6 (1 + 2 n_\mathrm{ph})^5 + 6162 \omega_0^2 \nonumber \\
& \quad + 13800 n_\mathrm{ph} \omega_0^2 + 13800 n_\mathrm{ph}^2 \omega_0^2 + 2734 \omega_0^4 + 5720 n_\mathrm{ph} \omega_0^4 + 5720 n_\mathrm{ph}^2 \omega_0^4 \nonumber \\
& \quad + 167 \omega_0^6 + 340 n_\mathrm{ph} \omega_0^6 + 340 n_\mathrm{ph}^2 \omega_0^6 - 2 (1 + 2 n_\mathrm{ph}) \omega_0 (1002 + 908 \omega_0^2 + 91 \omega_0^4) \varepsilon_{\mathbf{k}} \nonumber \\
& \quad + 2 (231 (1 + 2 n_\mathrm{ph})^2 + 12 (51 + 112 n_\mathrm{ph}(1 + n_\mathrm{ph})) \omega_0^2 + (95 + 196 n_\mathrm{ph}(1 + n_\mathrm{ph})) \omega_0^4) \varepsilon_{\mathbf{k}}^2 \nonumber \\
& \quad - 8 (1 + 2 n_\mathrm{ph}) \omega_0 (12 + 5 \omega_0^2) \varepsilon_{\mathbf{k}}^3 + 9 (6 + 5 \omega_0^2 + 12 n_\mathrm{ph}(1 + n_\mathrm{ph}) (2 + \omega_0^2)) \varepsilon_{\mathbf{k}}^4 \nonumber \\
& \quad + 18 (1 + 2 n_\mathrm{ph}) \omega_0 \varepsilon_{\mathbf{k}}^5 + 12 (1 + 2 n_\mathrm{ph})^2 \varepsilon_{\mathbf{k}}^6 \nonumber \\
& \quad + 140 g^2 \bigl( (1 + 2 n_\mathrm{ph}) (246 (1 + 2 n_\mathrm{ph})^2 + 11 (79 + 132 n_\mathrm{ph}(1 + n_\mathrm{ph})) \omega_0^2 \nonumber \\
& \quad + 3 (65 + 84 n_\mathrm{ph}(1 + n_\mathrm{ph})) \omega_0^4) - 4 \omega_0 (121 + 58 \omega_0^2 + 4 n_\mathrm{ph}(1 + n_\mathrm{ph}) (121 + 48 \omega_0^2)) \varepsilon_{\mathbf{k}} \nonumber \\
& \quad + 2 (1 + 2 n_\mathrm{ph}) (55 + 78 \omega_0^2 + 20 n_\mathrm{ph}(1 + n_\mathrm{ph}) (11 + 6 \omega_0^2)) \varepsilon_{\mathbf{k}}^2 \nonumber \\
& \quad - 36 (1 + 2 n_\mathrm{ph})^2 \omega_0 \varepsilon_{\mathbf{k}}^3 + 7 (1 + 2 n_\mathrm{ph})^3 \varepsilon_{\mathbf{k}}^4 \bigr) \nonumber \\
& \quad + 2100 g^4 (1 + 2 n_\mathrm{ph})^2 \bigl( 88 (1 + 2 n_\mathrm{ph})^2 + 50 (3 + 4 n_\mathrm{ph}(1 + n_\mathrm{ph})) \omega_0^2 \nonumber \\
& \quad - 100 (1 + 2 n_\mathrm{ph}) \omega_0 \varepsilon_{\mathbf{k}} + 25 (\varepsilon_{\mathbf{k}} + 2 n_\mathrm{ph} \varepsilon_{\mathbf{k}})^2 \bigr) \bigr) \cos 4k \nonumber \\
& - 70 g^6 \bigl( 4455 g^4 (1 + 2 n_\mathrm{ph})^5 + 3 (1 + 2 n_\mathrm{ph}) (96 (1 + 2 n_\mathrm{ph})^2 \nonumber \\
& \quad + 4 (79 + 132 n_\mathrm{ph}(1 + n_\mathrm{ph})) \omega_0^2 + (65 + 84 n_\mathrm{ph}(1 + n_\mathrm{ph})) \omega_0^4) \nonumber \\
& \quad - 8 \omega_0 (66 + 29 \omega_0^2 + 24 n_\mathrm{ph}(1 + n_\mathrm{ph}) (11 + 4 \omega_0^2)) \varepsilon_{\mathbf{k}} \nonumber \\
& \quad + 12 (1 + 2 n_\mathrm{ph}) (10 + 13 \omega_0^2 + 20 n_\mathrm{ph}(1 + n_\mathrm{ph}) (2 + \omega_0^2)) \varepsilon_{\mathbf{k}}^2 \nonumber \\
& \quad - 36 (1 + 2 n_\mathrm{ph})^2 \omega_0 \varepsilon_{\mathbf{k}}^3 + 7 (1 + 2 n_\mathrm{ph})^3 \varepsilon_{\mathbf{k}}^4 \nonumber \\
& \quad + 180 (g + 2 g n_\mathrm{ph})^2 \bigl( 3 (5 + 8 \omega_0^2) + 4 (n_\mathrm{ph}(1 + n_\mathrm{ph}) (15 + 8 \omega_0^2) \nonumber \\
& \quad - 4 (1 + 2 n_\mathrm{ph}) \omega_0 \varepsilon_{\mathbf{k}} + (\varepsilon_{\mathbf{k}} + 2 n_\mathrm{ph} \varepsilon_{\mathbf{k}})^2 \bigr) \bigr) \cos 6k \nonumber \\
& + 3150 g^8 (1 + 2 n_\mathrm{ph})^2 \bigl( 4 + 12 g^2 (1 + 2 n_\mathrm{ph})^3 + 6 \omega_0^2 + 8 n_\mathrm{ph}(1 + n_\mathrm{ph}) (2 + \omega_0^2) \nonumber \\
& \quad - 4 (1 + 2 n_\mathrm{ph}) \omega_0 \varepsilon_{\mathbf{k}} + (\varepsilon_{\mathbf{k}} + 2 n_\mathrm{ph} \varepsilon_{\mathbf{k}})^2 \bigr) \cos 8k \nonumber \\
& - 1890 g^{10} (1 + 2 n_\mathrm{ph})^5 \cos 10k
\end{align}

\section{Spectral functions in the Fr\"ohlich model}

In the main text, the results for the Fr\"ohlich model were presented in Sec.~IV~C, where a few CE and SCMA spectral functions were shown in Fig.~10 using a logarithmic scale. Here, we extend those results by presenting spectral functions at room temperature ($T = 300,\mathrm{K}$) for electron–phonon coupling strengths $\alpha$ representative of the conduction bands in the three polar semiconductors discussed in the main text: GaAs, GaN, and ZnO.

\begin{figure}[t!]
  \centering
  \includegraphics[width=0.95\linewidth]{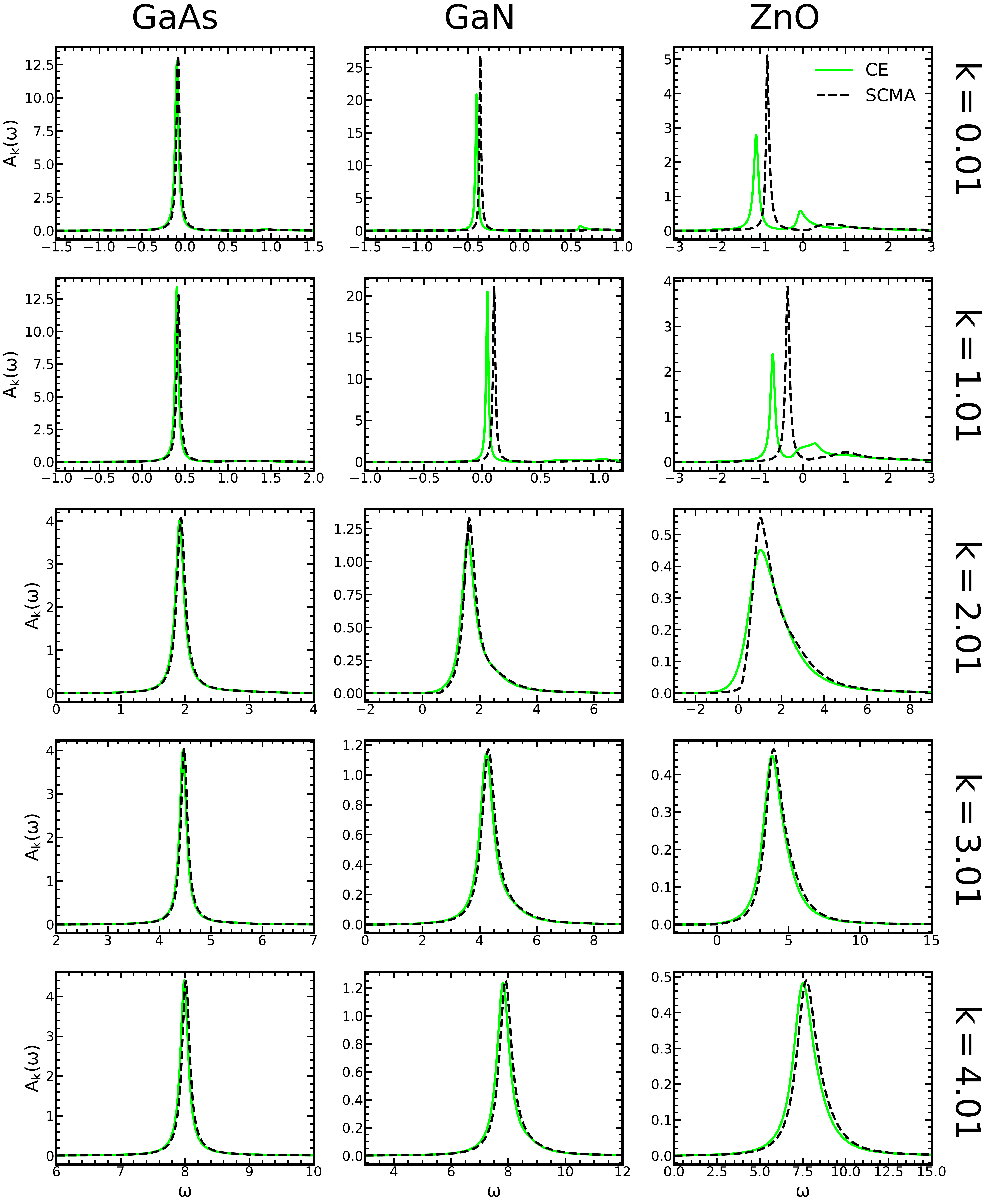}
  \caption{CE and SCMA spectral functions for the Fr\"ohlich model with $\alpha$ corresponding to the conduction bands of three polar semiconductors: GaAs, GaN, ZnO. Results are shown for five different momenta: $k=0.01$, $k=1.01$, $k=2.01$, $k=3.01$, $k=4.01$.}
  \label{fig:specf_frolih}
\end{figure}

\end{document}